\documentclass[aps,prx,twocolumn,footinbib,floatfix,superscriptaddress]{revtex4-2}

\usepackage{upgreek}

\usepackage{graphicx}
\usepackage{indentfirst}
\usepackage{braket}
\usepackage{float}
\usepackage{amsmath}
\usepackage{physics}
\usepackage{amssymb}
\usepackage{CJK}
\usepackage{esint}
\usepackage{color}
\usepackage[T1]{fontenc}
\usepackage{subfigure}
\usepackage{amsfonts}
\usepackage{footmisc}
\usepackage{scrextend}
\usepackage{multirow}
\usepackage[outdir=./]{epstopdf}
\usepackage{svg}

\usepackage{float}
\makeatletter
\let\newfloat\newfloat@ltx
\makeatother
\usepackage{algorithm}
\usepackage{algpseudocode}

\usepackage[english]{babel}
\usepackage{url}
\usepackage{comment}
\usepackage{array}
\usepackage{subfiles}
\usepackage{tikz}
\usetikzlibrary{shapes,arrows}
\usepackage{mathrsfs}
\usepackage[mathscr]{eucal}
\usepackage[hyperfootnotes=false]{hyperref}
\usepackage{cleveref}
\usepackage{stmaryrd}
\definecolor{darkblue}{rgb}{0,0,0.5}
\hypersetup{
colorlinks=true,
linkcolor=black,
filecolor=blue,
citecolor=darkblue,  
urlcolor=black,
}

\newtheorem{theorem}{Theorem}

\newtheorem{lemma}{Lemma}

\usepackage{trimclip}

\makeatletter
\DeclareRobustCommand{\shortto}{%
  \mathrel{\mathpalette\short@to\relax}%
}

\newcommand{\short@to}[2]{%
  \mkern2mu
  \clipbox{{.5\width} 0 0 0}{$\m@th#1\vphantom{+}{\shortrightarrow}$}%
  }
\makeatother

\newenvironment{proof}[1][Proof]{\noindent\textbf{#1.} }{\ \rule{0.5em}{0.5em}}

\newcommand\argmin{\mathop{\mathrm{argmin}}}

\usepackage{pict2e,picture,graphicx}

\makeatletter
\DeclareRobustCommand{\Arrow}[1][]{%
\check@mathfonts
\if\relax\detokenize{#1}\relax
\settowidth{\dimen@}{$\m@th\rightarrow$}%
\else
\setlength{\dimen@}{#1}%
\fi
\sbox\z@{\usefont{U}{lasy}{m}{n}\symbol{41}}%
\begin{picture}(\dimen@,\ht\z@)
\roundcap
\put(\dimexpr\dimen@-.7\wd\z@,0){\usebox\z@}
\put(0,\fontdimen22\textfont2){\line(1,0){\dimen@}}
\end{picture}%
}
\makeatother
\newcommand{\veryshortrightarrow}{\hspace{.2mm}\scalebox{.8}{\Arrow[.1cm]}\hspace{.2mm}}

\def\be{\begin{equation}}
\def\ee{\end{equation}}
\def\ba{\begin{eqnarray}}
\def\ea{\end{eqnarray}}
\def\bal{\begin{equation}\begin{aligned}}
\def\eal{\end{aligned}\end{equation}}

\def\bp{\begin{pmatrix}}
\def\ep{\end{pmatrix}}

\usepackage{bm}

\def\c2d{\rm C\veryshortrightarrow D}

\newcommand{\calI}{{\cal I}}

\newcommand{\calM}{{\cal M}}
\newcommand{\calN}{{\cal N}}

\newcommand{\calR}{{\cal R}}

\newcommand{\1}{^{(1)}}

\newcommand{\state}[1]{\ketbra{#1}{#1}}

\newcommand{\QZ}[1]{{{\textcolor{black}{#1}}}}

\newcommand{\JHS}[1]{{{\textcolor{green}{#1}}}}
\newcommand{\jhs}[1]{{{\textcolor{black}{#1}}}}
\usepackage[normalem]{ulem}

\begin{document}

\title{
Optimal entanglement-assisted electromagnetic sensing and communication in the presence of noise
}

%\author[1]{Haowei Shi}
\author{Haowei Shi}
\affiliation{Ming Hsieh Department of Electrical and Computer Engineering, University of Southern California, Los
Angeles, California 90089, USA
}

\author{Bingzhi Zhang}
\affiliation{Ming Hsieh Department of Electrical and Computer Engineering, University of Southern California, Los
Angeles, California 90089, USA
}
\affiliation{
Department of Physics and Astronomy, University of Southern California, Los
Angeles, California 90089, USA
}

\author{Jeffrey H. Shapiro}
\affiliation{Research Laboratory of Electronics, Massachusetts Institute of Technology, Cambridge, Massachusetts 02139, USA}

\author{Zheshen Zhang}
\affiliation{Department of Electrical Engineering and Computer Science,
University of Michigan, Ann Arbor, MI 48109, USA}

\author{Quntao Zhuang}
\email{qzhuang@usc.edu}
\affiliation{Ming Hsieh Department of Electrical and Computer Engineering, University of Southern California, Los
Angeles, California 90089, USA
}
\affiliation{
Department of Physics and Astronomy, University of Southern California, Los
Angeles, California 90089, USA
}

\begin{abstract}

High time-bandwidth product signal and idler pulses comprised of independent identically distributed two-mode squeezed vacuum (TMSV) states are readily produced by spontaneous parametric downconversion.  These pulses are virtually unique among entangled states in that they offer quantum performance advantages---over their best classical-state competitors---in scenarios whose loss and noise break their initial entanglement.   Broadband TMSV states' quantum advantage derives from its signal and idler having a strongly nonclassical phase-sensitive cross correlation, which leads to information bearing signatures in lossy, noisy scenarios stronger than what can be obtained from classical-state systems of the same transmitted energy.  Previous broadband TMSV receiver architectures focused on converting phase-sensitive cross correlation into phase-insensitive cross correlation, which can be measured in second-order interference.  In general, however, these receivers fail to deliver broadband TMSV states' full quantum advantage, even if they are implemented with ideal equipment.  This paper introduces the correlation-to-displacement receiver---a new architecture comprised of a correlation-to-displacement converter, a programmable mode selector, and a coherent-state information extractor---that can be configured to achieve quantum optimal performance in known sensing and communication protocols for which broadband TMSV provides quantum advantage that is robust against entanglement-breaking loss and noise.  
\end{abstract}

\date{\today}

\maketitle

\section{Introduction}

High time-bandwidth ($TW$) product signal and idler pulses whose quadrature components are maximally entangled, viz., pulses comprised of a large number of independent identically distributed (iid) two-mode squeezed vacuum (TMSV) states, are readily generated by pulse carving the signal and idler outputs from a continuous-wave (cw) pumped spontaneous parametric downconverter (SPDC)~\cite{zhang2013,hao2021entanglement,hao2022demonstration}.  These pulses are virtually unique among entangled states in that they offer optimal quantum advantages~\cite{nair2020fundamental,shi2020practical} over their best classical-state competitors in a variety of scenarios---including quantum illumination (QI) target detection~\cite{tan2008quantum}, entanglement-assisted (EA) phase estimation~\cite{shi2020practical}, EA classical communication~\cite{Bennett2002,shi2020practical}, and QI ranging~\cite{zhuang2020entanglement,zhuang2022ultimate}---whose loss and noise break their initial entanglement.  

Quantum illumination (QI) for detecting a weakly-reflecting target embedded in high-brightness background radiation was the first demonstrated example of broadband TMSV-enabled quantum advantage.  By evaluating information-theoretical limits, Tan~\emph{et al.}~\cite{tan2008quantum} showed that a high $TW$-product TMSV source offered a 6\,dB advantage in signal-to-noise ratio (SNR) over a coherent-state sensor of the same transmitted energy but failed to offer a receiver architecture capable of any quantum advantage. Subsequent optical~\cite{Zhang2015} and microwave~\cite{assouly2022demonstration} table-top experiments, however, used parametric amplifier (PA) receivers~\cite{Guha2009} to obtain $\sim$20\% SNR gains. Both the PA receiver and its closely related phase-conjugate (PC) receiver~\cite{Guha2009} only enable QI to achieve a sub-optimal 3\,dB quantum advantage; receiver nonidealities precluded the foregoing experiments from realizing that performance.
The same receiver design problem also haunts broadband TMSV's advantage in EA classical communication, where information-theoretical results promise a capacity advantage that grows without bound with decreasing signal brightness~\cite{Bennett2002} while an experimental demonstration only produced a $\sim16\%$ constant-rate advantage~\cite{hao2021entanglement}, due to the sub-optimality of PA and PC receivers~\cite{shi2020practical}.

In the entanglement-breaking scenarios considered to date, broadband TMSV's quantum advantage derives from the initial nonclassical phase-sensitive cross correlation between its transmitter's signal and idler leading to an information bearing signature in the phase-sensitive cross correlation between the returned radiation and the retained idler that is stronger than the signature preparable with classical resources of the same transmitted energy.  Phase-sensitive cross correlation cannot be measured in second-order interference~\cite{Shapiro2020}, so PA and PC receivers achieve their quantum advantage by converting phase-sensitive cross correlation into phase-insensitive cross correlation, which can be measured in second-order interference.   A later proposal, the feed-forward sum-frequency generation (FF-SFG) receiver~\cite{zhuang2017optimum}, was predicted to achieve QI target detection's full 6\,dB SNR advantage, but did so in a highly impractical manner \QZ{as it requires unit efficiency single-photon level SFG processes}.  Nevertheless, FF-SFG reception  introduced the kernel, converting correlation to coherence, by which a \QZ{different and more} practical architecture, namely the correlation-to-displacement (${\rm C}\to{\rm D}$) receiver proposed herein, can be realized.  Moreover, as we will show, ${\rm C}\to{\rm D}$ reception can be configured to provide quantum optimal performance in various known sensing and communication protocols for which broadband TMSV provides quantum advantage that is robust against entanglement-breaking loss and noise.   Furthermore, again as we will show, ${\rm C}\to{\rm D}$  reception enables exact performance analyses of these EA sensing and communication protocols. These analyses extend beyond the asymptotic operating regime---low-brightness TMSV, very lossy propagation, and high-brightness noise as treated in Refs.~\cite{Audenaert2007,Pirandola2008,nussbaum2011asymptotic,zhuang2017optimum,li2016discriminating}---thus permitting determination of the full range of scenario parameters in which quantum advantage exists.  The versatility of  ${\rm C}\to{\rm D}$ conversion as a theoretical tool is further demonstrated by its use, in this paper, to prove that QI's 6\,dB quantum advantage in error-probability exponent extends to channel pattern classification for arbitrary thermal-loss channels. 

The remainder of the paper is organized as follows.  Section~\ref{ChannelModel} describes the phase-shift thermal-loss channel that will be the entanglement-breaking presence in the broadband TMSV-enabled EA protocols introduced in Sec.~\ref{Protocols}.  Section~\ref{CtoD} describes the principles of ${\rm C}\to{\rm D}$  reception and shows how they are fulfilled by its ${\rm C}\to{\rm D}$ converter, programmable mode selector, and coherent-state information extractor.  Section~\ref{Performance} presents the ultimate performance limits for ${\rm C}\to{\rm D}$ reception in QI target detection, EA phase estimation, and EA classical communication under the assumption that noise from the undisplaced modes can be neglected.  Section~\ref{ExtractorDesign} describes the coherent-state information extractors for the preceding EA applications, and compares their performance to the limits established in Sec.~\ref{Performance}.  Section~\ref{SelectorDesign} is devoted to possible means for implementing ${\rm C}\to{\rm D}$ reception's programmable mode selector.  Section~\ref{Composite} extends the purview of ${\rm C}\to{\rm D}$ reception to quantum channel pattern classification, in which the information being sought is distributed across an array of sub-channels. Finally, Sec.~\ref{Discussion} provides concluding discussion to place our work in context. \QZ{Before proceeding, however, we direct the reader to Table~\ref{tab:symbols}'s glossary of the paper's principle acronyms and symbols, which can be referred to, as needed.}  

\section{Channel Model} 
\label{ChannelModel}

\begin{table}
\begin{minipage}[t]{\linewidth}
    \renewcommand{\arraystretch}{1.4}
    \centering
    % \caption{}
    \begin{tabular}{c c}
    \hline\hline
    \qquad Acronyms \qquad  ~ & Description\\
    \hline
    TW & Time-bandwidth\\
    TMSV & Two-mode squeezed vacuum\\
    QI & Quantum illumination\\
    SPDC & Spontaneous parametric downconverter \\
    EA & Entanglement-assisted \\
    SNR & Signal-to-noise ratio \\
    PA & Parametric amplifier\\
    PC & Phase conjugate \\
    FF-SFG & Feed-forward sum-frequency generation \\
    C$\to$D & Correlation-to-displacement   \\
    QCB & Quantum Chernoff bound \\
    PSK & Phase-shift keying \vspace{0.3em}\\
    LOCC &\shortstack{Local operations plus\\ classical communication} \\
    POVM &Positive operator-valued measurement \\
    HSW & Holevo-Schumacher-Westmoreland
    \\
    QCRB & Quantum Cram\'{e}r-Rao bound
    \\
    QFI & Quantum Fisher information
    \\
    PPM & Pulse-position modulation\\
    \end{tabular}
\end{minipage}%
\vspace{1em}
\begin{minipage}[t]{\linewidth}
    \renewcommand{\arraystretch}{1.4}
    \centering
    % \caption{}
    \begin{tabular}{c c}
    \hline\hline
    \qquad Symbols \qquad ~ & Description\\
    \hline
    $S,R$ &  Signal, returned signal\\
    $I,B$ &  Idler, background noise\\
    $\hat{E}_X$ & Field operator for system $X$    \vspace{0.5em}\\
    $S^{(\rm pi)}_{KK}(\cdot)$ & \shortstack{Phase-insensitive\\ spectrum for system\QZ{ $K$}, Eq.~\eqref{SKK_def}}    \vspace{0.3em} \\
    $S^{(\rm ps)}_{SI}(\cdot)$ & \shortstack{Phase-sensitive cross\\ spectrum between \QZ{systems} $S$ and $I$, Eq.~\eqref{SSI_def}}\vspace{0.3em}\\
    $\Phi_{\kappa,\theta}$ & \shortstack{Bosonic quantum channel of \\transmissivity $\kappa$ and phase shift $\theta$, Eq.~\eqref{ErChannelModel}} \vspace{0.3em}\\
    $N_S$& Signal brightness\\
    $N_B$& Background-noise \QZ{brightness}\\
    $M$ & Time-bandwidth product, \QZ{$M:=TW$}\\
    $C_p\equiv \sqrt{\kappa N_S(N_S+1)}$ & Return-idler cross-correlation amplitude \\
    $N_{I\mid R} \equiv \frac{N_S(N_B+1-\kappa)}{\kappa N_S + N_B +1}$ & \QZ{Idler's conditional noise brightness}, Eq.~\eqref{NIR_def}\\
    $N_R$ & \QZ{Heterodyne output's photon number} (Eq.~\eqref{N_R})\\
    $\zeta \equiv  \frac{\sqrt{\kappa N_S(N_S+1)}}{\kappa N_S + N_B+1}$ & Constant \\
    $\Pr(e)$& Error probability\\
    $r_{\rm CS}$& Error exponent of coherent-state protocol\\
     $r_{\rm C\to D}$& Error exponent of $\rm C\to D$ protocol\\
    $\calI_F$ & Quantum Fisher information\\
      $C_E$  & EA capacity\\
      $C_{\rm HSW}$& HSW capacity\\
      $\chi_{{\rm C}\to {\rm D}}$& Holevo information of $\rm C\to D$ protocol\\
         \hline\hline
    \end{tabular}
\end{minipage}
    \caption{\QZ{Glossary of the paper's principal acronyms and symbols.}}
    \label{tab:symbols}
\end{table}

The EA protocols to be considered all use the following transmitter.  A cw-pumped SPDC source produces signal ($S$) and idler ($I$)  outputs with center frequencies $\omega_S$ and $\omega_I$, and baseband $\sqrt{\mbox{photons/s}}$-units field operators $\hat{E}_S(t)$ and $\hat{E}_I(t)$ that are in a zero-mean jointly-Gaussian state which is fully characterized by its phase-insensitive (fluorescence) spectra and its phase-sensitive cross spectrum~\footnote{In both quantum and classical contexts we will use $\langle \cdot\rangle$ to denote unconditional expectation and $\mathbb{E}(\cdot \mid \cdot)$ to denote conditional expectation.}, viz.,
\begin{align}
\label{SKK_def}
S^{(\rm pi)}_{KK}(\omega) &\equiv \int\!{\rm d}\tau\,\langle \hat{E}_K^\dagger(t+\tau)\hat{E}_K(t)\rangle e^{-i\omega\tau}\nonumber\\
& = 
\left\{\begin{array}{ll}
N_S,& \mbox{for $|\omega|/2\pi \le W/2,$}\\[.05in]
0, & \mbox{otherwise}
\end{array}\right.\mbox{ for $K = S,I$},
\end{align}
and
\begin{align}
\label{SSI_def}
S^{(\rm ps)}_{SI}(\omega) &\equiv \int\!{\rm d}\tau\,\langle \hat{E}_S(t+\tau)\hat{E}_I(t)\rangle e^{-i\omega\tau}\nonumber\\
& = 
\left\{\begin{array}{ll}
\sqrt{N_S(N_S+1)},& \mbox{for $|\omega|/2\pi \le W/2,$}\\[.05in]
0, & \mbox{otherwise}.
\end{array}\right.
\end{align}
Because $|S^{(\rm ps)}_{SI}(\omega)|$'s classical limit is $\sqrt{S_{SS}^{(\rm pi)}(\omega)S_{II}^{(\rm pi)}(\omega)}$~\cite{Shapiro2020}, and cw-pumped SPDC sources produce low-brightness outputs, i.e., $N_S \ll 1$, the joint signal-idler state is highly nonclassical.  Furthermore, because of that low brightness, the protocols introduced below carve pulses of duration $T \gg 1/W$ from the signal and idler so that their total average photon number, $MN_S$, suffices to produce acceptable performance. Here $M \equiv TW \gg 1$ is the their time-bandwidth product. Taking those pulses to occupy the time interval $t\in \mathcal{T}_0$, and using $M\gg 1$ with $M$ assumed to be an odd integer, we have the mode expansions for the signal and idler,
\begin{equation}
\hat{E}_K(t) = \sum_{m=-\infty}^{\infty}\hat{a}_{K_m}\frac{e^{-i2\pi mt/T}}{\sqrt{T}},\mbox{ for $t\in \mathcal{T}_0$ and $K = S,I$},
\label{SImodes}
\end{equation}
which will prove useful in Sec.~\ref{CtoD}.  
Here, the excited mode pairs, $\{(\hat{a}_{S_m},\hat{a}_{I_m}): m= -(M-1)/2,\QZ{-(M-3)/2},\ldots,(M-1)/2\}$ are in iid, TMSV states with $\langle \hat{a}_{K_m}^\dagger\hat{a}_{K_m}\rangle = N_S$, for $K= S,I$, and $\langle \hat{a}_{S_m}\hat{a}_{I_m}\rangle = \sqrt{N_S(N_S+1)}$.  

For the EA protocols we will consider the signal is transmitted through a phase-shift thermal-loss channel $\Phi_{\kappa,\theta}$ that produces returned radiation whose baseband field operator is
\begin{equation}
\hat{E}_R(t) = \sqrt{\kappa}\,e^{i\theta}\hat{E}_S(t-\tau) + \sqrt{1-\kappa}\,\hat{E}_B(t).
\label{ErChannelModel}
\end{equation}
In this expression: $\tau$ is the channel's propagation delay, which we will presume to be known so that $\tau=0$ will be used for convenience in what follows; $\kappa<1$ is the channel's transmissivity; $\theta$ is its phase shift; 
$\hat{E}_B(t)$ is a background-noise field operator which is in a bandwidth-$W$ thermal state with brightness $N_B/(1-\kappa)$, i.e., a zero-mean circulo-symmetric Gaussian state characterized by the fluorescence spectrum
\begin{equation}
S^{(\rm pi)}_{BB}(\omega) = \left\{\begin{array}{ll}
N_B/(1-\kappa),& \mbox{for $|\omega|/2\pi \le W/2$,}\\[.05in]
0, & \mbox{otherwise},
\end{array}\right.
\end{equation}
where filtering has been used implicitly to limit the noise's bandwidth to that of the signal, and the $1/(1-\kappa)$ factor makes $N_B$ the noise brightness at the channel's output.  Mode expansions for $\hat{E}_R(t)$ and $\hat{E}_B(t)$, given by
\begin{equation}
\hat{E}_K(t) = \sum_{m=-\infty}^{\infty}\hat{a}_{K_m}\frac{e^{-i2\pi mt/T}}{\sqrt{T}},\mbox{ for $t\in \mathcal{T}_0$ and $K = R,B$},
\label{RBmodes}
\end{equation}
where
\begin{equation}
\hat{a}_{R_m} = \sqrt{\kappa}\,e^{i\theta}\hat{a}_{S_m} + \sqrt{1-\kappa}\,\hat{a}_{B_m},
\label{ArChannelModel}
\end{equation}
will be of use, together with those from \eqref{SImodes}, in Sec.~\ref{CtoD}.  Toward that end we note that 
the excited background modes, $\{\hat{a}_{B_m} : m=-(M-1)/2,\QZ{-(M-3)/2},\ldots,(M-1)/2\}$, are in iid thermal states with $\langle \hat{a}_{B_m}^\dagger\hat{a}_{B_m}\rangle = N_B/(1-\kappa)$.  

Depending on the protocol, the idler will either be stored (losslessly) at the transmitter location or delivered (again losslessly) to another location.    In either case the ${\rm C}\to{\rm D}$ receiver will use a cascaded positive operator-valued measurement (POVM)---to be described in Sec.~\ref{CtoD}---in which the returned radiation is heterodyne detected and the classical data obtained therefrom dictates the POVM that will be performed on the idler.  

\section{Entanglement-Assisted Protocols} \label{Protocols}
\begin{figure}
    \centering
    \includegraphics[width=0.5\textwidth]{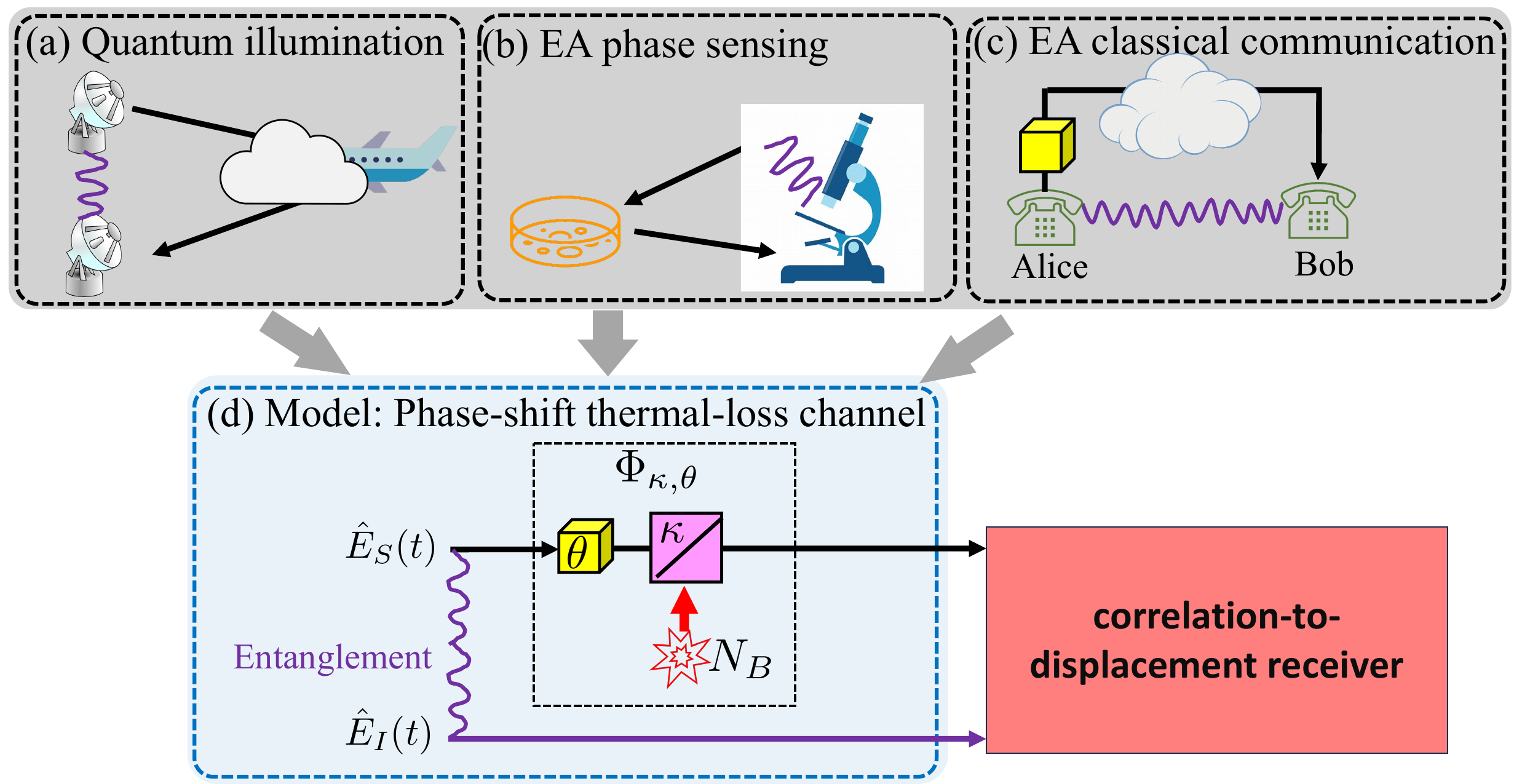}
    \caption{Schematic of the entanglement-assisted (EA) protocols under consideration: (a) quantum illumination, (b) EA phase sensing and (c) EA classical communication.  All the physical processes in these protocols can be modeled by the phase-shift thermal-loss channel sketched in (d).  For each protocol a correlation-to-displacement receiver, proposed in this paper, provides optimum reception capability. 
    }
    \label{fig:conversion_concept_scenarios} 
\end{figure}

The three protocols we shall consider for broadband TMSV-enabled entanglement-assisted quantum sensing and communication over an entanglement-breaking lossy, noisy channel are illustrated in Fig.~\ref{fig:conversion_concept_scenarios}.  Figure~\ref{fig:conversion_concept_scenarios}(a) shows QI target detection.  Here, $\hat{E}_S(t)$ interrogates a region that may or may not contain a weakly-reflecting target.  When the target is absent, the receiver observes the output of a $\Phi_{0,0}$ channel, i.e., only background noise.  Alternatively, when the target is present, the receiver observes the output of a $\Phi_{\kappa,\theta}$ channel, viz., reflected signal embedded in background noise.  As in prior studies of QI target detection, see, e.g., Ref.~\cite{tan2008quantum}, we will assume $\theta$ to be known and set it to 0.  A ${\rm C}\to{\rm D}$ receiver---shown as a simple block in Fig.~\ref{fig:conversion_concept_scenarios}(d)---can be used to achieve quantum-optimal target-detection error probability by applying the appropriate cascaded POVM to $\hat{E}_R(t)$ and $\hat{E}_I(t)$.  Note that broadband TMSV states are known to be optimal for QI target detection~\cite{DePalma2018,nair2020fundamental,Bradshaw2021} and EA communication~\cite{shi2020practical}, so finding a practical receiver for realizing their full quantum advantage is of considerable value.

Figure~\ref{fig:conversion_concept_scenarios}(b) shows EA phase sensing.  Here, refractive-index differences in biosensing, gravitational-wave detection ~\cite{abbott2016observation}, or other applications impose an unknown phase shift $\theta$ on the radiation returned from transmission of $\hat{E}_S(t)$.  EA estimation of this phase via a cascaded POVM realized by a ${\rm C}\to{\rm D}$ receiver can achieve quantum-optimal mean-squared error in the little studied bright-noise operating conditions, unlike other EA phase-sensing paradigms~\cite{nagata2007beating,dowling2008quantum,daryanoosh2018experimental} that work in the better understood low-noise regime.

Figure~\ref{fig:conversion_concept_scenarios}(c) shows EA classical communication, in which Alice has pre-shared entanglement with Bob by losslessly delivering her idler to him.  She then transmits a classical message to Bob via phase-shift keying (PSK) of $\hat{E}_S(t)$~\cite{shi2020practical}. For example, take binary phase-shift keying (BPSK). Here, Alice transmits each bit $b$, over a known-phase (taken to be 0) thermal-loss channel, as a $T$-s-duration pulse of broadband TMSV radiation $e^{i\theta}\hat{E}_S(t)$, where $\theta = 0$ for $b = 1$ and $\theta = \pi$ for $b=0$.  In each $T$-s-long bit interval, Bob's receiver must decide between Alice's signal $\hat{E}_S(t)$ having arrived as the output of a $\Phi_{\kappa,0}$ channel or a $\Phi_{\kappa,\pi}$ channel.  For continuous PSK, Bob's decoding involves extracting phase information about the $\Phi_{\kappa,\theta}$ channel.  Using PSK \QZ{modulation}, and a properly configured ${\rm C}\to{\rm D}$ receiver, EA classical communication provides an optimal classical information capacity that exceeds the Holevo-Schumacher-Westmoreland (HSW) capacity~\cite{hausladen1996classical,schumacher1997sending,holevo1998capacity,giovannetti2014ultimate}
for unassisted (coherent-state) communication at the same average photon flux over the $\Phi_{\kappa,0}$ channel

The quantum advantages afforded by the entanglement assistance in Figs.~\ref{fig:conversion_concept_scenarios}(a)--(c) are quantified by comparing the performance of their ${\rm C}\to{\rm D}$ receivers to those of corresponding classical systems which use coherent-state radiation of the same average photon number, $MN_S$, as the EA systems' $T$-s-duration $\hat{E}_S(t)$ transmission. 

\section{Correlation-to-Displacement Receiver} \label{CtoD}

Before delving into ${\rm C}\to{\rm D}$ reception's three constituent modules, it is worth understanding both the motivation behind this architecture and the essence of why it is so powerful.  

We focus on QI as an application to provide the intuition, as it has been most extensively studied.
When Guha and Erkmen proposed the PA and PC receivers for QI target detection~\cite{Guha2009}, they knew that conventional direct detection, homodyne detection, and heterodyne detection failed to deliver any quantum advantage. They also recognized that this failure was due to the phase-sensitive cross correlation between $\hat{E}_R(t)$ and $\hat{E}_I(t)$, which is the signature of target presence, being undetectable in second-order interference.  Thus their receivers use parametric interactions to convert the foregoing signature to one of phase-insensitive cross correlation, which can be measured in second-order interference.  That ideal implementation of these receivers realized only half of QI target detection's predicted quantum advantage was later ascribed~\cite{zhuang2017optimum} to their being local operations plus classical communication (LOCC) strategies \QZ{between the different return-idler mode pairs}.  Indeed, accepted wisdom until now has been that general mixed-state discrimination, of which QI target detection is an example, requires a joint measurement coupling \emph{all} the modes of the returned and retained radiation~\cite{LOCC_NO_GO,cheng2021discrimination}, i.e., LOCC is insufficient.  That wisdom led to Zhuang~\emph{et al}.'s FF-SFG receiver~\cite{zhuang2017optimum}, in which single-photon sensitive SFG was used to extract single-mode coherence from the $M$ excited mode pairs in $\hat{E}_R(t)$ and $\hat{E}_I(t)$.  Unfortunately, FF-SFG reception required many cycles of yet to be developed single-photon sensitive SFG, and it converted only half the phase-sensitive cross correlation into single-mode coherence.  

\begin{figure}
    \centering
    \includegraphics[width=0.5\textwidth]{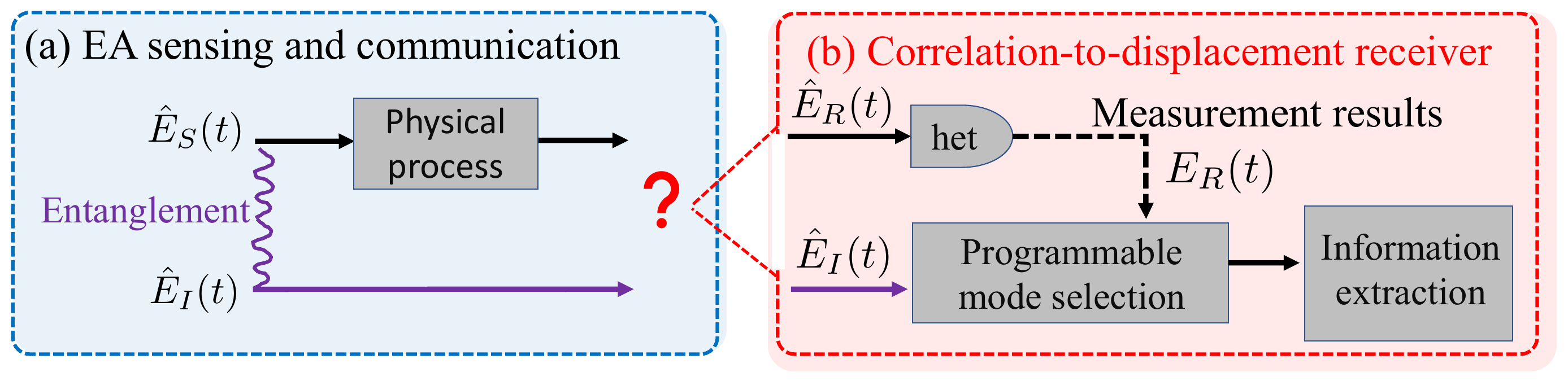}
    \caption{Schematic of EA sensing and communication using the ${\rm C}\to{\rm D}$ receiver.  The receiver's correlation-to-displacement converter heterodynes (Het) the returned radiation, $\hat{E}_R(t)$, thereby  converting the phase-sensitive cross correlation between it and the retained idler, $\hat{E}_I(t)$, into a displacement of the idler mode proportional to $E^*_R(t)$, the conjugate of the heterodyne detector's complex-envelope output.  Programmable mode selection then transfers the resulting noisy coherent-state idler mode to a near-vacuum state noise bath.  Finally, well known coherent-state information extraction, appropriate to the EA application of interest, is used to obtain quantum-optimal performance.}
    \label{fig:conversion_concept} 
\end{figure}

Recently, Shi~\emph{et al}.~\cite{shi2022fulfilling}, in a preliminary version of this paper, recognized the importance of correlation-to-coherence conversion \emph{and} that it could be done, more easily and more completely, by means of heterodyne detection.  The receiver, shown schematically in Fig.~\ref{fig:conversion_concept}, performs ${\rm C}\to{\rm D}$ conversion, programmable mode selection, and coherent-state information extraction.  It achieves quantum-optimal performance in broadband TMSV-enabled EA sensing and communication as follows:  (1) ${\rm C}\to{\rm D}$ conversion transfers the phase-sensitive cross correlation between the returned radiation and the retained idler to the mean field of a coherent-state idler mode---whose mode shape is proportional to $E_R^*(t)$, the conjugate of the heterodyne detector's complex-envelope output---embedded in broadband thermal noise; (2) programmable mode selection removes the idler's noisy coherent-state mode from its thermal-noise bath and embeds that mode in a near-vacuum noise environment; and (3) coherent-state information extraction applies well known optimum processing for coherent-state signals in near-vacuum noise environments to realize quantum-optimal performance for the EA sensing or communication application at hand. 

The ${\rm C}\to{\rm D}$ receiver is a cascaded POVM in that heterodyne detection of the returned radiation determines what POVM will be applied to the retained idler.  That this receiver, as will be seen explicitly in Sec.~\ref{Performance}, achieves quantum optimal performance refutes the previously accepted wisdom that a joint quantum measurement of the returned and retained radiation is required for accessing broadband TMSV-enabled EA sensing and communication's full quantum advantage. \QZ{However, the previously accepted wisdom dictates that quantum-optimum performance cannot be obtained from LOCC measurements on the $M$ returned and retained \emph{mode pairs}, $\{\hat{a}_{R_m},\hat{a}_{I_m}\}_{m=1}^M$. Our ${\rm C}\to{\rm D}$ receiver's cascaded POVM thus \emph{can} realize quantum-optimum performance because it is an LOCC measurement on the returned and retained \emph{fields}, $\{\hat{E}_R(t),\hat{E}_I(t)\}$, which is joint in mode pairs as it performs operations jointly on all $\{\hat{a}_{I_m}\}_{m=1}^M$.  Furthermore, that ${\rm C}\to {\rm D}$ reception's} constituent modules are more immediately practical than FF-SFG reception, and applicable to known broadband TMSV-enabled EA sensing and communication protocols, \QZ{has already} led to follow-on work~\cite{Reichert2023,Angeletti2023,chen2023}  \QZ{which builds} on the foundation established in Ref.~\cite{shi2022fulfilling}.  The relationship between those papers and this one will be discussed in Sec.~\ref{Discussion}.

Turning now to a more complete characterization of the ${\rm C}\to{\rm D}$ receiver, we start from the Wigner cross covariance of one of the $M$ iid $(\hat{a}_{R_m},\hat{a}_{I_m})$ excited mode pairs from the expansions given in Eqs.~(\ref{SImodes}) and (\ref{RBmodes}),
\begin{equation}
\label{eq:CM_RI}
\Lambda = \frac{1}{4}\left[\begin{array}{cc}
(2\kappa N_S + 2N_B + 1)\mathbb{I}_2 & 2C_p\mathbb{R}\mathbb{Z}\\[.05in]
 2C_p\mathbb{R}\mathbb{Z}& (2N_S +1)\mathbb{I}_2\end{array}\right],
 \end{equation}
 where $\mathbb{I}_2$ is the $2\times 2$ identity matrix, $C_p\equiv \sqrt{\kappa N_S(N_S+1)}$, $\mathbb{R} \equiv \cos(\theta) \mathbb{I}_2-i \sin(\theta) \mathbb{Y}$, $\mathbb{Y}$ is the Pauli Y matrix, and $\mathbb{Z}$ is the Pauli Z matrix. We assume ideal heterodyne detection, in which case we can take its complex-envelope output, $E_R(t)$ for $t\in \mathcal{T}_0$, to be the result of applying the $\hat{E}_R(t)$ POVM to the returned radiation for $t\in \mathcal{T}_0$.  Next, using the mode expansion
 \begin{equation}
 E_R(t) = \sum_{m=-(M-1)/2}^{(M-1)/2}a_{R_m}\frac{e^{-i2\pi mt/T}}{\sqrt{T}},\mbox{ for $t\in\mathcal{T}_0$,}
 \label{hetRmodes}
\end{equation}
where we have assumed the heterodyne detector's output has been filtered to match the bandwidth of Alice's transmitted signal, conditional Gaussian-map analyses~\cite{weedbrook2012gaussian,genoni2016conditional} imply the $\{\hat{a}_{I_m} : m= -(M-1)/2,\QZ{-(M-3)/2},\ldots,(M-1)/2\}$ are in independent Gaussian states with conditional means
\begin{equation}
\mathbb{E}(\hat{a}_{I_m}\mid a_{R_m}) = \frac{\sqrt{\kappa N_S(N_S+1)}}{\kappa N_S + N_B +1}a_{R_m}^*e^{i\theta},
\end{equation}
and conditional Wigner covariance matrix
\begin{equation}
\Lambda_{I_m\mid R_m} = \frac{2N_{I\mid R}+1}{4}\mathbb{I}_2,
\end{equation}
where 
\begin{equation}
\label{NIR_def}
N_{I\mid R} \equiv \frac{N_S(N_B+1-\kappa)}{\kappa N_S + N_B +1} \le N_S.
\end{equation}
Substituting these results back into \eqref{hetRmodes} we have that, conditioned on $E_R(t)$, the idler field is in a coherent state with mean field 
\begin{equation}
\mathbb{E}[\hat{E}_I(t)\mid E_R(t)] = \frac{\sqrt{\kappa N_S(N_S+1)}}{\kappa N_S+N_B + 1}E_R^*(t)e^{i\theta}, \mbox{ for $t\in\mathcal{T}_0$},
\end{equation}
that is embedded in a thermal-noise bath,
\begin{equation}
\hat{E}_n(t) \equiv \hat{E}_I(t) - \mathbb{E}[\hat{E}_I(t) \mid E_R(t)],
\end{equation}
with fluorescence spectrum
\begin{equation}
S^{(\rm pi)}_{nn}(\omega) = \left\{\begin{array}{ll}
N_{I\mid R} , & \mbox{for $|\omega|/2\pi \le W/2$},\\[.05in]
0, & \mbox{otherwise}.\end{array}\right.
\label{condxbath}
\end{equation}

The ${\rm C}\to{\rm D}$ receiver needs a programmable mode selector because the thermal-noise bath's fluorescence spectrum from \eqref{condxbath} is too strong to enable information extraction with near-optimal performance.  Consider broadband TMSV-enabled EA sensing and communication's usual asymptotic regime, wherein $N_S \ll 1, \kappa \ll 1$, and $N_B \gg 1$.   There, the $\hat{E}_n(t)$ thermal bath contains approximately $MN_S$ photons, on average, over the time interval $\mathcal{T}_0$.  For broadband TMSV-enabled EA sensing and communication in the asymptotic regime, $M\kappa N_S/N_B>1$ is required for reasonable performance, which necessarily leads to $MN_S \gg 1$, thus severely degrading performance if coherent-state information extraction is performed directly on $\hat{E}_I(t)$.  The purpose of the ${\rm C}\to{\rm D}$ receiver's programmable mode selector, therefore, is transferring the idler's post-heterodyne coherent-state mode---whose shape is proportional to $E_R^*(t)$---to a near-vacuum noise bath, $\hat{E}_v(t)$, i.e., one with fluorescence spectrum 
\begin{equation}
S^{(\rm pi)}_{vv}(\omega) = \left\{\begin{array}{ll}
N_v, & \mbox{for $|\omega|/2\pi \le W/2$},\\[.05in]
0, & \mbox{otherwise},\end{array}\right.
\label{nvbath}
\end{equation}
for which $MN_v \ll 1$.  

\begin{figure}
    \centering
    \includegraphics[width=0.5\textwidth]{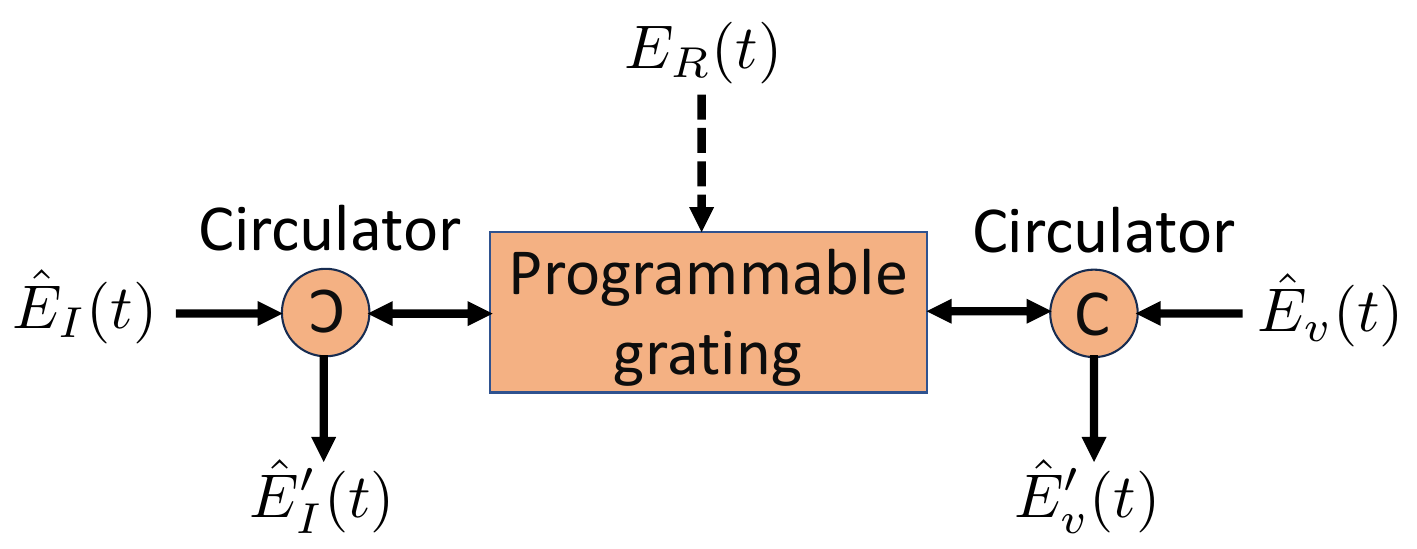}
    \caption{Notional scheme for the ${\rm C}\to{\rm D}$ receiver's programmable mode selector.  The idler field $\hat{E}_I(t)$ and the near-vacuum field $\hat{E}_v(t)$ enter a programmable grating through a pair of circulators.  The grating, programmed by the heterodyne detector's output $E_R(t)$, perfectly reflects the idler and near-vacuum fields' modes that are proportional to $E^*_R(t)$, and it perfectly transmits the remaining field modes.  See the text for characterizations of the mode selector's output fields, $\hat{E}'_I(t)$ and $\hat{E}'_v(t)$, that emerge from the second pass through the mode selector's circulators. 
        \label{fig:notionalmodeselect}
    }
\end{figure}

A notional scheme for programmable mode selection is shown in Fig.~\ref{fig:notionalmodeselect}.  The idler field $\hat{E}_I(t)$ and the near-vacuum field $\hat{E}_v(t)$ enter a programmable grating through a pair of circulators.  The grating, programmed by the heterodyne detector's output $E_R(t)$, perfectly reflects the idler and near-vacuum fields' modes that are proportional to $E^*_R(t)$, and it perfectly transmits the remaining field modes.  To characterize the mode selector's output fields, $\hat{E}'_I(t)$ and 
$\hat{E}'_v(t)$, that emerge from \QZ{a second pass through Fig.~\ref{fig:notionalmodeselect}'s} circulators, we define a complete orthonormal set of mode shapes, $\{\phi_m(t): -\infty < m < \infty\}$, on $t\in \mathcal{T}_0$, where $\phi_0(t) = E_R^*(t)/\sqrt{N_R}$, with
\begin{equation}
N_R \equiv \int_{\mathcal{T}_0}\!{\rm d}t\,|E_R(t)|^2.
\label{N_R}
\end{equation}
Given $E_R(t)$, we then have that
\begin{equation}
\hat{E}'_K(t) = \sum_{m=-\infty}^\infty\hat{a}'_{K_m}\phi_m(t), \mbox{ for $t\in \mathcal{T}_0,K=I,v$},
\label{EK_expansion}
\end{equation}
where the $\{\hat{a}'_{I_m}, \hat{a}'_{v_m}\}$ modes are in independent states.  Specifically, the $\hat{a}'_{I_0}$ mode is in a displaced thermal state with mean $\sqrt{\kappa N_S(N_S+1)N_R}\,e^{i\theta}/(\kappa N_S + N_B+1)$ and average noise photon number $N_{I\mid R}$, i.e., the desired coherent state embedded in \QZ{weak} thermal noise.  The remaining modes of $\hat{E}'_I(t)$ are in iid thermal states with average photon numbers $N_v$, i.e., the desired near-vacuum noise bath.  Conversely, the $\hat{a}'_{v_0}$ mode is in a thermal state with \QZ{near-vacuum} average photon number $N_v$, while the remaining modes of $\hat{E}'_v(t)$ are in iid thermal states with average photon numbers $N_{I\mid R}$. 

The ${\rm C}\to{\rm D}$ receiver's architecture is completed by the coherent-state information extractor.  This module takes the form of whatever is needed, for the chosen broadband TMSV-enabled EA sensing or communication application, to achieve quantum optimal performance.  To the extent that the near-vacuum average photon number, $N_v$, is low enough that it does not impact the desired error probability (for QI target detection), mean-squared estimation error (for EA phase estimation), or capacity (for EA classical communication), then this extractor becomes what would be used in conventional coherent-state, $\Phi_{\kappa,\theta}$-channel versions of those applications.  We postpone providing details of the coherent-state information extractors and designs for the programmable mode selector until Secs.~\ref{ExtractorDesign} and \ref{SelectorDesign}, respectively, and devote the next section to the ultimate performance limits of ${\rm C}\to{\rm D}$ reception.

\section{Performance Limits of C-to-D Reception} \label{Performance}
To get at ${\rm C}\to{\rm D}$ reception's ultimate performance limits, we will assume the previous section's near-vacuum noise brightness $N_v$ is so low that its effect on receiver performance can be entirely neglected.  We will discuss the degree to which this assumption is warranted in Sec.~\ref{Discussion}.

\subsection{QI Target Detection \label{QIdetectionPerformance}}
QI target detection using ${\rm C}\to{\rm D}$ reception whose programmable mode selector's near-vacuum thermal noise is negligible has an $\hat{E}'_I(t)$ for $t\in \mathcal{T}_0$ with only one excited mode, viz., $\phi_0(t)$.  Moreover, the task for that receiver's coherent-state information extractor is to decide between target absence (hypothesis $H_0$) and target presence (hypothesis $H_1$) given knowledge of $\{E_R(t) : t\in \mathcal{T}_0\}$ by means a POVM on $\hat{E}'_I(t)$, the programmable mode selector's idler field output.  The relevant conditional density operators for the extractor's hypothesis test are those for $\hat{E}'_I(t)$'s $\phi_0(t)$ mode:
\begin{equation}
\hat{\rho}_{0\mid R} = \sum_{n=0}^\infty \frac{N_S^n}{(N_S+1)^{n+1}}|n\rangle\langle n|,
\end{equation}
and
\begin{equation}
\label{rho1R_def}
\hat{\rho}_{1\mid R} = \int\!{\rm d}^2\alpha\,\frac{e^{-|\alpha-\zeta\sqrt{N_R}|^2/N_{I\mid R}}}{\pi N_{I\mid R}}|\alpha\rangle\langle \alpha|,
\end{equation}
where $\{|n\rangle\}$ and $\{|\alpha\rangle\}$ are the $\phi_0(t)$ mode's number states and coherent states, and 
$\zeta \equiv  \sqrt{\kappa N_S(N_S+1)}/(\kappa N_S + N_B+1).$ 

With equally-likely target absence or presence, the minimum conditional error probability given $\{E_R(t): t\in \mathcal{T}_0\}$ is then the Helstrom bound~\cite{Helstrom1967,Helstrom_1976}  $\Pr(e\mid R)_{{\rm C}\to {\rm D}}$ between states $\hat{\rho}_{1\mid R}$ and $\hat{\rho}_{0\mid R}$, which can be numerically evaluated efficiently as the states are single-mode and finite energy~\cite{supp}.  
To find the ${\rm C}\to{\rm D}$ receiver's \emph{unconditional} error probability we must average the foregoing result over $N_R$'s probability density function (pdf), $p(N_R)$, i.e.,
\begin{equation}
\Pr(e)_{{\rm C}\to {\rm D}}  = \int\!{\rm d}N_R\,p(N_R)\Pr(e\mid R)_{{\rm C}\to {\rm D}}\,.
\label{PeCD}
\end{equation}  
Using the mode expansion from \eqref{RBmodes}, we have that $N_R = \sum_{m=-(M-1)/2}^{(M-1)/2}|a_{R_m}|^2$, where the $\{|a_{R_m}|^2\}$ are iid exponential random variables with mean values $\kappa N_S + N_B + 1$.  
It follows that $N_R$'s pdf is
\begin{equation}
p(N_R) = \frac{N_R^{M-1}e^{-N_R/(\kappa N_S + N_B + 1)}}{(\kappa N_S + N_B + 1)^M(M-1)!}u(N_R),
\label{NRpdf}
\end{equation}
where $u(\cdot)$ is the unit-step function. As we shall see below, Eqs.~ (\ref{PeCD}) and (\ref{NRpdf}) provide a powerful new tool for evaluating the performance of QI target detection.

To benchmark QI target detection using the ${\rm C}\to{\rm D}$ receiver we will compare its error probability to the Nair-Gu lower bound~\cite{nair2020fundamental},
\begin{equation}
\Pr(e)_{\rm NG} = e^{M\ln[1-\kappa/(N_B+1)]N_S}/4,
\end{equation}
which sets a lower limit on the error probability of any system that transmits $MN_S$ photons on average, regardless of whether or not it employs entanglement.  We shall also compare $\Pr(e)_{{\rm C}\to {\rm D}}$ to the Helstrom limit, $\Pr(e)_{\rm CS}$, for a coherent-state system that transmits $MN_S$ photons on average and must decide between the equally-likely possibilities of its returned radiation being the output of a $\Phi_{0,0}$ channel or a $\Phi_{\kappa,0}$ channel.  

\begin{figure}[t]
    \centering
    \includegraphics[width=0.45\textwidth]{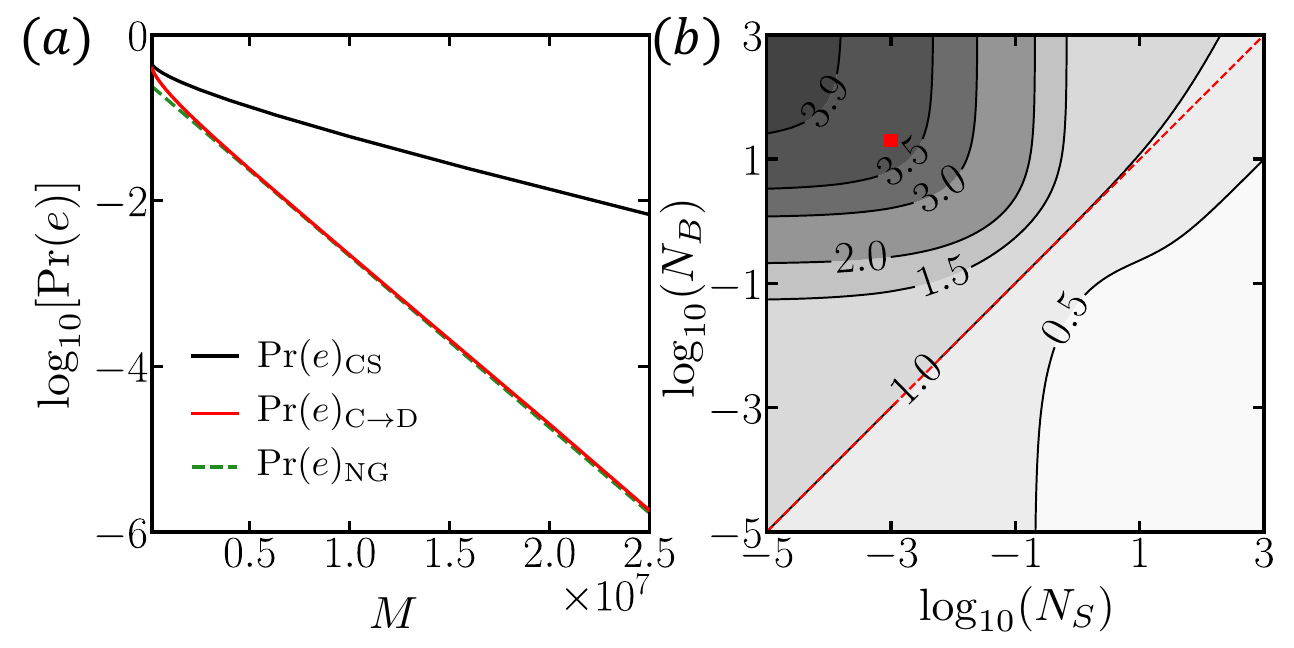}
    \caption{QI target detection in the asymptotic regime for $\kappa = 0.01$: (a) error probability versus time-bandwidth product $M$ with $N_S=0.001, N_B = 20$. 
    (b) Lower bound of the error exponent ratio $r_{{\rm C}\to{\rm D},\rm LB}/r_{\rm CS}$ versus $N_S, N_B$. Red dashed line indicates $N_S=N_B$.  The red dot indicates the parameters chosen in (a).  
        \label{fig:illumination_exponent}
    }
\end{figure}

We begin our benchmarking in the $N_S \ll 1,\kappa\ll 1,N_B\gg 1,M\gg 1$ asymptotic regime, as previously studied in Refs.~\cite{tan2008quantum,zhuang2017optimum}.  
In this scenario, we can approximate $\hat{\rho}_{0\mid R}$ as the vacuum state $|0\rangle\langle 0|$ and $\hat{\rho}_{1\mid R}$ as the coherent state $|\zeta\sqrt{N_R}\rangle\langle \zeta\sqrt{N_R}|$. These approximations make the hypothesis test quantum limited, in that there is no thermal noise.
\begin{align}
\Pr(e\mid R)_{{\rm C}\to {\rm D}} &\approx  (1-\sqrt{1-e^{-\zeta^2N_R}})/2 \label{PeCDcondx}\\[.05in] &\approx e^{-\zeta^2N_R}/4, \mbox{ for $N_R \gg 1$}. \label{PeCDcondxApprox}
\end{align}
The ${\rm C}\to{\rm D}$ receiver can achieve this performance when, as assumed here, both conditional density operators are coherent states and the information extractor is a Dolinar receiver~\cite{Dolinar1973}; see Sec.~\ref{ExtractorDesign} for more information\QZ{.}
Using the low error-probability approximation from \eqref{PeCDcondxApprox}, we find
\begin{align}
\Pr(e)_{{\rm C}\to {\rm D}} &\approx \int_0^\infty\!{\rm d}N_R\, \frac{N_R^{M-1}e^{-N_R/N_B}}{N_B^M(M-1)!}e^{-\kappa N_SN_R/N_B^2}/4 \\[.05in]
&= \frac{1}{4}\!\left(\frac{1}{1+\kappa N_S/N_B}\right)^M \approx e^{-M\kappa N_S/N_B}/4,
\end{align}
which equals half the asymptotic regime's quantum Chernoff bound (QCB) for QI target detection~\cite{tan2008quantum}, and matches the asymptotic-regime version of the Nair-Gu bound.  We verify the latter statement in Fig.~\ref{fig:illumination_exponent}(a), which shows close agreement between the numerically-evaluated exact $\Pr(e)_{{\rm C}\to {\rm D}}$ and $\Pr(e)_{\rm NG}$.    Figure~\ref{fig:illumination_exponent}(a) also shows the enormous error-probability advantage afforded by QI target detection with ${\rm C}\to{\rm D}$ reception as opposed to coherent-state operation at the same transmitted energy, whose error probability was numerically evaluated from the exact Helstrom limit.

To get further insight into QI's quantum advantage, we now examine its error exponent more closely.
In Ref.~\cite{supp}, we derive the following lower bound on QI's error exponent,
\begin{align}
r_{{\rm C}\to{\rm D}}&\ge r_{{\rm C}\to{\rm D},\rm LB} \nonumber \\[.05in]
&\equiv
\ln[1+\zeta^2(\kappa N_S + N_B+1) (\sqrt{N_S+1}-\sqrt{N_S})^2],
\label{r_LB}
\end{align}
while the coherent state's error exponent is~\cite{tan2008quantum,supp}.
\begin{equation}
r_{\rm CS}=\kappa N_S(\sqrt{N_B+1}-\sqrt{N_B})^2
\end{equation}
When $r_{{\rm C}\to{\rm D}}\ll1$, we have 
\be 
\frac{r_{{\rm C}\to{\rm D},\rm LB} }{r_{\rm CS}}\approx
\frac{(\sqrt{N_S+1}-\sqrt{N_S})^2}{(\sqrt{N_B+1}-\sqrt{N_B})^2} 
\frac{  \left(N_S+1\right)}{\kappa  N_S+N_B+1}\JHS{.}
\label{QI_adv_cond}
\ee 
It is then easy to see that the condition for advantage (${r_{{\rm C}\to{\rm D}} }/{r_{\rm CS}}>1$) is true for $N_S< N_B$, as  confirmed in Fig.~\ref{fig:illumination_exponent}(b). At the same time, we see $\kappa\ll1$ is not a necessary \QZ{condition} for QI to offer quantum advantage. Via the conversion module, we obtained \eqref{QI_adv_cond} as the precise condition for QI's quantum advantage, a result that was not previously known.

\begin{figure}[t]
\centering
    \includegraphics[width=0.45\textwidth]{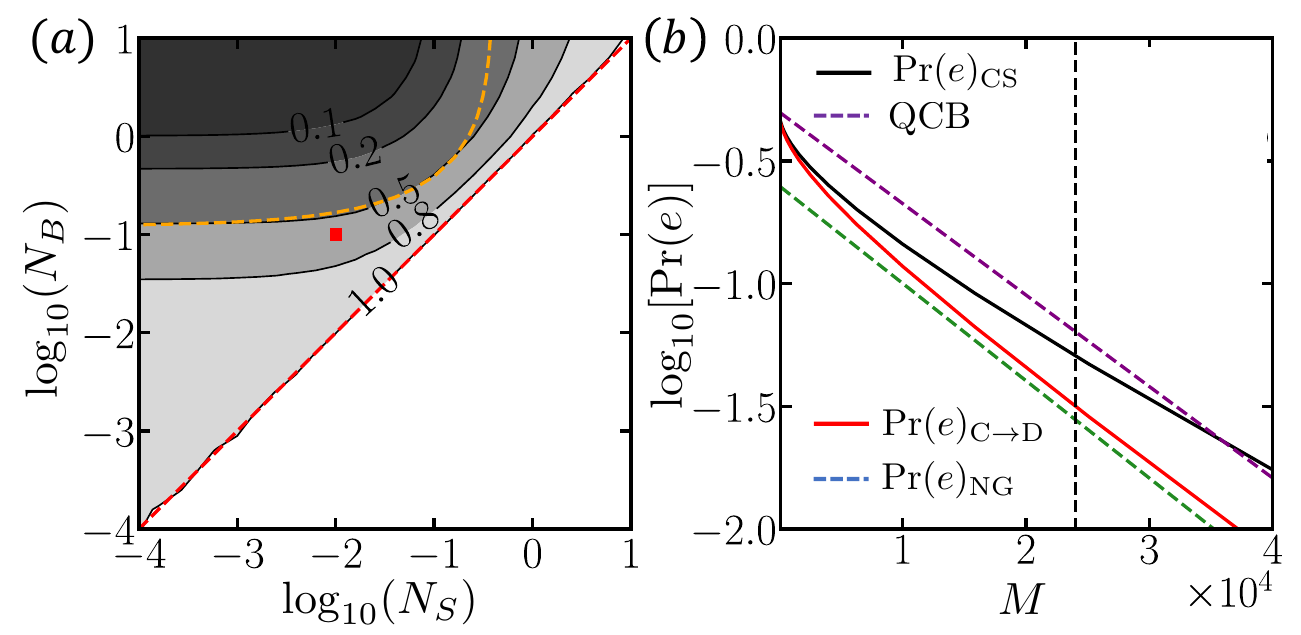}
    \caption{QI outside the $N_S \ll 1, \kappa\ll 1, N_B \gg 1$ asymptotic regime.   (a) Error-probability ratio $\Pr(e)_{{\rm C}\to {\rm D}}/\Pr(e)_{\rm CS}$ versus $N_S, N_B$ with $\kappa = 0.01$ and $M$  chosen to fix $\Pr(e)_{\rm CS}$ at 0.05.  The red dot indicates the $N_S = 0.01, N_B=0.1$ values used in (b).  In the region above the red dashed $N_S = N_B$ line ${\rm C}\to{\rm D}$ reception offers quantum advantage, but QI's QCB only predicts quantum advantage in the region above the orange curve .  (b) Error probability versus time-bandwidth product, $M$, with $N_S = 0.01, \kappa = 0.01, N_B = 0.1$.  The vertical line indicates the $M$ value for which $\Pr(e)_{\rm CS} = 0.05$.  \label{fig:illumination_peratio}}
\end{figure}

Finally, let us examine QI's performance beyond its asymptotic regime.  In this case, Eqs.~~(\ref{PeCD}) and (\ref{NRpdf}) provide an efficiently calculated and achievable error-probability lower bound for QI, in contrast to QI's QCB~\cite{Audenaert2007,Pirandola2008,tan2008quantum}, which is an exponentially-tight upper bound on the error probability. 
Thus, as shown in Fig.~\ref{fig:illumination_peratio}(a), fixing $\Pr(e)_{\rm CS}$ at 0.05 we have $\Pr(e)_{{\rm C}\to {\rm D}} < \Pr(e)_{\rm CS}$ when $N_S< N_B$, i.e., above Fig.~\ref{fig:illumination_peratio}(a)'s dashed red line. However, QI's QCB only shows quantum advantage in a strictly smaller region, viz.,  above Fig.~\ref{fig:illumination_peratio}(a)'s orange dashed curve. Figure~\ref{fig:illumination_peratio}(b) plots the error probability versus time-bandwidth product, $M$, assuming $N_S = 0.01, \kappa = 0.01$, and $N_B = 0.1$.  Whereas our $\Pr(e)_{{\rm C}\to {\rm D}}$ calculations show quantum advantage over the entire range of $M$ values in Fig.~\ref{fig:illumination_peratio}(b), QI's QCB fails to do so until the very high end of the figure's $M$ range.  

\subsection{Entanglement-Assisted Phase Estimation}

EA phase estimation is a well-studied problem in the low-noise limit~\cite{nagata2007beating,dowling2008quantum,daryanoosh2018experimental}, i.e., $N_B \ll 1$ in the case of our $\Phi_{\kappa,\theta}$ channel. Much less is known, however, about EA phase estimation in the noisy limit, viz.,  $N_B \gg 1$ in our case. Shi~\emph{et al}.~\cite{shi2020practical} paralleled the QI target detection protocol in their treatment of EA phase estimation.  Specifically, the signal field from \eqref{SImodes} was applied to the input of the phase-shift thermal-loss channel $\Phi_{\kappa,\theta}$, while the idler field from \eqref{SImodes} was retained for use in conjunction with the returned radiation, $\hat{E}_R(t)$, in estimating the unknown phase $\theta$. Their analysis revealed that broadband TMSV-enabled EA phase estimation affords a quantum-optimal 3\,dB phase-estimation Fisher information advantage over its coherent-state competitor of the same transmitted energy, and it does so despite the channel's loss and noise destroying the initial entanglement. For $N_S \ll 1, \kappa \ll 1$, Ref.~\cite{shi2020practical} also showed that phase-estimation versions of Guha and Erkmen's PA and PC receivers obtain the optimal EA advantage, a result which led to the experimental demonstration in Ref.~\cite{hao2022demonstration}. 

To assess the ultimate performance of broadband TMSV-enabled EA phase estimation using ${\rm C}\!\!\to\!\! {\rm D}$ reception, we rely on the quantum Cram\'{e}r-Rao bound (QCRB)~\cite{Braunstein1994}.  Given $\{E_R(t): t\in \mathcal{T}_0\}$, the QCRB provides a universal lower bound,
\begin{equation}
\delta\theta^2 \ge 1/\mathcal{I}_F(\hat{\rho}_{I\mid R,\theta}),
\label{QCRB}
\end{equation}
on the mean-squared error, $\delta\theta^2$, of any unbiased estimator of the unknown phase $\theta$ in terms of the observation's---here $\{\hat{E}_R(t), \hat{E}'_I(t) : t\in \mathcal{T}_0\}$---conditional quantum Fisher information (QFI), 
\begin{equation}
\mathcal{I}_F(\hat{\rho}_{I\mid R, \theta}) = \lim_{\epsilon\rightarrow 0}\left\{8\!\left[1-\sqrt{\mathcal{F}[\hat{\rho}_{I\mid R, \theta},\hat{\rho}_{I\mid R, \theta+\epsilon}]}\right]\right\}/\epsilon^2,
\label{QFisher}
\end{equation}
where 
\begin{equation}
\mathcal{F}(\hat{\rho},\hat{\rho}') \equiv 
\left[{\rm Tr}\!\left(\!\sqrt{\sqrt{\hat{\rho}}\,\hat{\rho}'\sqrt{\hat{\rho}}}\,\right)\right]^2
\end{equation}
is the Uhlmann fidelity between density operators $\hat{\rho}$ and $\hat{\rho}'$,  
and $\hat{\rho}_{I\mid R,\theta}$ is the conditional density operator of $\hat{E}'_I(t)$'s $\phi_0(t)$ mode:
\begin{equation}
\hat{\rho}_{I\mid R,\theta} = \int\!{\rm d}^2\alpha\,\frac{e^{-|\alpha-\zeta\sqrt{N_R}e^{i\theta}|^2/N_{I\mid R}}}{\pi N_{I\mid R}}|\alpha\rangle\langle \alpha|,
\label{PSKdensityoperator}
\end{equation}
with $\zeta, N_R, N_{I\mid R},$ and $|\alpha\rangle$ as defined in Sec.~\ref{Performance}A.  From Ref.~\cite{supp} we have that
\begin{equation}
\mathcal{I}_F(\hat{\rho}_{I\mid R,\theta}) = \frac{4\zeta^2N_R}{2N_{I\mid R}+1},
\end{equation}
and averaging over $p(N_R)$ from \eqref{NRpdf} gives us the ${\rm C}\!\!\to\!\! {\rm D}$ receiver's \emph{unconditional} quantum Fisher information,
\begin{align} 
\mathcal{I}_F(\hat{\rho}_{RI\mid \theta})_{{\rm C}\to {\rm D}} &= \langle \mathcal{I}_F(\hat{\rho}_{I\mid R,\theta})_{{\rm C}\to {\rm D}}\rangle \\[.05in] &=\frac{4 M\kappa  N_S \left(N_S+1\right)}{1+N_B+N_S \left(2 N_B+2-\kappa \right)}.
\label{eq:qfi_c2d}
\end{align}

The unconditional QCRB for ${\rm C}\!\!\to\!\! {\rm D}$ reception is asymptotically tight as the time-bandwidth product grows without bound,  $M\rightarrow \infty$. It is given by the average of the conditional QCRB, i.e.,
\begin{equation}
\delta\theta^2 \ge \langle 1/\mathcal{I}_F(\hat{\rho}_{I\mid R,\theta})\rangle =M/[(M-1)\mathcal{I}_F(\hat{\rho}_{RI\mid \theta})_{{\rm C}\to {\rm D}}],
\end{equation}
which agrees with $1/\mathcal{I}_F(\hat{\rho}_{RI\mid \theta})_{{\rm C}\to {\rm D}}$ in the usual $M \gg 1$ case.

% \\[.05in]
% &= \frac{1+N_B+N_S \left(2 N_B+2-\kappa \right)}{4 (M-1)\kappa  N_S \left(N_S+1\right)},
% \end{align}
% for which $1/\mathcal{I}_F(\hat{\rho}_{RI\mid \theta})_{{\rm C}\to {\rm D}}$ is an excellent approximation in the usual $M \gg 1$ case.

To establish the optimality of broadband TMSV-enabled EA phase estimation, we compare $\mathcal{I}_F(\hat{\rho}_{RI\mid \theta})_{{\rm C}\to{\rm D}}$ to Gagatsos~\emph{et al.}'s universal Fisher-information upper bound~\cite{gagatsos2017bounding},
\bal 
&\mathcal{I}_F(\hat{\rho}_{RI\mid \theta})_{\rm UB}=\\
&{\frac{4M\kappa N_S\!\left[\kappa N_S+\left(1-\kappa\right)N_B'+1\right]}{(1-\kappa)\!\left[\kappa N_S \left(2N_B'+1\right)-\!\kappa N_B'\!\left(N_B'+1\right)+\!\left(N_B'+1\right)^2\right]}\,,}
\label{eq:qfi_UB}
\eal 
with $N_B' \equiv N_B/(1-\kappa)$, which applies to estimating $\theta$ from the returned radiation and retained idler originating from $M$ iid arbitrarily entangled signal-idler mode pairs whose signals each contain $N_S$ photons on average.

At low source brightness, $N_S\ll1$---where TMSV radiation has a strongly nonclassical phase-sensitive cross correlation between its signal and idler so that quantum advantage is to be expected---we find 
\begin{equation}
\mathcal{I}_F(\hat{\rho}_{RI\mid \theta})_{{\rm C}\to{\rm D}}\approx
[1-\kappa/(N_B+1)]\,\mathcal{I}_F(\hat{\rho}_{RI\mid \theta})_{\rm UB}.
\end{equation}
Hence TMSV's low-brightness optimality for EA phase estimation is established when either $\kappa\ll 1$ or $N_B\gg 1$.  

\begin{figure}[t]
    \centering
    \includegraphics[width=0.45\textwidth]{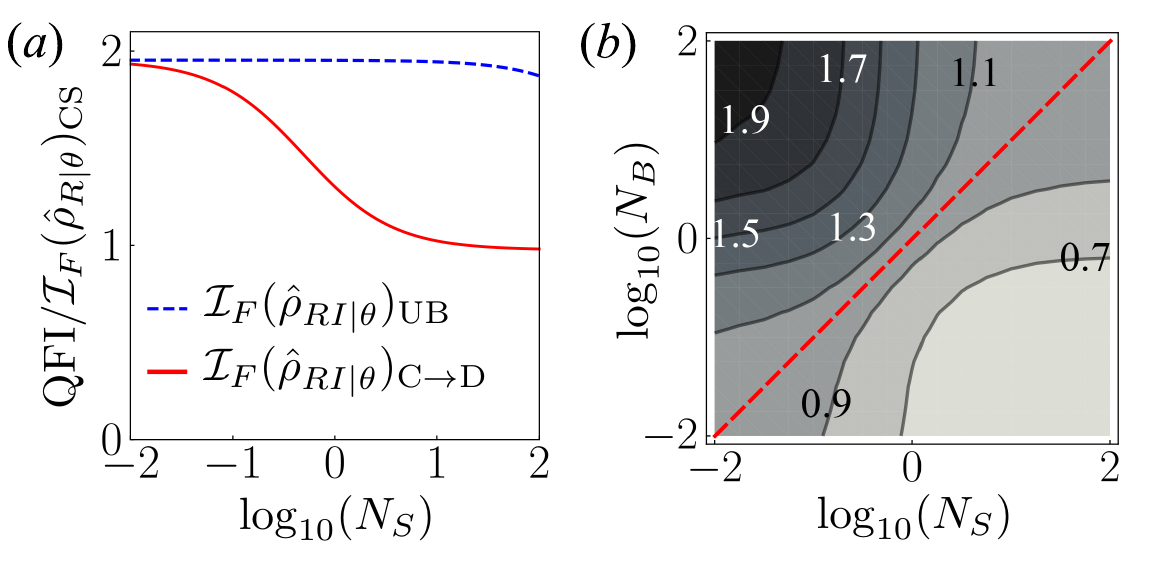}
    \caption{Normalized quantum Fisher information for $\kappa = 0.01$.  (a) $\mathcal{I}_F(\hat{\rho}_{RI\mid \theta})_{{\rm C}\to {\rm D}}/\mathcal{I}_F(\hat{\rho}_{R\mid \theta})_{\rm CS}$ (red) and $\mathcal{I}_F(\hat{\rho}_{RI\mid \theta})_{\rm UB}/\mathcal{I}_F(\hat{\rho}_{R\mid \theta})_{\rm CS}$ (blue dashed) versus $N_S$ with $N_B = 20$.  (b) Contour of $\mathcal{I}_F(\hat{\rho}_{RI\mid \theta})_{{\rm C}\to {\rm D}}/\mathcal{I}_F(\hat{\rho}_{R\mid \theta})_{\rm CS}$ versus $N_S, N_B$.  The red dashed line indicates $N_S = N_B/(1-\kappa)$.  
    \label{fig:qfi_limits_NbNs}}
\end{figure}

To assess the quantum advantage afforded by broadband TMSV-enabled EA phase estimation when $N_S \ll 1$, we compare its QFI to that of a coherent-state system using $MN_S$ photons on average~\cite{supp},
\begin{equation}
\mathcal{I}_F(\hat{\rho}_{R\mid \theta})_{\rm CS}=4M\kappa N_S/(2N_B+1).
\end{equation}
For high-brightness noise, $N_B\gg1$, this comparison shows broadband TMSV-enabled EA phase estimation has a 3\,dB QFI advantage---equivalently a 3\,dB QCRB advantage---over its coherent-state rival, as seen in 
Fig.~\ref{fig:qfi_limits_NbNs}(a).  Indeed, it is easily shown that the foregoing full 3\,dB quantum advantage only prevails when $N_S \ll 1$ and $N_B \gg 1$.  Finally, a general comparison of
$\mathcal{I}_F(\hat{\rho}_{RI\mid \theta})_{{\rm C}\to{\rm D}}$ and $\mathcal{I}_F(\hat{\rho}_{R\mid \theta})_{\rm CS}$ reveals that quantum advantage is present when $N_S\le N_B/(1-\kappa)$, as illustrated in Fig.~\ref{fig:qfi_limits_NbNs}(b). 

\subsection{Entanglement-Assisted Communication}
Since the proposal of superdense coding~\cite{bennett1992,mattle1996dense}, it has been known that pre-shared entanglement can increase the classical information capacity of quantum channels. Subsequent work~\cite{Bennett2002,bennett1999entanglement,holevo02} derived 
the per-mode classical capacity, $C_{\rm E}$, of EA communication with unlimited pre-shared entanglement. Of particular interest here is the $C_{\rm E}$ result~\cite{Bennett2002} for the thermal-loss channel $\Phi_{\kappa,0}$,  which can greatly exceed the per-mode HSW (no entanglement) capacity, $C_{\rm HSW}$~\cite{hausladen1996classical,schumacher1997sending,holevo1998capacity}, when both systems's transmitted signals have the same average photon number, i.e., the energy of the pre-shared entanglement is not included.  
Remarkably, with a low-brightness signal, $N_S \ll 1$, and high-brightness noise, $N_B \gg 1$, there is a logarithmic divergence between the two capacities, viz.,  $C_{\rm E}/C_{\rm HSW}\sim \ln (1/N_S)\rightarrow \infty$ as $N_S \rightarrow 0$.

This seminal EA capacity result, albeit encouraging, did not elucidate practical encoding strategies and receiver structures that reap the promised EA advantage. Various attempts, including protocols transmitting qubits over noisy channels~\cite{prevedel_2011,chiuri_2013} and continuous-variable superdense coding~\cite{ban1999quantum,braunstein2000,ban2000quantum},
were unable to beat the HSW capacity of a noisy bosonic channel; see Ref.~\cite{hao2021entanglement}'s supplemental materials, and Refs.~\cite{sohma2003,mizuno2005experimental,barzanjeh2013,li2002quantum}. Recent work by Shi~\emph{et al.}~\cite{shi2020practical} showed that phase encoding of broadband TMSV states does achieve the $\Phi_{\kappa,0}$ channel's EA classical capacity. Experiments to date, however, used sub-optimal reception~\cite{hao2021entanglement}, which only achieves a small constant-rate advantage over the HSW capacity. We now explain how ${\rm C}\to{\rm D}$ reception can be used with a phase-encoded transmitter to approach the EA capacity and thereby gain an enormous quantum advantage, at low signal brightness, over the HSW capacity.  

Suppose, as shown in Fig.~\ref{fig:conversion_concept_scenarios}(c), Alice has (losslessly) delivered her idler beam to Bob, thus pre-sharing its entanglement with her signal beam.  Further suppose that they also share a codebook whose codeword symbols are phase shifts randomly chosen from a uniform distribution over $[0,2\pi)$.  To communicate, Alice sends Bob her encoded classical message---as a sequence of phase-shifted $T$-s-duration signal-pulse symbols---over the $\Phi_{\kappa,0}$ channel from Sec.~\ref{ChannelModel}.  After ${\rm C}\to{\rm D}$ conversion and programmable mode selection, and given $\{E_R(t) : t\in \mathcal{T}_0\}$, the non-vacuum idler mode entering the information extractor from a single symbol with phase shift $\theta$ has conditional density operator $\hat{\rho}_{I\mid R,\theta}$ given by \eqref{PSKdensityoperator}.
The achievable per-mode classical information rate for this EA communication protocol is then
\be 
\chi_{{\rm C}\to {\rm D}}=\frac{1}{M}\int\!{\rm d}N_R\, p(N_R)\chi(\{\hat{\rho}_{\I\mid R,\theta}\}),
\label{Holevo_module}
\ee 
where $p(N_R)$ is the pdf from \eqref{NRpdf}, and 
$\chi(\{\hat{\rho}_{I\mid R,\theta}\})$ is the Holevo information~\cite{holevo1973bounds,wilde2013quantum} of the $\{\hat{\rho}_{I\mid R,\theta}\}$ state ensemble generated by $\theta$'s being uniformly distributed on $[0,2\pi)$.
The Gaussian nature of each $\hat{\rho}_{\I\mid R, \theta}$ and the uniform distribution of $\theta$ enable $\chi_{{\rm C}\to {\rm D}}$ to be evaluated efficiently from \eqref{Holevo_module}.

\begin{figure}[t]
    \centering
    \includegraphics[width=0.45\textwidth]{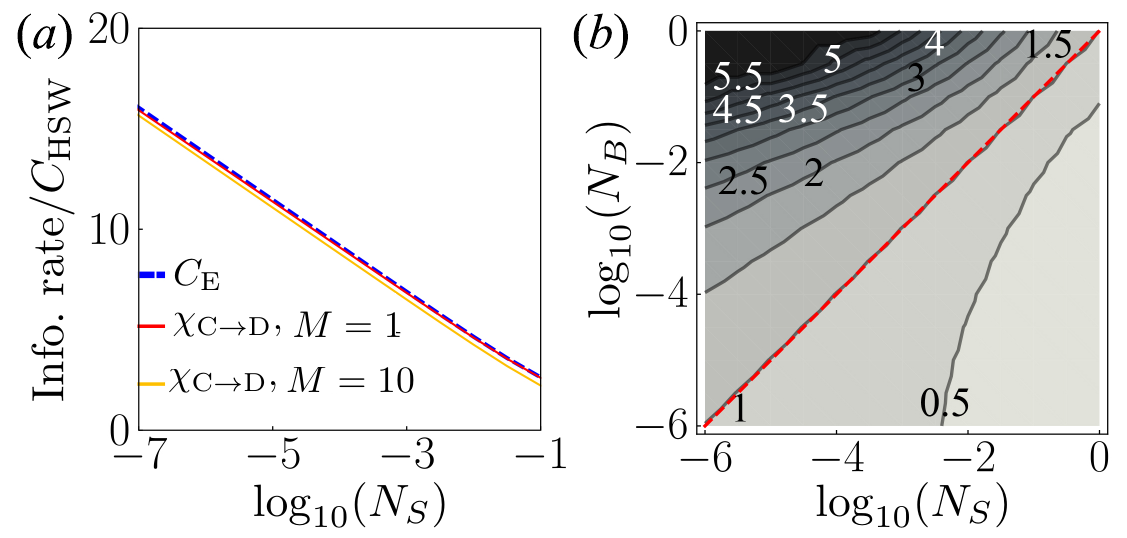}
    \caption{Per-mode information rates, $C_{\rm E}$ and $\chi_{{\rm C}\to {\rm D}}$, normalized by the per-mode HSW capacity, $C_{\rm HSW}$, for $\kappa = 0.01$.  (a) $\chi_{{\rm C}\to {\rm D}}$ for $M=1$ (red) and $M = 10^4$ (yellow) compared with $C_{\rm E}$ (blue dashed); $N_B = 100$ is assumed.  The $M=1$ curve is only included to illustrate the insensitivity of $\chi_{{\rm C}\to {\rm D}}$ to time-bandwidth product; EA classical communication will operate with $M \gg 1$. (b) Contour of $\chi_{{\rm C}\to {\rm D}}/C_{\rm HSW}$ versus $N_S, N_B$.  The red dashed line indicates $N_S = N_B$.  
    \label{fig:EACOMM_c2d_limits}}
\end{figure}

Figure~\ref{fig:EACOMM_c2d_limits}(a) plots $C_{\rm E}/C_{\rm HSW}$ and $\chi_{{\rm C}\to {\rm D}}/C_{\rm HSW}$ versus $N_S$ for $\kappa = 0.01$ and $N_B = 100$; see Ref.~\cite{supp} for details. It shows that 
$\chi_{{\rm C}\to {\rm D}}$ closely matches the EA capacity.  Indeed, as $N_S\to 0$, we obtain $\chi_{{\rm C}\to {\rm D}}
   \sim \kappa  N_S \ln(1/N_S)/(N_B+1),$
which achieves the scaling of the EA capacity. The same optimality result holds when Alice's modulation is restricted to BPSK, i.e., $\hat{\rho}_{I\mid R,\theta}$ from \eqref{PSKdensityoperator} maximizes the per-mode Holevo information for BPSK modulation.   

To exhibit the huge advantage enjoyed by EA classical communication with ${\rm C}\to{\rm D}$ reception over the $\Phi_{\kappa,0}$ channel's HSW capacity, ${\rm C}_{\rm HSW}$ from Giovannetti~\emph{et al.}~\cite{giovannetti2014ultimate},  Figure~\ref{fig:EACOMM_c2d_limits}(b) plots their ratio versus $N_S,N_B$ for $\kappa = 0.01$. When $N_S\ll1, N_B\gg1$, we indeed see a huge advantage.  Moreover, we find that TMSV-enabled EA communication offers quantum advantage whenever $N_S< N_B$, similar to what we found for QI target detection and EA phase estimation. Note that the $N_S\ll1$ region is relevant to covert communication, where signal brightness is kept low to avoid detection by an adversary~\cite{bash2015quantum,shi2020entanglement}.

\section{Coherent-State Information Extractors} \label{ExtractorDesign}
\begin{figure*}
    \centering
    \includegraphics[width=0.7\textwidth]{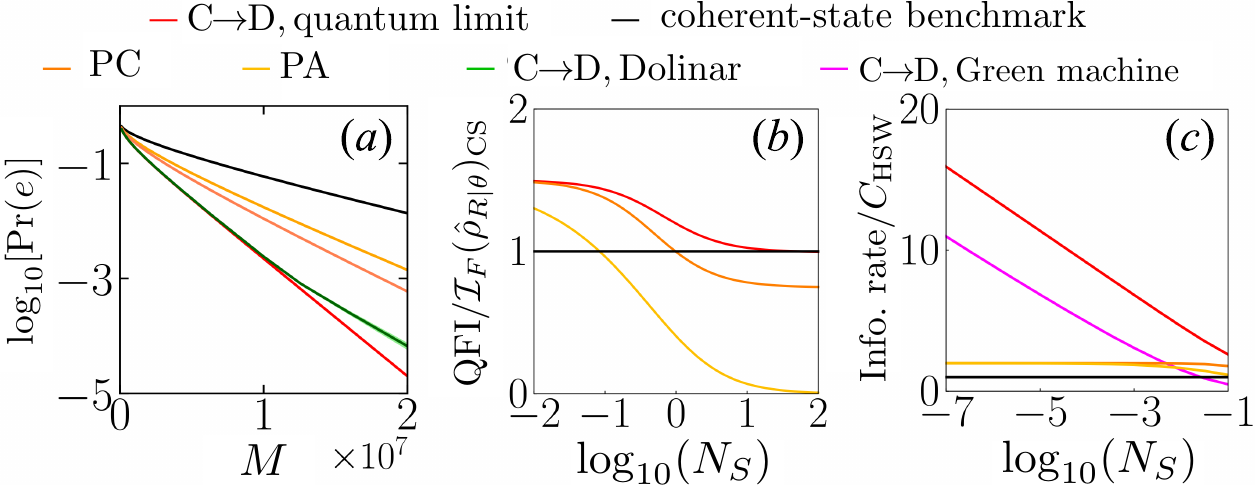}
    \caption{(a) QI error probability versus time-bandwidth product, $M$, for $N_S = 0.001$, $\kappa = 0.01$, and $N_B = 20$.  The quantum limit (red) of ${\rm C}\to{\rm D}$ reception and its practical performance achieved by the Dolinar receiver (green) are shown in comparison with those of PC (orange) and PA (yellow) reception, as well that of a classical system using a coherent-state transmitter  with a homodyne receiver (black).  The \QZ{narrow light-green} shading shows the simulation's numerical precision\QZ{, which is quite high.}  (b) Normalized quantum Fisher information, ${\rm QFI}/\mathcal{I}_F(\hat{\rho}_{R\mid \theta})_{\rm CS}$, versus $N_S$ for phase estimation with $\kappa = 0.98$ and $N_B = 1$.  All receivers use homodyne detection. (c) Normalized information rates versus $N_S$ for $\kappa = 0.01$ and $N_B=100$: $\chi_{{\rm C}\to{\rm D}}/C_{\rm HSW}$ for both quantum-limited and Green machine implementations, $\chi_{\rm PC}/C_{\rm HSW}$, $\chi_{\rm PA}/C_{\rm HSW}$, and $C_{\rm HSW}/C_{\rm HSW}=1$.    \label{fig:receivers}}
\end{figure*}

In this section we provide some details about the ${\rm C}\to{\rm D}$ receiver's coherent-state information extractor. 
As in Sec.~\ref{Performance}, we assume that the noise brightness, $N_v$, of the $\{\phi_m(t) : m\neq 0\}$ modes in $\hat{E}_I'(t)$ is low enough to be inconsequential.  Thus, under quantum-limited conditions, i.e., when the noise brightness, $N_{I\mid R}$, of that field's $\phi_0(t)$ mode is also inconsequentially low, the information extraction tasks for our EA scenarios reduce to those of detection, estimation, or communication for coherent states.  As such, we can draw upon an extensive library of theoretical and experimental results for coherent-state receivers~\cite{Dolinar1973,tsujino2011quantum,chen2012optical,becerra2013experimental,becerra2015photon,ferdinand2017multi,burenkov2018quantum,izumi2020experimental,guha2011structured,cui2022quantum} in fleshing out our optimal information extractors.  We also benchmark them against PA and PC receivers, which are state-of-the-art systems for broadband TMSV-enabled EA   protocols~\cite{Guha2009,shi2020practical,hao2021entanglement}, and classical (coherent-state) systems. See Ref.~\cite{supp} for a review of PA and PC reception.

In QI target detection, under the ideal (quantum-limited) conditions assumed in Sec.~\ref{Performance}A, the information extractor needs to distinguish between the $\phi_0(t)$ mode's vacuum state $|0\rangle$ and its non-vacuum coherent state $|\zeta\sqrt{N_R}\rangle$.   Consequently, the well-known Dolinar receiver~\cite{Dolinar1973} saturates the Helstrom bound---via adaptive control of linear displacement operations and photon counting---and completes the ${\rm C}\to{\rm D}$ receiver design for QI target detection~\cite{supp}. Owing to $0 < N_S\ll 1$, however, QI's conditional density operator $\hat{\rho}_{0\mid R}$ is primarily, but not entirely, a vacuum state.  Likewise, because $0 < N_{I\mid R} \ll 1$, QI's conditional density operator $\hat{\rho}_{1\mid R}$ is approximately, but not exactly, the $|\zeta\sqrt{N_R}\rangle$  coherent state.  To capture the effects of those noises, we evaluated the Dolinar receiver's performance, for equally-likely target absence or presence, by numerical simulation~\cite{supp}.  

Figure~\ref{fig:receivers}(a) plots, versus time-bandwidth product, the resulting error probability for ${\rm C}\to{\rm D}$ reception with \JHS{the} Dolinar receiver (green) assuming $N_S = 0.001, N_B=20$ and $\kappa = 0.01$.   Also included are the PA and PC receiver's error probabilities along with the ideal result assuming optimal noisy coherent state discrimination in \eqref{PeCD}.  Below $M = 10^7$ the Dolinar receiver reaches its quantum limit. At higher $M$ values, however, its performance diverges therefrom owing to the emerging effect of having $N_S, N_{I\mid R} >0$.  This divergence may be ameliorated using optimized feed-forward control from machine learning~\cite{cui2022quantum}. As expected, PA  and PC reception perform worse than ${\rm C}\to{\rm D}$ reception, although they still outperform the classical benchmark of a coherent-state transmitter and homodyne receiver (black)~\cite{Guha2009}.

In EA phase estimation, the ${\rm C}\to{\rm D}$ receiver's information extractor is confronted with estimating the unknown $\theta$ embedded in $\hat{E}'_I(t)$'s $\phi_0(t)$ mode, specifically in that mode's coherent-state component $|\zeta\sqrt{N_R}e^{i\theta}\rangle$.  Cued by a preliminary measurement that provides an initial estimate $\theta_c$---or equivalently, by means of an adaptive protocol---the ${\rm C}\to{\rm D}$ receiver will achieve the QFI from \eqref{eq:qfi_c2d} when it uses a homodyne-detection information extractor that measures ${\rm Im}(\hat{a}'_{I_0}e^{-i\theta_c})$ and $|\theta-\theta_c| \ll 1$; see Ref.~\cite{supp} for details.   Although PA and PC reception, using $\theta_c$-cued homodyne detection, match ${\rm C}\to{\rm D}$ reception's 3\,dB quantum advantage in the $N_S \ll 1, \kappa \ll 1, N_B\gg 1$, asymptotic regime~\cite{shi2020practical}, ${\rm C}\to{\rm D}$ reception has higher QFI outside the asymptotic regime.  This QFI advantage for ${\rm C}\to{\rm D}$ reception is especially evident when $\kappa \sim 1$ and $N_B \le 1$, as illustrated in Fig.~\ref{fig:receivers}(b).  Interestingly, for EA phase estimation, the ${\rm C}\to{\rm D}$ receiver's programmable mode-selection and coherent-state information extraction modules can be combined into a single homodyne-detection operation, using a strong classical-state local oscillator $E_{\rm LO}(t) = \sqrt{N_{\rm LO}}\,\phi_0(t)e^{i\theta_c}$.

The information extractor's task for EA communication reduces to achieving the Holevo information of an $\hat{E}'_I(t)$ whose sole excited mode takes on an ensemble of phase-modulated coherent states which are embedded in weak thermal noise~\cite{guha2011structured,wilde2012explicit}. To enable a near-term measurement design, we assume Alice's transmitter uses BPSK modulation with the Hadamard code.  Bob's information extractor then uses a Green machine---a structured array of beam splitters---that transforms the phase-modulated Hadamard code to pulse-position modulation (PPM)~\cite{guha2011structured, guha2020infinite,supp} followed by photon counting to identify the PPM time slot containing the coherent-state excitation.  
Figure~\ref{fig:receivers}(c) (magenta) shows that this Green-machine extractor achieves the optimal $\ln (1/N_S)$ scaling of $\chi_{\rm C\to D}$ while only relying on linear optics and photon counting.  See Ref.~\cite{supp} for details.  Note that the ${\rm C}\to{\rm D}$ Green-machine performance (magenta) is constant factor off from the quantum limit (red) and even falls below the classical capacity at high $N_S$. This behavior is not due to the conversion module, which achieves optimal performance as shown in Fig.~\ref{fig:EACOMM_c2d_limits}(a). Instead, it is due to the Green machine's coherent-state processor having room for improvement.

\section{Mode-Selector Designs} \label{SelectorDesign}

In this section we present two high-level concepts for the ${\rm C}\to{\rm D}$ receiver's programmable mode selector, along the lines suggested in Fig.~\ref{fig:notionalmodeselect}, and a more detailed treatment of two near-optimal but more practical mode selectors.  Ideal realization  of Fig.~\ref{fig:notionalmodeselect}-style mode selection is critical to ${\rm C}\to{\rm D}$ reception's providing the full quantum advantage of broadband TMSV-enabled EA detection, estimation, and classical communication, as explained in Sec.~\ref{CtoD}.  In the case of QI, ideal realizations of our near-optimal mode selectors yield SNRs that are $\sim$2\,dB better than those of ideal PC and PA receivers, hence $\sim$1\,dB worse than what is obtained from optimal mode selection in the $N_S \ll 1, \kappa \ll 1, N_B \gg 1$ asymptotic regime.

\subsection{Architectures for Optimal Mode Selection}
\begin{figure}[t]
    \centering    \includegraphics[width=0.425\textwidth]{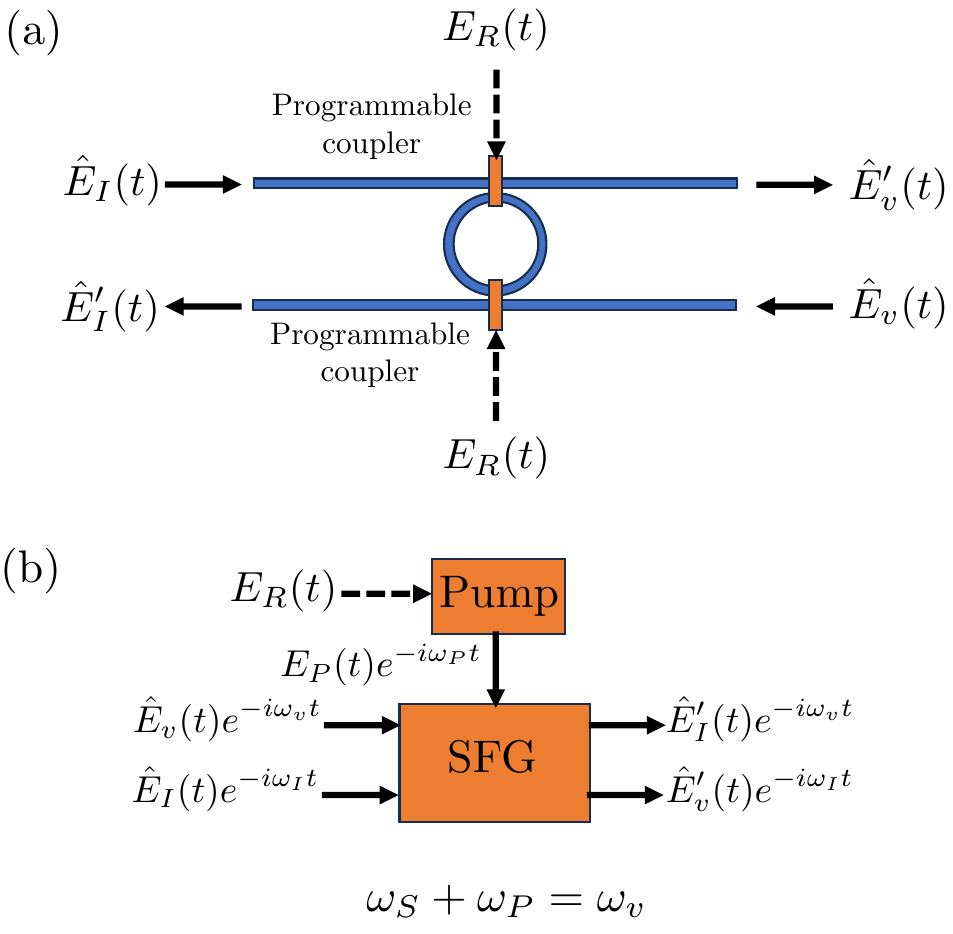}
    \caption{Two possible approaches to realizing the optimal programmable mode-selection module. (a) Programmable add-drop multiplexer in which a ring resonator, with couplers programmed by the ${\rm C}\to{\rm D}$ receiver's heterodyne-detector output $\hat{E}_R(t)$, switches the $\phi_0(t)$ mode in $\hat{E}_I(t)$ from the upper input rail to the lower output rail, and the $\phi_0(t)$ mode in $\hat{E}_v(t)$ from the lower input rail to the upper output rail, while the $\{\phi_m(t) : m \neq 0\}$ modes on each input rail are transmitted through to their respective output rail. (b) Programmable sum-frequency generation (SFG) in which a strong frequency-$\omega_P$ pump laser is programmed by  $E_R(t)$ to upconvert the $\phi_0(t)$ mode of the frequency-$\omega_I$ idler to $\omega_v = \omega_P+\omega_S$, and downconvert that mode of the frequency-$\omega_v$ near-vacuum field $\hat{E}_v(t)$ to $\omega_I$, while leaving the other modes of both fields unaffected.   }
    \label{fig:adddrop} 
\end{figure}

Figure~\ref{fig:adddrop}(a) shows the first of our concepts for optimal programmable mode selection.  Generalizing from fiber-network technology's reconfigurable optical add-drop multiplexers (ROADMs)~\cite{Gringieri2010,Strasser2010}, we propose a programmable add-drop multiplexer in which programmable couplers on two rails connected to a ring resonator are controlled by the ${\rm C}\to{\rm D}$ receiver's heterodyne-detector output $\hat{E}_R(t)$, so that the $\phi_0(t)$ mode in $\hat{E}_I(t)$ is transferred from the upper input rail to the lower output rail, and conversely the $\phi_0(t)$ mode in $\hat{E}_v(t)$ is transferred from the lower input rail to the upper output rail, while the $\{\phi_m(t) : m \neq 0\}$ modes on each input rail are transmitted through to their respective output rail.  There will be considerable challenges in implementing this modified-ROADM concept, e.g., it is highly likely that waveform needed to drive the couplers will be a complicated function of $E_R(t)$ that must be computed on the fly.  

Figure~\ref{fig:adddrop}(b) shows our second concept for optimal programmable mode selection.  Unlike our first concept, this one has been demonstrated in the optical domain for extracting simple mode shapes~\cite{kowligy2014quantum,kumar2019mode,manurkar2016multidimensional}.  Here the positive-frequency idler field, $\hat{E}_I(t)e^{-i\omega_It}$, and a positive-frequency near-vacuum state field, $\hat{E}_v(t)e^{\omega_vt}$ where $\omega_v > \omega_I$, are inputs to an SFG crystal. A strong  (and hence classical) positive-frequency pump field, $E_P(t)e^{-i\omega_Pt}$ where $\omega_S+\omega_P = \omega_v$, is also applied to that crystal. 
The pump's mode shape is controlled by $E_R(t)$ to upconvert the idler field's $\phi_0(t)$ mode to center frequency $\omega_v$, and downconvert the near-vacuum field's $\phi_0(t)$ mode to center frequency $\omega_I$, while leaving the other modes of both fields unaffected. 

At microwave frequencies, for which idler operations must be done in a dilution refrigerator, our SFG mode-selection architecture has a significant upside: having $\hat{E}_I'(t)$ be at a significantly higher frequency than $\hat{E}_I(t)$ drives down $N_v$ exponentially.  A significant downside of SFG mode selection is the same as what we cited for the programmable add-drop architecture, i.e., it is highly likely that pump waveform needed will be a complicated function of $E_R(t)$ that must be computed on the fly.  

\subsection{Architectures for Near-Optimal Mode Selection} 

\begin{figure}[t]
    \centering
   \includegraphics[width=0.45\textwidth]{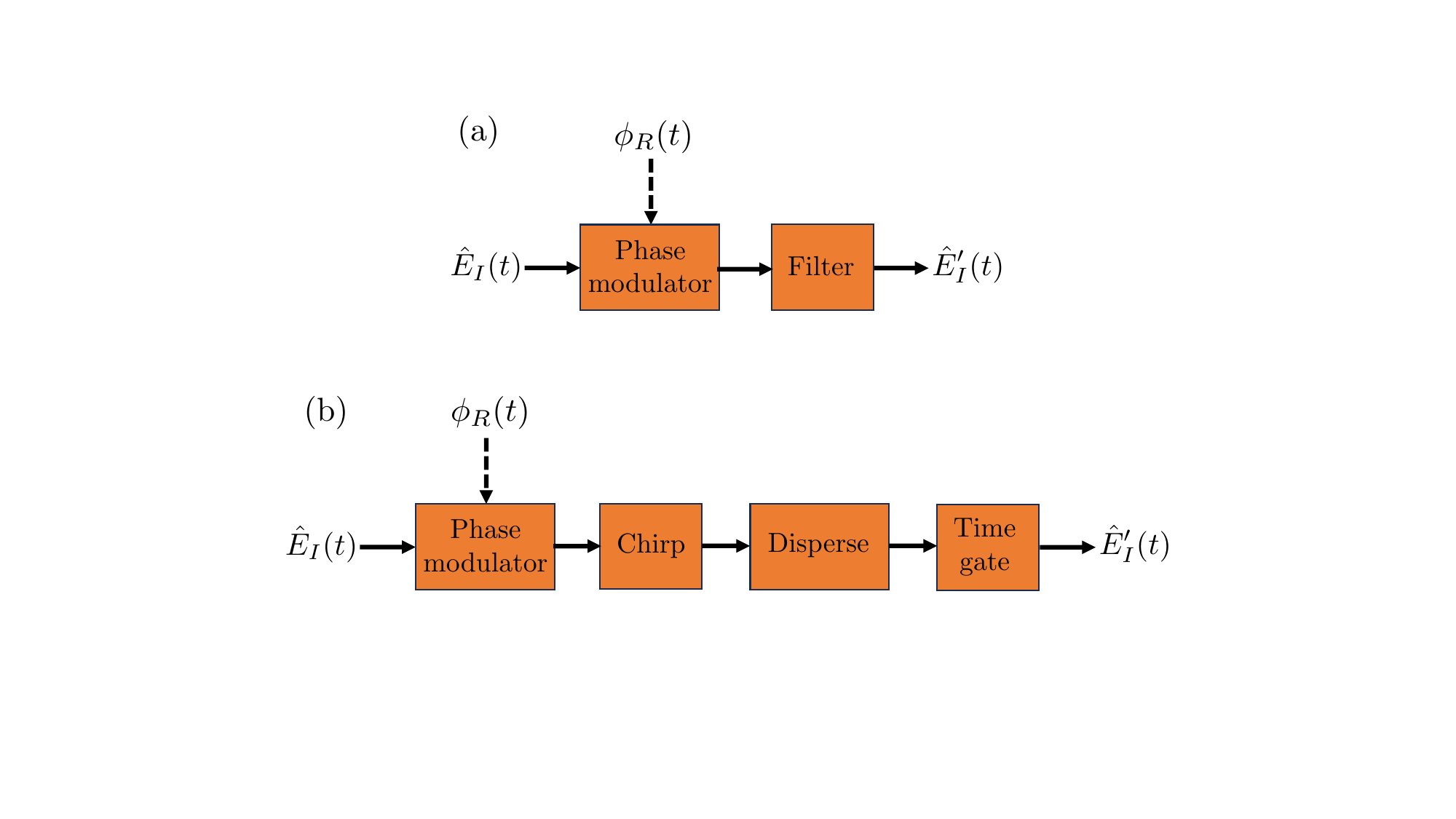}
    \caption{Two possible approaches to realizing a near-optimal programmable mode-selection module. Both use a phase demodulator (not shown) to extract $\phi_R(t)$ from the amplitude and phase decomposition of the heterodyne detector's output, i.e., $E_R(t) = A_R(t)e^{i\phi_R(t)}$ with $A_R(t) \ge 0$ and $\phi_R(t)$ real valued, and both phase modulate $\hat{E}_I(t)$ with $\phi_R(t)$.  (a) In the near-optimal mode selector's filter architecture, $\hat{E}_I(t)e^{i\phi_R(t)}$ undergoes brickwall filtering to achieve the best compromise between passing its coherent-state component while suppressing its thermal-noise bath.  (b) In the near-optimal mode selector's chirp-disperse-gate architecture, $\hat{E}_I(t)e^{i\phi_R(t)}$ is multiplied by a chirp waveform, then passes through a dispersion module, and is time gated to achieve the best compromise between passing its coherent-state component while suppressing its thermal-noise bath.  See the text for details.}
    \label{fig:nearoptselection} 
\end{figure}

In view of the substantial challenges in realizing either of our optimal mode-selection architectures, we have two near-optimal architectures, shown in 
Fig.~\ref{fig:nearoptselection}, that have far simpler paths to realization.  Both use a phase demodulator (not shown) to extract $\phi_R(t)$ from the amplitude and phase decomposition of the heterodyne detector's output, i.e., $E_R(t) = A_R(t)e^{i\phi_R(t)}$ with $A_R(t) \ge 0$ and $\phi_R(t)$ real valued, and both phase modulate $\hat{E}_I(t)$ with $\phi_R(t)$.  In the near-optimal mode selector's filter architecture, $\hat{E}_I(t)e^{i\phi_R(t)}$ undergoes brickwall filtering, whereas in its chirp-disperse-gate architecture it is multiplied by a chirp waveform, then passed through a dispersive compressor and time gated.  In both cases the goal is to preserve as much of the average photon number in $\hat{E}_I(t)$'s $\phi_0(t)$ mode while reducing the average photon number of the  accompanying noise component to an inconsequential level.  The following  results illustrate the extent to which the preceding architectures achieve their dual goals of displacement preservation and noise suppression. Additional details are given in \cite{supp}. 

Conditioned on knowledge of $\{E_R(t): t\in \mathcal{T}_0\}$, the phase-compensated idler, $\hat{E}_I(t)e^{i\phi_R(t)}$, consists of the coherent state $|\zeta A_R(t)e^{i\theta}\rangle$ embedded in a zero-mean, conditionally-Gaussian noise field $\hat{E}'_n(t) \equiv \hat{E}_n(t)e^{i\phi_R(t)}$ that is completely characterized by its auto-correlation function,
\begin{align}
\mathbb{E}[\hat{E}&^{\prime \dagger}_n(t)\hat{E}'_n(t')\mid E_R(t)] 
= \nonumber \\[.05in] &e^{-i[\phi_R(t)-\phi_R(t')]}N_{I\mid R}W\,{\rm sinc}[\pi W(t-t')],
\end{align}
where ${\rm sinc}(x) \equiv \sin(x)/x$.  The unconditional behavior of $\hat{E}'_n(t)$ is still broadband, but it is no longer Gaussian.  On the other hand, the unconditional behavior of $\zeta A_R(t)e^{i\theta}$ benefits from $A_R(t)$ having a nonzero mean,
\begin{align}
\langle A_R(t)\rangle &= \int_0^\infty\!{\rm d}A_R\,\frac{2A_R^2}{W(\kappa N_S+N_B+1)}e^{-A_R^2/W(\kappa N_S+N_B+1)} \nonumber \\[.05in]
&= \sqrt{\pi W(\kappa N_S + N_B + 1)/4},\mbox{ for $t\in \mathcal{T}_0,$}
\label{ArBar}
\end{align} 
where the first equality follows from $E_R(t)$ being a zero-mean, circulo-complex Gaussian random process with $\langle |E_R(t)|^2\rangle = W(\kappa N_S + N_B + 1)$.  Equation~(\ref{ArBar}) implies that 
\begin{equation}
N_{\bar{A}_R} \equiv \int_{t\in \mathcal{T}_0}\!{\rm d}t\,\zeta^2\langle A_R(t)\rangle^2 = \frac{\pi}{4}\int_{t\in \mathcal{T}_0}\!{\rm d}t\,\zeta^2\langle A^2_R(t)\rangle,
\label{NarBar}
\end{equation}
showing that $\{\zeta \langle A_R(t)\rangle : t\in \mathcal{T}_0\}$ carries 78.5\% of the displacement energy in $\hat{E}_I(t)$. Moreover, it does so in bandwidth much \emph{narrower} than that of $\hat{E}'_n(t)$.  Armed with these statistics, we now obtain some displacement and noise results for our two near-optimal architectures.  

\begin{figure}[t]
    \centering
    \includegraphics[width=.8\linewidth]{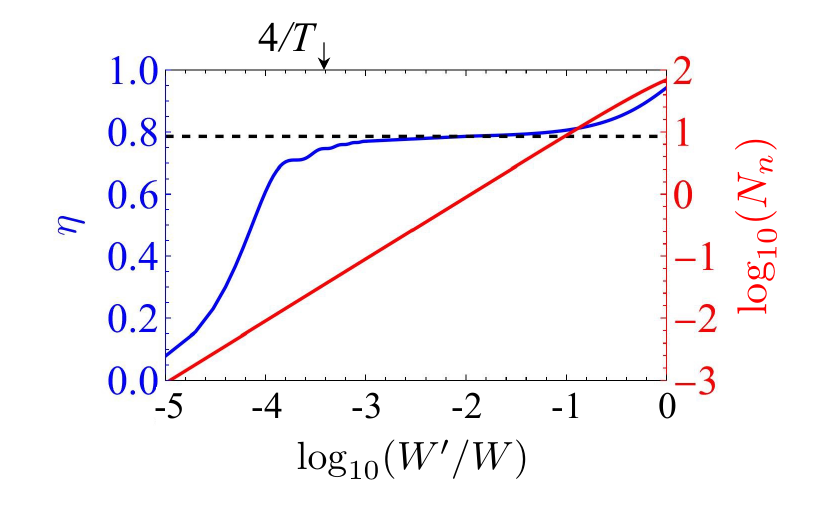}
    \caption{The filter architecture's filtering efficiency, $\eta \equiv N_d/N_I$, versus $W'/W$ (blue) with $M = TW = 10^4$, where $N_d$ is the average displacement photon number at the filter's output and $N_I = M\kappa N_S(N_S+1)/(\kappa N_S + N_B + 1)$ is the average displacement photon number at its input. The average noise photon number, $N_n$, at the filter's output versus $W'/W$ (red) with $N_S = 0.01, \kappa = 0.01, N_B = 20$, and $T=1\,\upmu$s.  The black dashed line is $\pi/4 \approx 78.5\%$, the fraction of the $\hat{E}_I(t)$'s displacement energy carried by $\{\zeta \langle A_R(t)\rangle : t\in \mathcal{T}_0\}$.
}
    \label{fig:NdPe_filter}
\end{figure}

For the filter architecture shown in Fig.~\ref{fig:nearoptselection}(a), the filter block passes a $W'$\,Hz bandwidth of the positive-frequency phase-compensated idler field, $\hat{E}_I(t)e^{-i[\omega_It-\phi_R(t)]}$, centered on $\omega_I/2\pi$ and suppresses all other frequencies.  At baseband, this amounts to passing $\hat{E}_I(t)e^{i\phi_R(t)}$ through the brickwall low-pass filter whose frequency response is
\begin{equation}
H(\omega) = \left\{\begin{array}{ll}
1, & \mbox{for $|\omega|/2\pi \le W'/2,$}\\[.05in]
0, & \mbox{otherwise,}\end{array}\right.
\end{equation}
with $W'\ll W$.  Because of the result we obtained in \eqref{NarBar}, we expect that strong noise suppression should be obtainable with good displacement preservation.  Figure~\ref{fig:NdPe_filter}, using bandwidths relevant to microwave operation, shows that this is indeed the case.  Here, the blue curve is a plot of the filtering efficiency, $\eta \equiv N_d/N_I$, versus $W'/W$ with $M= TW = 10^4$, where $N_d$ is the average displacement photon number at the filter's output and $N_I = M\kappa N_S(N_S+1)/(\kappa N_S + N_B + 1)$ is the average displacement photon number at its input.  Figure~~\ref{fig:NdPe_filter}'s red curve is a plot of $N_n$, the average noise photon number at the filter's output, versus $W'/W$ with $N_S = 0.01,\kappa =0.01,$ and $N_B=20$.  We see that at $W'= 4/T = 40\,$MHz  the filter approach captures $\sim$75\% of the input displacement energy while suppressing noise in its output to an inconsequential average of $\sim$0.03 photons. In this asymptotic-regime example, the filter architecture's SNR in QI, given by ${\rm SNR} = \eta M\kappa N_S/N_B$ is 1.27\,dB inferior to the $M\kappa N_S/N_B$ SNR achieved with optimal mode selection, but it is 1.73\,dB better than those of PC or PA reception.

Our second architecture for near-optimal mode selection is the chirp-disperse-gate setup shown in Fig.~\ref{fig:nearoptselection}(b).  Chirp-disperse processing was first proposed for pulse compression~\cite{kolner1988active}, but now it is  widely used in temporal-spectral photon manipulation~\cite{sosnicki2023interface,mittal2017temporal} and as a time lens~\cite{kolner1989temporal}. As we shall see below, this mode selector's output photon flux is an image of its input's fluorescence spectrum. Time gating that output is thus equivalent to the filter architecture's low-pass filtering.  The specifics are as follows. \QZ{In particular, this sub-optimal architecture can potentially be made optimal via cascading a few dispersion and phase modulator elements, as analyzed in Ref.~\cite{joshi2022picosecond}.}

First, the phase-compensated idler field, $\{\hat{E}_I(t)e^{i\phi_R(t)} : t\in \mathcal{T}_0\}$, is multiplied by the bandwidth-$W_c$ chirp waveform, $e^{-i\pi W_ct^2/T}$, using a phase modulator.  The compensated and chirped idler is then passed through a dispersive filter with impulse response $h_d(t) = \sqrt{W_c/iT}\,e^{i\pi W_ct^2/T}$ to compress the chirp.  The chirp modulation and dispersive filtering are both unitary linear operations, so the filter's output, $\hat{E}_{\rm out}(t)$, is given by
\begin{eqnarray}
\lefteqn{\hat{E}_{\rm out}(t) = \int\!{\rm d}u\,\hat{E}_I(u)e^{i\phi_R(u)}e^{-i\pi W_cu^2/T}h_d(t-u),}\nonumber \\[.05in]
&=& \sqrt{\frac{W_c}{iT}}\,e^{i\pi W_ct^2/T}\int\!{\rm du}\,\hat{E}_I(u)e^{i\phi_R(u)}e^{-i2\pi W_cut/T} \\[.05in]
&=& \sqrt{\frac{W_c}{iT}}\,e^{i\pi W_ct^2/T}\left.\hat{\mathcal{E}}_{\rm in}(\omega)\right|_{\omega = 2\pi W_ct/T},
\end{eqnarray}
where $\hat{E}_{\rm in}(t) \equiv \hat{E}_I(t)e^{i\phi_R(t)}$ and 
$\hat{\mathcal{E}}_{\rm in}(\omega) \equiv \int\!{\rm d}t\,\hat{E}_{\rm in}(t)e^{-i\omega t}$ is its Fourier transform.  At this point it should be clear that time gating $\hat{E}_{\rm out}(t)$ to a $T'$-s-long interval centered on the midpoint of $\mathcal{T}_0$ is equivalent to the filter architecture's frequency filtering.  Indeed, see \cite{supp}, the chirp-disperse-gate architecture's filtering efficiency and average output-noise photon number match those of the filtering architecture when they filter down to the same time-bandwidth product, i.e., $TW' = T'W_c$.  

For both our near-optimal mode selectors, a final note is in order.  Our analysis of their performance assumes single-pulse transmission.  However, the operations they perform on $\{\hat{E}_I(t)e^{i\phi_R(t)} : t\in \mathcal{T}_0\}$ produce outputs that extend beyond $t \in \mathcal{T}_0$.  In EA classical communication this can lead to intersymbol interference that can impair communication performance.  No such problem occurs for single-pulse QI target detection or EA phase estimation.

\section{Quantum Channel Pattern Classification} \label{Composite}

\begin{figure}[t]
    \centering
    \includegraphics[width=0.45\textwidth]{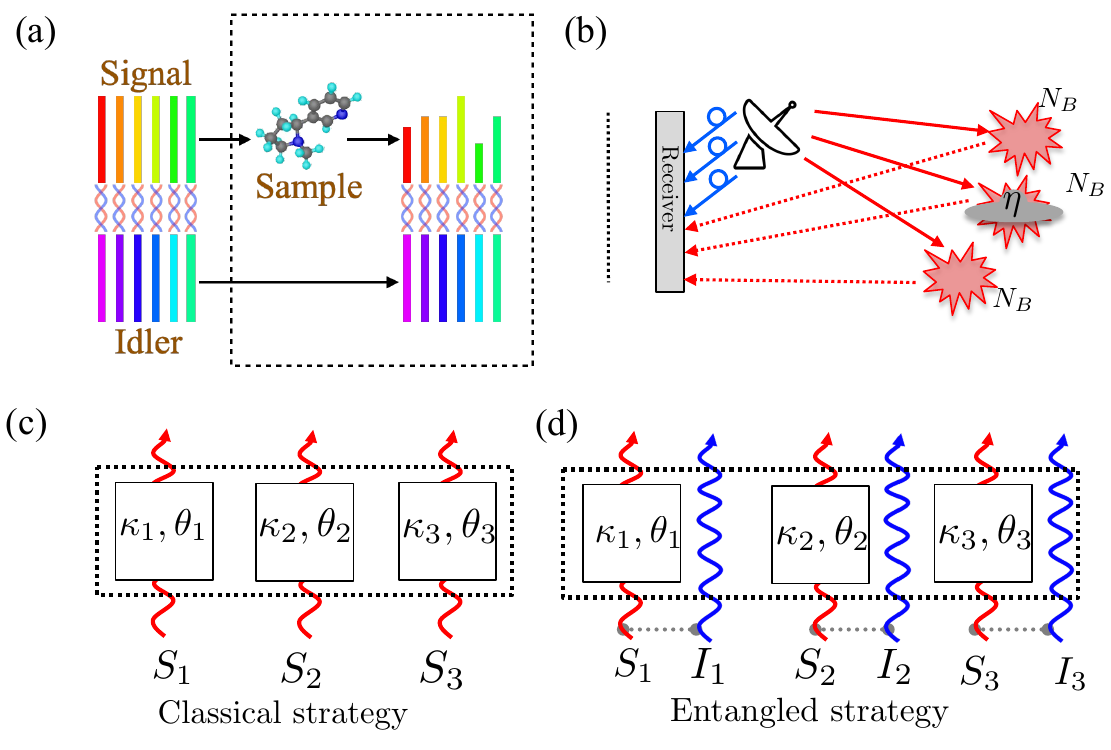}
    \caption{Quantum sensing over a composite channel.  (a) EA spectroscopy scenario.  (b) EA target-location scenario.  (c) Classical model for $K=3$ sub-channels. (d) EA model for $K=3$ sub-channels.}
    \label{fig:scheme_pattern}
\end{figure}

So far we have considered sensing for a single phase-shift thermal-loss channel, $\Phi_{\kappa, \theta}$. Complex sensing problems often involve composite channels, in which the information being sought is distributed across $K >1$ sub-channels. For example, in a spectroscopy measurement, as shown in Fig.~\ref{fig:scheme_pattern}(a), the transmissivities $\{\kappa_k\}$ at different frequencies (indexed by $k$) jointly provide information about the sample's composition~\cite{shi2020spectroscopy}. In a target location problem, the target may be in one of many possible positions all of which must be probed to determine its location~\cite{zhuang2020entanglement}. 

These and other complex-sensing applications, such as barcode recognition~\cite{banchi2020quantum} and quantum ranging~\cite{zhuang2020entanglement,zhuang2022ultimate}, fall under the rubric of quantum channel pattern classification. A $K$-fold composite channel is a tensor product of $K$ sub-channels,
$
\Phi_{\bm \kappa, \bm \theta}=\otimes_{k=1}^K \Phi_{\kappa_k, \theta_k},
$
where  $\bm \kappa=\{\kappa_k\}_{k=1}^K$, $\bm \theta=\{\theta_k\}_{k=1}^K$, and we have assumed that the background-noise brightness, $N_B$, is identical across all sub-channels. Figure~\ref{fig:scheme_pattern} shows an example of quantum sensing over a $K=3$ composite channel.  

Figure~\ref{fig:scheme_pattern}(c)'s classical strategy uses a coherent-state probe for each sub-channel.  In contrast,  Fig.~\ref{fig:scheme_pattern}(d)'s quantum strategy divides the signal beam from a broadband TMSV source to address all the sub-channels, while retaining that source's idler for a subsequent joint measurement with the radiation returned from them.  For hypothesis testing between $M$ general composite channels, each comprised of $K$ sub-channels, we have the following theorem.  

\begin{theorem}
\label{theorem1}
Let $\{\Phi_{{\bm \kappa}_m,{\bm \theta}_m}: m = 1,2,\dots,M\}$ be a collection of composite channels, each containing $K$ sub-channels with high-brightness ($N_B \gg1$) noise.  Further suppose that the hypothesis test of interest is to determine which composite channel is active, given they occur with known prior probabilities $\{\pi_m\}$.  Then, distribution of a low-brightness ($N_S \ll 1$) broadband TMSV signal beam across the $\{\Phi_{{\bm \kappa}_m,{\bm \theta}_m}\}$, followed by optimal joint measurement with the retained idler beam, provides a $6\,$dB quantum advantage in error-probability exponent over the best coherent-state competitor of the same transmitted energy.
\end{theorem}

Theorem~1's proof, given in Ref.~\cite{supp}, relies on the ${\rm C}\to{\rm D}$ conversion and programmable mode-selection modules, followed by an information extractor that does optimal discrimination among $M$ coherent states. Hence, Theorem~1 immediately implies that ${\rm C}\to{\rm D}$ reception with $M$-ary information extraction is  optimal for the quantum-channel position sensing~\cite{zhuang2020entanglement}.

\section{Discussion} \label{Discussion}
We have proposed a ${\rm C}\to{\rm D}$ receiver for broadband TMSV-enabled EA detection, estimation, and communication and showed that it realizes optimal quantum advantage in all those applications.  This receiver, comprised of a ${\rm C}\to{\rm D}$ converter, a programmable mode selector, and a coherent-state information extractor, is a cascaded POVM \QZ{that is LOCC between the returned and retained \emph{fields}, $\{\hat{E}_R(t),\hat{E}_I(t)\}$, but not between the returned and retained \emph{mode pairs}, $\{\hat{a}_{R_m},\hat{a}_{I_m}\}$.  Hence it is not bound by no-go theorems that forbid obtaining full quantum advantage from mode-pair LOCC strategies.}  Neither does it require a joint measurement between the returned radiation and the retained idler.  Overall, these features confer three advantages on ${\rm C}\to{\rm D}$ reception: (1) its structure is far more practical and less complicated than the FF-SFG receiver for optimal QI target detection, and it strictly outperforms the FF-SFG receiver in EA communication; (2) it outperforms PC and PA reception in phase estimation outside of the $N_S \ll 1, \kappa \ll 1, N_B \gg 1$ asymptotic regime; and (3) it can detect the returned radiation at one location and perform mode selection and information extraction on the idler at another location. Future applications of the ${\rm C}\to{\rm D}$ architecture include receiver design for quantum ranging~\cite{zhuang2022ultimate} and EA classical network communication~\cite{shi2021entanglement}. 
\QZ{In terms of ${\rm C}\to{\rm D}$ reception's near-term implementation, the principal challenge is building a high-efficiency programmable mode selector for high time-bandwidth product random mode shapes.}

Two final topics are worthy of discussion:  the prospects for a ${\rm C}\to{\rm D}$ receiver's mode selector having a near-vacuum auxiliary field, $\hat{E}_v(t)$, whose noise brightness, $N_v$, can be so low as to be ignored; and the relationship between the present paper and the works of Reichert~\emph{et al.}~\cite{Reichert2023}, Angeletti~\emph{et al.}~\cite{Angeletti2023} and Chen and Zhuang~\cite{chen2023}, which built on the foundation established in Shi~\emph{et al.}~\cite{shi2022fulfilling}, i.e., the preliminary version of our present work.  

Broadband TMSV-enabled EA detection, estimation, and communication offers its best results in the asymptotic regime wherein $N_S \ll 1, \kappa \ll 1,$ and $N_B \gg 1$.  Thus, although optical channels have been considered from the start, see, e.g., Tan~\emph{et al.}~\cite{tan2008quantum}, noise brightness at infrared and visible wavelengths is very low, e.g., $N_B \sim 10^{-6}$ at 1.55\,$\upmu$m in daytime conditions and orders of magnitude lower at night~\cite{Shapiro2005}.  Thus the principal application interest for broadband TMSV-enabled EA detection, estimation, and communication will be for  microwave wavelengths, where $N_B \gg 1$ is the norm.  Not surprisingly, the microwave QI work of Barzanjeh~\emph{et al.}~\cite{Barzanjeh2015} ignited enormous (and continuing) interest in the microwave radar community.  

The microwave region's high-brightness ambient noise dictates that microwave QI and other microwave EA protocols have their source, detectors, and quantum memory in a dilution refrigerator.  As noted in Sec.~\ref{SelectorDesign}, the $\hat{E}_v(t)$ in our SFG optimal-mode selection architecture could easily have a center frequency whose $N_v$ would entirely negligible.  Other mode selector architecture's would benefit in the same way if they used non-degenerate SPDC sources with $\omega_I > \omega_S$.  Note the above mention of quantum memory.  Quantum memory capable of faithfully storing the high time-bandwidth product, low-brightness idler beam for the duration of the $\Phi_{\kappa,\theta}$ channel's propagation delay---introduced in \eqref{ErChannelModel}  and then immediately suppressed---is required by all of our EA protocols.    

The relationship between this paper and Refs.~\cite{Reichert2023,Angeletti2023,chen2023} is as follows.  Reference~\cite{Reichert2023} is an outgrowth of Refs.~\cite{shi2022fulfilling} and \cite{Shapiro2022} in that it used the former's ${\rm C}\to{\rm D}$ conversion module, and the latter's application of sequential detection to QI, to obtain 6\,dB additional quantum advantage in microwave target detection  
over a non-sequential coherent-state radar of the same transmitted energy. It did not use optimal mode selection, but combined its mode selection with its information extraction in its hetero-homodyne receiver, which is sub-optimal for QI.  We note in passing that a sequential version of our ${\rm C}\to{\rm D}$ receiver will also gain that 6\,dB of extra advantage in QI target detection.

Angeletti~\emph{et al.}~\cite{Angeletti2023} also considered microwave QI. Following Ref.~\cite{shi2022fulfilling}, they proposed mitigating the experimental difficulty of phase stabilizing the  ${\rm C}\to{\rm D}$ receiver by using a  Kennedy receiver~\cite{Kennedy1973} for the coherent-state information extractor because it provides near-optimum performance while being phase insensitive.  That said, they did not provide any realistic suggestions for their mode selector. 

Chen and Zhuang.~\cite{chen2023} analyzed the ${\rm C}\to{\rm D}$ receiver's QI performance for Rayleigh fading, random-phase targets and found that quantum optimal performance is retained, but with reduced advantage compared to the case of quiescent (non-fading, known-phase) targets.

\begin{acknowledgements}
This project was supported by Cisco Systems, Inc, the NSF CAREER Award CCF-2142882 and Office of Naval Research Grant No. N00014-23-1-2296. QZ also acknowledges support from the Defense Advanced Research Projects Agency (DARPA) under Young Faculty Award (YFA) Grant No. N660012014029, the National Science Foundation (NSF) Engineering Research Center for Quantum Networks Grant No. 1941583, and support from Raytheon Missiles and Defense and Halliburton. during the completion of the paper. JHS acknowledges support from the MITRE Corporation's Quantum Moonshot Program and the National Science Foundation (NSF) Engineering Research Center for Quantum Networks Grant No. 1941583. ZZ acknowledges support from NSF Grant No. 2304118 and 2317471.

QZ conceived the project. QZ and HS proposed the correlation-to-displacement conversion module, with calculations also verified by BZ. BZ analyzed quantum illumination with inputs from HS, and HS analyzed phase estimation and communication, both under the supervision of QZ. QZ proved Sec.~\ref{Composite}'s theorem with inputs from BZ. JHS and ZZ contributed to the near-optimal mode selectors in Sec~\ref{SelectorDesign}.  JHS performed the continuous-time analysis and wrote the manuscript, with inputs from all authors.
\end{acknowledgements}

%\bibliography{myref}
%apsrev4-2.bst 2019-01-14 (MD) hand-edited version of apsrev4-1.bst
%Control: key (0)
%Control: author (8) initials jnrlst
%Control: editor formatted (1) identically to author
%Control: production of article title (0) allowed
%Control: page (0) single
%Control: year (1) truncated
%Control: production of eprint (0) enabled
%

\appendix

\section*{Supplementary materials}

These supplementary materials present detailed analyses and additional numerical results in support of the main text. In \jhs{Sec.~\ref{sec:review}}, we review known results that were utilized in the main text and will be utilized here. \jhs{Section~\ref{sec:details_C2D}} then contains our analyses and proofs, presented in the order they are mentioned in the main text. Readers can also start directly from \jhs{Sec.~\ref{sec:details_C2D}} and refer back to \jhs{Sec.~\ref{sec:review}} whenever needed.

To begin, we introduce new notation that will be used here for some quantities appearing in the main text.
First, the main text uses $N_{I\mid R}$ for the conditional brightness, given the heterodyne detector's output, of the idler beam's noise component, but here we will use $E$ for that quantity.  Second, regarding the idler mode's average displacement photon number, $\zeta^2 N_R \equiv \zeta^2\int_{\mathcal{T}_0}\!{\rm d}t\,|E_R(t)|^2$ for $\zeta \equiv \sqrt{\kappa N_S(N_S+1)}/(\kappa N_S + N_B+1)$, the main text gives  the distribution of $N_R$ in its \eqref{NRpdf}.  Here, however, we will work with $x \equiv \zeta^2\int_{\mathcal{T}_0}\!{\rm d}t\,|E_R(t)|^2$, which is $\chi^2$ distributed with $2M$ degrees of freedom, 
\be 
P_{\rm disp}^{(M)}(x) = \frac{x^{M-1}}{(2\xi)^M (M-1)!}\,e^{-x/2\xi},
\label{p_overall}
\ee 
with mean $2M\xi$ and variance $4M\xi^2$, where 
\be 
\xi\equiv \frac{\kappa N_S(N_S+1)}{2(N_B+\kappa N_S+1)}=\zeta^2(\kappa N_S + N_B+1)/2\,.
\label{xi_definition}
\ee 
  We also note that, with the exception of Sec.~\ref{sec:mode_selection}, the material presented here will use modal analysis rather than the continuous-time treatment that is dominant in the main text.  Connections between the modal results presented here and the main text's continuous-time results can be made via the main text's Eqs.~(\ref{SImodes}),  (\ref{RBmodes}), (\ref{hetRmodes}), and (\ref{EK_expansion}).

\section{Review of known results}
\label{sec:review}

In this section, we review known results mentioned in the main text and/or utilized in \jhs{Sec.~\ref{sec:details_C2D}}.

\subsection{Helstrom limit and the quantum Chernoff bound}
\label{app:QCB}

In general, given $K$ quantum states \jhs{$\{\hat{\rho}_k\}_{k=1}^{K}$} with prior probabilities \jhs{$\{p_k\}_{k=1}^K$}, the Helstrom limit is an achievable lower bound on the error probability for disciminating between them~\cite{Helstrom1969, Helstrom_1967,Helstrom_1976}, 
\begin{equation}
    P_{\rm H}(\{p_k,\hat{\rho}_k\}_{k=1}^{K}) = 1 - \max_{\{\hat{\Pi}_k\}} \sum_{k=1}^{K} p_k \Tr{\hat{\Pi}_k \hat{\rho}_k},
\end{equation}
where $\{\hat{\Pi}_k\}$ is a positive operator-valued measurement (POVM) whose $k$th element corresponds to state $\hat{\rho}_k$ and $\sum_k \hat{\Pi}_k = \hat{\mathbb{I}}$, where $\mathbb{I}$ is the identity operator.  In the binary case with equal prior probabilities, the Helstrom limit is 
\begin{equation}
    P_{\rm H}(\hat{\rho}_1, \hat{\rho}_2) = \frac{1}{2}\left(1-\frac{1}{2}\Tr{|\hat{\rho}_1 - \hat{\rho}_2|}\right),
\end{equation}
and for pure states $\ket{\psi_1}, \ket{\psi_2}$, it can be further simplified to $P_{\rm H} = \left(1-\sqrt{1-|\bra{\psi_1}\ket{\psi_2}|^2}\right)/2$. 

In general, the Helstrom limit is hard to evaluate. A useful upper bound for the the Helstrom limit is the quantum Chernoff bound (QCB)~\cite{Audenaert2007}. \jhs{For binary} state discrimination between $M$ identical copies of states, \jhs{$\hat{\rho_1}$ and $\hat{\rho_2}$}, we have
\begin{equation}
    P_{\rm H}(\hat{\rho}_1^{\otimes M}, \hat{\rho_2}^{\otimes M})\le P_{\rm QCB} =\frac{1}{2}\left({\rm inf}_{s\in[0,1]}Q_s\right)^M,
\end{equation}
with $Q_s(\hat{\rho}_1,\hat{\rho}_2) = \Tr{\hat{\rho}_1^s\hat{\rho}_2^{1-s}}$.

The QCB can be evaluated efficiently for Gaussian states~\cite{Pirandola2008}.
Of interest to us is the case of two $N$-mode Gaussian states $\{\hat{\rho}_h\}_{h=1}^2$ with mean quadrature vectors $\overline{\bm x}_h$ and quadrature covariance matrices $V_h$. Here we can find the symplectic decompositions of the covariance matrices, i.e., $V_h = S_hV_h^{\oplus}S_h^T$, where $V_h^{\oplus} = \oplus_{n=1}^N \nu_n^{(h)}\mathbb{I}$ with $\{\nu_n^{(h)}\}_{n=1}^N$ being their symplectic spectra~\cite{weedbrook2012gaussian}.
In this case, the QCB can be evaluated via 
\begin{equation}
    Q_s = \overline{Q}_s\exp{-\frac{1}{2}{\bm d}^T\left(\tilde{V}_1(s)+\tilde{V}_2(1-s)\right)^{-1}{\bm d}},
\end{equation}
where ${\bm d}=\overline{\bm x}_1 - \overline{\bm x}_2$ and $\overline{Q}_s$ is defined as
\begin{equation}
    \overline{Q}_s = \frac{2^N\prod_{n=1}^N G_s(\nu_n^{(1)})G_{1-s}(\nu_n^{(2)})}{\sqrt{\det\left[\tilde{V}_1(s)+\tilde{V}_2(1-s)\right]}},
\end{equation}
with $\tilde{V}_h(s)=S_h\left[\oplus_{n=1}^N \Lambda_s(\nu_n^{(h)})\mathbb{I}\right]S_h^T$. The $G_p(\nu)$ and $\Lambda_p(\nu)$ functions are
\begin{subequations}
\begin{align}
    G_p(\nu) &\equiv \frac{2^p}{(\nu+1)^p-(\nu-1)^p},
    \\
    \Lambda_p(\nu) &\equiv \frac{(\nu+1)^p+(\nu-1)^p}{(\nu+1)^p-(\nu-1)^p}.
\end{align}
\end{subequations}

\subsection{Review of classical communication capacity. }

For the $\Phi_{\kappa,\theta}$ chsannel described by the main text's \eqref{ArChannelModel}, the classical capacity with energy constraint $\expval{\hat a^\dagger \hat a}\le N_S$, without entanglement assistance, is known to be~\cite{giovannetti2004}
\be 
C=g(\kappa N_S+N_B)-g(N_B)\,.
\ee
Here, $g(n)=(n+1)\log_2(n+1)-n \log_2 n$ is the entropy of a thermal state with mean photon number $n$. 
When $N_S\to 0$, we can expand $C$ to \jhs{its} leading order and obtain
\be 
C=\kappa N_S\log_2(1+\frac{1}{N_B})+O(N_S^2)\,.
\ee
In the noisy scenario $N_B \gg 1$, the classical capacity $C\simeq \kappa N_S/\ln(2)N_B$ is saturated by a heterodyne or a homodyne receiver~\cite{shi2020practical}. 

\begin{figure}
    \centering
    \includegraphics[width=0.25\textwidth]{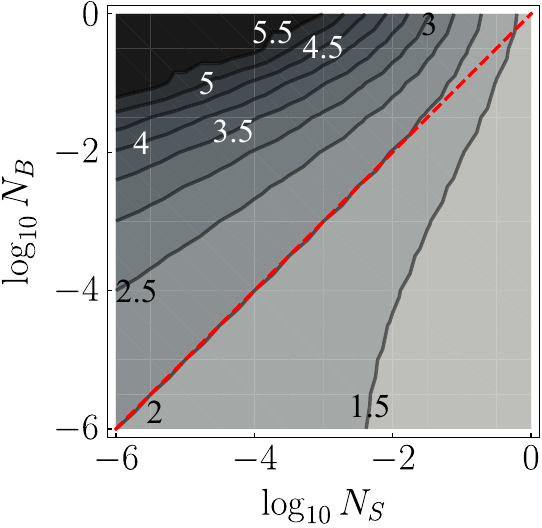}
    \caption{The ultimate EA capacity over the  $\Phi_{\kappa,\theta}$ channel for $\kappa = 0.01$, normalized by the unassisted capacity $C$, versus  $N_S$ and $N_B$. The red dashed diagonal line indicates $N_S=N_B$.}
    \label{fig:EACOMM_NbNs_CE}
\end{figure}

Entanglement assistance boosts the communication capacity to \cite{Bennett2002}
\be
C_E=g(N_S)+g(N_S^\prime)-g(A_+)-g(A_-),
\label{CE_exact}
\ee 
where 
$A_\pm=(D-1\pm(N_S^\prime-N_S))/2$, $N_S^\prime=\kappa N_S+N_B$ and $D=\sqrt{(N_S+N_S^\prime+1)^2-4\kappa N_S(N_S+1)}$. 
When $N_S\to 0$, the leading order can be obtained as
\bal 
C_E&= \frac{\kappa N_S}{N_B+1} \left[\log_2 \left(\frac{1}{N_B N_S \left(N_B-\kappa+1\right)}\right)+\calR
   \right]\!+\!O\left(N_S^2\right)\\[.05in]
   &= \frac{\kappa N_S\log_2(1/N_S)}{N_B+1}+O(N_S)\,,
\label{CE_expansion}
\eal 
where 
\ba
\calR&=\big[\left(N_B+1\right) \log_2 \left(N_B-\kappa +1\right)\nonumber\\[.05in]&+\kappa +\left(-N_B+2 \kappa -1\right) \log_2
   \left(N_B+1\right)\big]/\kappa
\ea  
is independent of $N_S$.
We see a diverging advantage $
C_E/C\sim \ln\left(1/N_S\right),
$ in the $ N_S\ll1$ limit. Remarkably, such an advantage is not limited to the region $N_S\ll 1, N_B\gg 1$ considered in Refs.~\cite{Bennett2002,shi2020practical}, as shown in main text's Fig.~\ref{fig:EACOMM_NbNs_CE}. The main text also shows that such an extension of \jhs{the} advantageous region holds for our correlation-to-displacement receiver as well.
The entanglement-assisted (EA) capacity is known to be achieved by the Holevo information of phase encoding on broadband two-mode squeezed vacuum (TMSV) states in the $N_S \ll 1, N_B \gg 1$ limit~\cite{shi2020practical}.

\subsection{Review of the Green machine for achieving optimal EA communication scaling}
\label{app:GM}

The Green machine is a receiver that attains a rate very close to the classical communication capacity with phase modulation of coherent
states~\cite{guha2011structured}. Here we show that a design concatenating it with our $\rm C\to  D$ receiver (see Supplementary~Fig.~\ref{fig:GM} for a schematic plot) achieves the same scaling \jhs{as} the ultimate EA capacity.
The sender jointly encodes a block of $n$ signal modes by BPSK modulation according to an $n$-codeword Hadamard code, where $n$ is a power of 2 and each codeword contains $n$ symbols $\{\theta_k\}_{k=1}^n$. Supplementary Fig.~\ref{fig:GM} shows an example with $n=8$. The information rate can be further improved by repetitive encoding over $M$ independent identically-distributed (iid) copies of signal modes with identical symbols, where $M$ is to be optimized. After an $n$-block of signals goes through the channel $\Phi_{\kappa,0}$, the receiver obtains an $n$-block of returned signals. Then the receiver applies the $\rm C\to  D$ conversion module to the $n$-block of return-idler pairs, which yields $n$ displaced thermal states, with quadrature phases set by the $n$-codeword Hadamard code. We note that, the $\rm C\to  D$ converter combines the $M$ copies together, thereby the brightness of the $n$ displaced thermal states is increased such that the thermal background $E$ is negligible and the states resemble coherent states. The $n$ quasi-coherent-state outputs are input into the Green machine. The Green machine consists of a beam splitter array. Denote the input modes as a vector $\bm {\hat a}=[\hat a_1,\ldots \hat a_n]^T$, with mean $\expval{\bm{\hat a}}\propto [e^{i\theta_1},\ldots,e^{i\theta_n}]^T$, where we have left out the amplitude to focus on the phase. The beam splitter array performs a Bogoliubov transform $\bm {\hat a}\to S\bm {\hat a}$ with the unitary matrix
\be 
S=\frac{1}{\sqrt{n}}
\bp 
1 & 1\\
-1 & 1
\ep ^{\otimes \log_2 n}\,.
\label{eq:GM_BS}
\ee
Finally the receiver \jhs{does} zero-or-not photon counting on the $n$ output modes of the Green machine individually. In the limit of weak thermal background $E\to 0$, each output mode yields non-zero photon count if and \jhs{only} if the input modes are constructively interfered by the beam splitter array.
\jhs{Therefore}, the Green machine converts one of the $n$ coherent-state codewords of the binary phase-shift keying (BPSK) Hadamard code  into one of the $n$ codewords of coherent-state pulse position modulation (PPM) (in the photon count), i.e,. the photons of $n$ input modes are merged into one output mode.

Now we evaluate the performance of the above protocol. The magnitudes of the means at the output modes of the $\rm C\to  D$ converter depend on the squared mean $x$ of the random heterodyne readout, which has the $\chi^2$ distribution $P_\kappa (x)$ as defined in \eqref{p_overall}. 
% with mean 
% \be 
% \overline x=M\kappa N_S(N_S+1)/(N_B+\kappa N_S+1)\,.
% \ee 
For an $M$-copy, $n$-codeword Green machine, the per-symbol rate given the squared mean $x$ is \cite{guha2020infinite}
\small 
\begin{align} 
R_{\rm GM}(x)&=
\frac{1}{ M n \ln 2}\left[\left(n-1\right) P_d\left(x\right) \ln \left(\frac{n P_d\left(x\right)}{P_{\rm E}\left(x\right)}\right)
\right.
\nonumber
\\
&
-\Bigg(\left(n-1\right) P_d\left(x\right)+P_{\rm E}\left(x\right)\Bigg) \ln \left(\frac{\left(n-1\right) P_d\left(x\right)}{P_{\rm E}\left(x\right)}+1\right)
\nonumber
\\
&
\left.+P_{\rm E}\left(x\right) \ln \left(n\right)\right],
\label{rate_formula}
\end{align}
\normalsize
where
$P_d(x)=[1-P_c(x)]P_b(1-P_b)^{n-2}$, $P_{\rm E}(x)=P_c(x)(1-P_b)^{n-1}$ and
\bal 
P_c(x)=1-\frac{e^{-\frac{x}{1+E}}}{1+E}\,\,\,,P_b=1-\frac{1}{1+E}\,.
\eal 
Here the thermal background is 
\be
E={N_S \left(N_B+1-\kappa \right)}/{ (\kappa N_S+N_B+1)}\le N_S,
\ee 
\jhs{which is given in the main text by $N_{I\mid R}$}.
Thus, the overall rate is
\be 
R_{\rm GM}=\int_{0}^\infty dx P_\kappa (x) R_{\rm GM}(x).
\label{eq:Rgm_exact_supp}
\ee

\begin{figure*}
    \centering
    \includegraphics[width=0.75\textwidth]{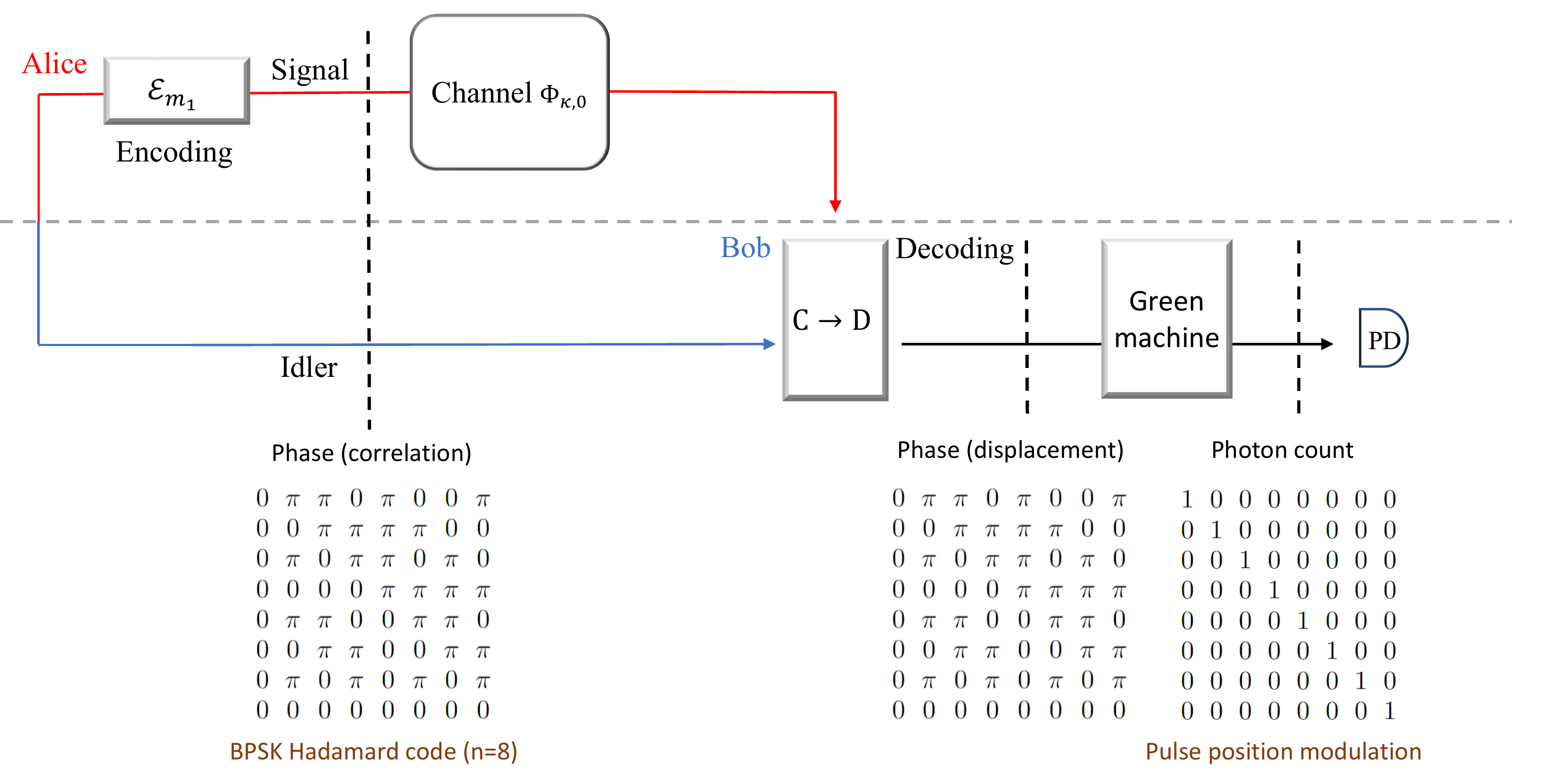}
    \caption{Entanglement-assisted communication protocol using the BPSK Hadamard code and the Green machine. The Green machine, a beams plitter array defined in \eqref{eq:GM_BS}, converts the BPSK Hadamard code in phase to pulse position modulation in photon count.}
    \label{fig:GM}
\end{figure*}
% \begin{figure}
%     \centering
%     \includegraphics[width=0.3\textwidth]{EACOMM_c2d_GM_M.eps}
%     \caption{The numerically optimized copy number $M^*$. The codeword length $n$ is chosen asymptotically optimal as in \eqref{eq:optn}. $\kappa=0.01, N_B=100$.}
%     \label{fig:GM_M}
% \end{figure}

% \begin{figure}
%     \centering
%     \includegraphics[width=0.3\textwidth]{EACOMM_c2d_GM_ratio.eps}
%     \caption{The ratio of the rate of Green machine over the Holevo information of the output ensemble of the correlation-to-displacement module. $\kappa=0.01, N_B=100$.}
%     \label{fig:GM_ratio}
% \end{figure}

Observe that $R_{\rm GM}$ depends on $M, n$. We numerically optimized $R_{\rm GM}$ over integer $M$ for each value of $N_S$ in Supplementary~Fig.~\ref{fig:EACOMM_NsNb_GM}, while choosing the block size $n^\star$ to be the asymptotic optimum from \eqref{eq:optn}. The results in the main text were obtained similarly.

Below, we provide asymptotic analyses to obtain more insight. Note that our numerical results above are evaluated via the exact formula \eqref{eq:Rgm_exact_supp} without asymptotic approximations, using only the value of $n^\star$ derived below, and therefore represents an exact achievable rate.
\begin{figure}
    \centering
    \includegraphics[width=0.25\textwidth]{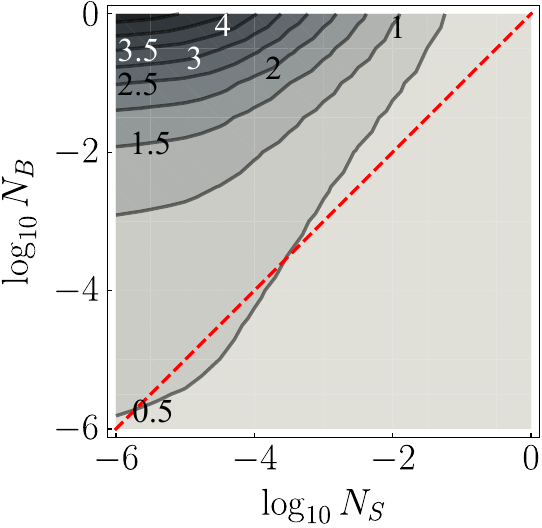}
    \caption{The information rate per symbol of the Green machine, normalized by the unassisted capacity $C$, using BPSK-encoded TMSV states under various $N_B,N_S$, with $\kappa = 0.01$  It is plotted with numerically-optimized repetitions $M$ and asymptotically optimal codeword length $n$ from \eqref{eq:optn}. The black diagonal line indicates $N_S=N_B$.}
    \label{fig:EACOMM_NsNb_GM}
\end{figure}
The optimized $M^*\equiv{\rm argmax}_M R_{\rm GM}$ is numerically found $\sim 10^4$, sufficiently large for all parameters being considered. Thus it is fair to invoke the law of large numbers, that the squared mean $x$ converges in probability to the expectation value $\overline x=M\kappa N_S(N_S+1)/(N_B+\kappa N_S+1)$ of the $\chi^2$ distribution. Hence, the rate converges to \jhs{its} value at $x=\overline x$,
\be 
R_{\rm GM}\to R_{\rm GM}(\overline x)=R_{\rm GM}\left(\frac{M\kappa N_S(N_S+1)}{N_B+\kappa N_S+1}\right)
\ee 
Below we optimize $R_{\rm GM}$ with respect to $n$, in the asymptotic regime $N_S\to 0$. Let the optimal $n$ be
\be 
n^*\equiv {\rm argmax}_{n\in \calN} R_{\rm GM},
\ee
where $\calN=\{2^k| k\in \{1,2,\ldots\}\}$ is the set of positive powers of 2. 
In this case $P_{\rm E}\gg P_d$, thus the rate in \eqref{rate_formula} is dominated by $P_{\rm E}\ln (n)/Mn$.
Then one can obtain the second-order expansion of the rate
\be 
MR_{\rm GM}\simeq \frac{P_{\rm E}\ln(n)}{n  \ln 2}=(u +v n) \log_2(n)+O(N_S^3)
\label{eq:GM_R_1}
\ee
where 
\bal 
u&=\frac{\!- N_S \big[\!-\!\kappa  \left(N_B\!+\!1\right) \left(M\!+\!2 N_S\right) \!+\! \left(N_B\!+\!1\right){}^2 N_S\!+\! \kappa ^2 N_S\big]}{\left(N_B+1\right){}^2}\,,\\
v&=-\frac{\kappa  M N_S^2 \big[2 \left(N_B+1\right)+\kappa  (M-2)\big]}{2 \left(N_B+1\right)^2}\,.
\eal 
By solving ${\rm d} R_{\rm GM}/{\rm d} n=0$, we obtain
\be 
n^*\simeq -\frac{u}{v W(-ue/v)},
\label{eq:optn}
\ee
where $W$ is the principle branch of the Lambert $W$ function which satisfies $W(xe^x)=x$ for $x\ge -1$. Using the relation $\ln W(x)=\ln (x)-W(x)$ and the asymptotic expansion $W(x)=\ln(x)-\ln\ln(x)+O(1)$ as $\ln(x)\to \infty$, we get
\small 
\bal 
&\ln n^*\simeq -1+W(\frac{-ue}{v})\\
=&-\ln \left[\ln \left(-\frac{2 e\left(-\kappa  \left(N_B+1\right) \left(M+2 N_S\right)+\left(N_B+1\right){}^2 N_S+\kappa ^2 N_S\right)}{\kappa  M N_S \left(2 N_B-2 \kappa +\kappa  M+2\right)}\right)\right]\\
&+\ln \left(-\frac{2  \left(-\kappa  \left(N_B+1\right) \left(M+2 N_S\right)+\left(N_B+1\right){}^2 N_S+\kappa ^2 N_S\right)}{\kappa  M N_S \left(2 \left(N_B+1\right)+\kappa  (M-2)\right)}\right).
\label{eq:optlogn}
\eal 
\normalsize
Plugging \eqref{eq:optn} and~\eqref{eq:optlogn} into \eqref{eq:GM_R_1}, we have the asymptotic rate
\bal 
R_{\rm GM}&= \frac{\kappa  N_S \left(\ln \left(\frac{2 \left(N_B+1\right)}{N_S \left(2 \left(N_B+1\right)+\kappa 
   (M-2)\right)}\right)-1\right)}{(N_B+1) \ln 2}+O\left(N_S^2\right).
\label{eq:GM_R_fin}
\eal 
In the limit of $N_S\to 0$, the preceding rate achieves the optimal scaling of the ultimate EA capacity $C_{\rm E}$~\cite{Bennett2002}
\be 
R_{\rm GM}=\frac{\kappa N_S}{(N_B+1) \ln 2}(\ln(1/N_S)-O(1))\propto N_S\ln (1/N_S)\,.
\label{eq:GM_R_asym}
\ee
% It is verified in Fig.~\ref{fig:GM_ratio} by investigating the rate ratio 
Remarkably, we see that
\be 
\frac{R_{\rm GM}}{C_{\rm E}}= 1- O(\frac{1}{\ln (1/N_S)}),
\ee
which goes to 1 as $N_S\to 0$.

\subsection{Review of \jhs{coherent-state} discrimination}
\label{app:receivers_summary}

In this section, we summarize coherent-state discrimination, including the Helstrom limit~\cite{Helstrom1969, Helstrom_1967,Helstrom_1976}, homodyne detection, heterodyne detection, the Kennedy receiver~\cite{Kennedy_1972} and the Dolinar receiver. 

In what follows we consider discriminating between the  vacuum state $\ket{0}$ and the coherent state $\ket{\alpha}$, where $\alpha=\alpha_R+i\alpha_I$ is the state's quadrature decomposition and states are equally likely to occur. The noisy version of this discrimination problem is exactly the sub-task performed by ${\rm C}\!\!\to\!\! {\rm D}$ reception's information extractor.

We begin with the noiseless version, in which case the Helstrom-limit error probability has the closed form 
\begin{equation}
    P_{\rm H}(\ket{0},\ket{\alpha}) = \frac{1}{2}\left(1-\sqrt{1-e^{-|\alpha|^2}}\right).
\end{equation}

Now we discuss the performance of homodyne detection.
Homodyne detection consists of measuring a single quadrature of the mode. Its error probability is minimized when the $\hat{a}_\theta \equiv {\rm Re}(\hat{a}e^{-i\theta})$ quadrature is measured, where $\hat{a}$ is the mode's annihilation operator and $\alpha = |\alpha|e^{i\theta}$. The error probability is then
\begin{equation}
\begin{split}
    &P_{\rm E, homo} = \frac{1}{2}{\rm Erfc}\left(\frac{|\alpha|}{\sqrt{2}}\right).
\end{split}
\end{equation}
When the $|\alpha|\gg1$, as ${\rm Erfc}(x)\sim e^{-x^2}/\sqrt{\pi}x$ for $x > 0$, we have $P_{\rm E, homo}\sim \exp\left(-|\alpha|^2/2\right)$. 

Similarly, heterodyne detection measures the $\hat{a}$ POVM, leading to the error probability 
\begin{equation}
    P_{\rm E, het} = \frac{1}{2}{\rm Erfc}\left(\frac{|\alpha|}{2}\right).
\end{equation}
When $|\alpha|\gg1$, we have $P_{\rm E, het}\sim \exp\left(-|\alpha|^2/4\right)$.

For coherent state discrimination, there exists a well-known nulling receiver called the Kennedy receiver~\cite{Kennedy_1972}. For two arbitrary coherent states $\ket{\alpha_1},\ket{\alpha_2}$, the Kennedy receiver performs a displacement $-\alpha_1$, transforming the discrimination problem to between the vacuum state and the coherent state $\ket{\alpha_2-\alpha_1}$. Ideal photon counting, with the decision rule decide $|\alpha_1\rangle$ was present if and only if no photons are counted, results in an error probability given by
\begin{equation}
    P_{\rm E, Kennedy} = \frac{1}{2}|\bra{0}\ket{\alpha_2-\alpha_1}|^2 = \frac{1}{2}e^{-|\alpha_2-\alpha_1|^2},
\end{equation}
since there is no error in predicting the state $\ket{\alpha_1}$. For the case under consideration, $\ket{0}$ versus $\ket{\alpha}$, as one of the state is always in vacuum state, the Kennedy receiver reduces to a photon-counting receiver whose error probability is 
$
    P_{\rm E, Kennedy} = \frac{1}{2}e^{-|\alpha|^2}
$
which, for high photon number ($|\alpha|^2 \gg 1$), satisfies $P_{\rm E, Kennedy} \simeq 2P_{\rm H}$.

Note that Kennedy receiver is sub-optimal for coherent-state discrimination. An adaptive receiver, Dolinar receiver~\cite{Dolinar1973}, approaches the Helstrom limit in the noiseless case. It splits the input coherent state into $S$ slices and makes a decision in terms of the prior probability of the each slice where displacement and a Bayesian updating rule are applied. To explain how that works, we introduce $h$ to denote the true state, $g$ as the current decision ($h,g\in [0,1]$) and $p_h^{(k)}$ as the prior probability for $k$th slice to be state $\hat{\rho}_h$. The number of photons measured from $k$th slice is denoted as $N^{(k)}$, whose conditional distribution, $p_N^{(k)}(N^{(k)},g|h)$, is the Bayesian conditional probability for obtaining $N^{(k)}$ photons when the $k$th slice is determined to be $\hat{\rho}_g$ while it is actually $\hat{\rho}_h$.
For $\ket{0}$ and $\ket{\alpha}$ with equal prior probability $p_0^{(0)}=p_1^{(1)}$, the Dolinar receiver's algorithmic representation is \jhs{given} below in Algorithm~\ref{Dolinar_alg}:

\begin{algorithm}
\begin{algorithmic}
\State $S$, $h$, $\gamma = \sqrt{|\alpha|^2}/2\sqrt{S}$
\State $k \gets 1$, $g \gets None$
\While{$k \le S$}
\State $u^{(k)} = \gamma/\sqrt{1-\exp{-|\alpha|^2(k-1/2)/S}}$
\If{$p_0^{(k)} > p_1^{(k)}$}
    \State $g \gets 0$
\ElsIf{$p_0^{(k)} < p_1^{(k)}$}
    \State $g \gets 1$
\Else
    \State $g \gets \{0,1\}$ with equal probability
\EndIf
\If{$g = 0$}
    \State Perform displacement $-\gamma+u^{(k)}$
\Else
    \State Perform displacement $-\gamma-u^{(k)}$
\EndIf
\State Measure the photon number $N^{(k)}$ with probability $p_N(N^{(k)},g|h)$
\State Update prior probability \\
$p_0^{(k+1)} \gets p_0^{(k)} p(N^{(k)},g|0)/\sum_{h^\prime=0}^1 p_{h^\prime}^{(k)} p(N^{(k)},g|h^\prime)$\\
$p_1^{(k+1)} \gets p_1^{(k)} p(N^{(k)},g|1)/\sum_{h^\prime=0}^1 p_{h^\prime}^{(k)} p(N^{(k)},g|h^\prime)$
\EndWhile
\If{$p_0^{S+1} > p_1^{S+1}$}
    \State $g \gets 0$
\ElsIf{$p_0^{S+1} < p_1^{S+1}$}
    \State $g \gets 1$
\Else
    \State $g \gets \{0,1\}$ with equal probability
\EndIf\\
\Return $g$
\end{algorithmic}
\caption{Dolinar receiver \label{Dolinar_alg}}
\end{algorithm}

The Dolinar receiver gives a prediction on the unknown state and we perform Monte-Carlo simulation to evaluate the error probability.
For noiseless case $\ket{0}$ versus $\ket{\alpha}$, the measured photon distribution on $k$th slice follows \jhs{the} Poisson distribution
\begin{equation}
    N^{(k)}\sim p_N(n,g|h)=\begin{cases}
    {\rm Pois}(n;(\gamma-u^{(k)})^2), & \textit{if $g=h$}\\
    {\rm Pois}(n;(\gamma+u^{(k)})^2), & \textit{otherwise}
    \end{cases}
\end{equation}
where \jhs{${\rm Pois}(n;\lambda)$} is the Poisson probability mass function \jhs{with mean $\lambda$}. The conditional Bayesian probability of getting $N^{(k)}$ photons is $p(N^{(k)},g|h) = p_N(N^{(k)},g|h)$.  The above formulation of the Dolinar receiver realizes Helstrom-limit performance in the limit $S\rightarrow \infty$.  Its continuous-time analysis leads to Helstrom-limit performance assuming use of an ideal photodetector, i.e., unit quantum efficiency, no dark counts, no dead time, no thermal noise, and infinite electrical bandwidth.  

\subsection{Known sub-optimal receivers for quantum illumination target detection, phase sensing, and communication}
\label{app:OPA_PC receiver}

\subsubsection{Parametric amplifier (PA) receiver}

\begin{figure}
    \centering
    \includegraphics[width=0.45\textwidth]{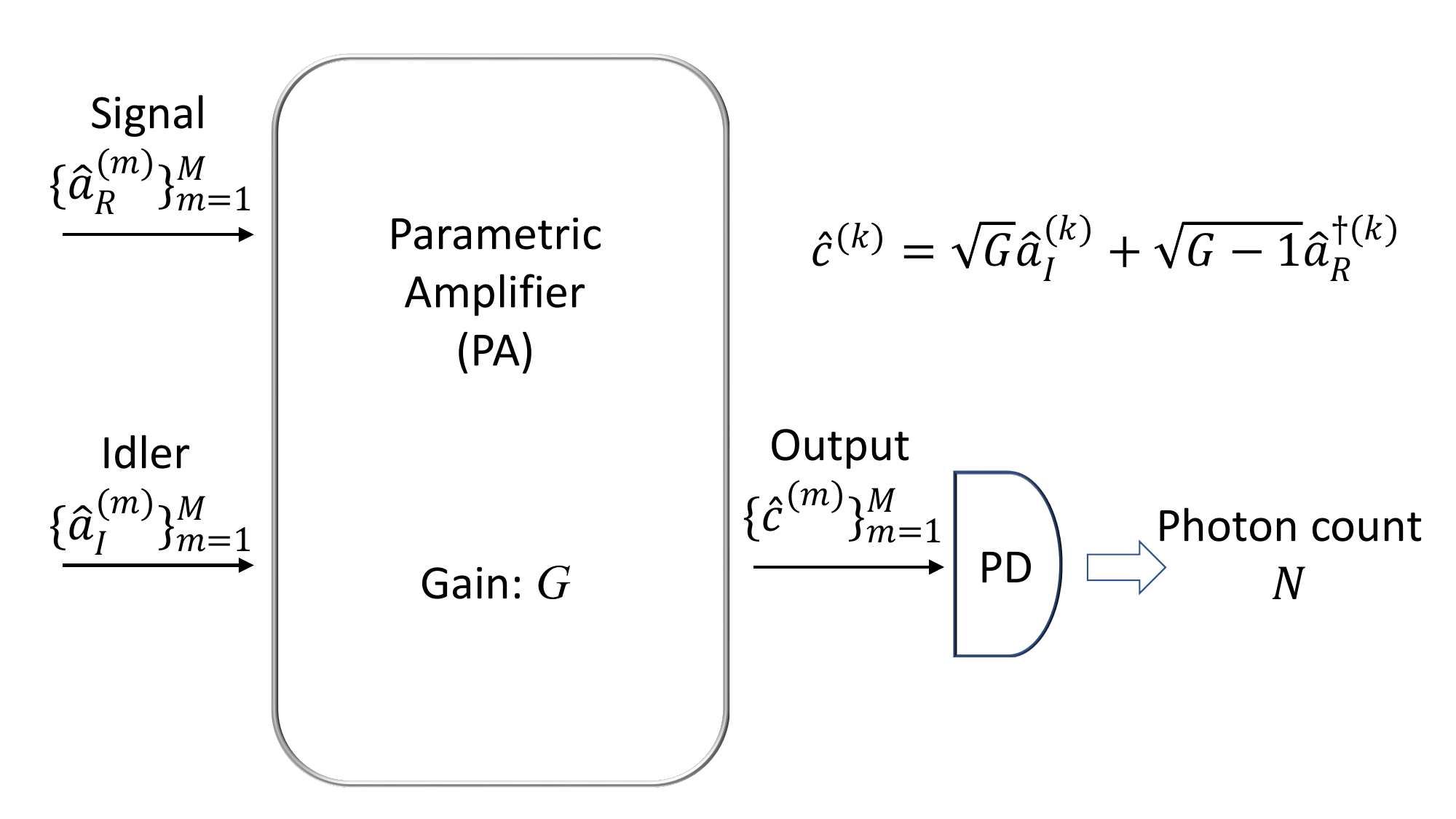}
    \caption{Schematic of the parametric amplifier (PA) receiver. $M$ iid return-idler mode pairs $\{\hat a_R^{(m)},\hat a_I^{(m)}\}_{m=1}^M$ are input to the receiver. A parametric amplifier combines the idler and the conjugate of the signal in the
    output that is photodetected. The total photon count over the $M$ pairs is the sufficient statistic for completing the \jhs{EA sensing} or communication task at hand. }
    \label{fig:schematicOPA}
\end{figure}

Supplementary Fig.~\ref{fig:schematicOPA} is a schematic of the parametric amplifier (PA) receiver. The PA receiver applies parametric amplification to the $M$ return-idler mode pairs $\{\hat a_{R}^{(m)},\hat a_{I}^{(m)}\}$ thus converting their phase-sensitive cross correlations into photon-number differences. The amplification produces $M$ output modes $\hat c^{(m)}=\sqrt{G}\, \hat a_{I}^{(m)}+\sqrt{G-1}\,\hat a_{R}^{(m)\dagger}, 1\le m\le M$. For two-mode Gaussian states with zero mean and covariance matrix specified by the main text's \eqref{eq:CM_RI}, each output mode is in a thermal state with mean photon number 
\bal
\overline N(\theta,\kappa)\equiv & \expval{\hat c^{\dagger(m)} \hat c^{(m)}}\\
=&G N_S\!+\!(G\!-\!1)(\kappa N_S\!+\!N_B\!+\!1)\\
&+2\sqrt{G(G-1)\kappa N_S(1+N_S)}\cos\theta \,.
\label{eq:OPA_mean}
\eal
We collect the total photon number $\hat N=\sum_{m=1}^M \hat c^{\dagger(m)} \hat c^{(m)} $ across the $M$ modes. The probability mass function of the random-variable readout $N$ is~\cite{shi2020practical}
\begin{align}
&P_{N|\theta,\kappa}^{(M)}(n|\theta,\kappa)=
\nonumber
\\
&\!\binom {n\!\!+\!\!M\!\!-\!\!1}{n}\!\left(\frac{\overline N(\theta,\kappa)}{1\!+\!\overline N(\theta,\kappa)}\right)^{\!n}\!\!\left(\frac{1}{1\!+\!\overline N(\theta,\kappa)}\right)^{\!M},
\label{eq:Pn_OPA}
\end{align}
where $\binom{a}{b}$ is the binomial coefficient $a$ choose $b$.
Below we utilize the above measurement statistics to evaluate the performance of quantum illumination, EA phase sensing, and EA communication.

In quantum illumination, the task is to discriminate between two channel hypotheses, $H_0:\Phi_{0,0}$ and $H_1:\Phi_{\kappa,0}$. When $M \gg 1$, the central limit theorem allows us to approximate \eqref{eq:Pn_OPA} by a Gaussian distribution with mean and variance $\mu_0=G N_S+(G-1)(1+N_B)$, $\sigma_0^2 = \mu_0(\mu_0+1)$ for $H_0$, and $\mu_1 = G N_S+(G-1)(1+N_B+\kappa N_S)+2\sqrt{G(G-1)\kappa N_S(N_S+1)}$, $\sigma_1^2 = \mu_1(\mu_1+1)$ for $H_1$. A near-optimum decision using a threshold detector that decides in favor of hypothesis $H_0$ if $N<N_{\rm th}$, and $H_1$ otherwise, with $N_{\rm th}\equiv\lceil M(\sigma_1\mu_0+\sigma_0\mu_1)/(\sigma_0+\sigma_1)\rceil$~\cite{Guha2009}. The error probability for target detection is then
\be
P_{E,{\rm PA}} = \frac{1}{2}{\rm Erfc}\left(\sqrt{R_{\rm PA}^{\rm QI}M}\right)
\label{eq:pe_OPA}
\ee
where $R_{\rm PA}^{\rm QI} = (\mu_1 - \mu_0)^2/2(\sigma_0+\sigma_1)^2$. In the $N_S\ll 1, \kappa\ll 1, N_B\gg 1$ asymptotic regime, $R_{\rm PA}^{\rm QI}\simeq \kappa N_S/2N_B$. Note that the exact optimal decision threshold is \jhs{a} lengthy \jhs{expression that} only \jhs{changes} the results \jhs{slightly}.

In the phase estimation scenario, the task is to estimate the parameter $\theta$ of the  $\Phi_{\kappa,\theta}$ channel.  The PA receiver's Fisher information is
\be 
\mathcal{I}_{F,\rm PA} \equiv \sum_{n=0}^\infty\left(\partial_\theta\mathrm{log}P_{N|\theta,\kappa}^{(M)}(n|\theta,\kappa)\right)^2P_{N|\theta,\kappa}^{(M)}(n|\theta,\kappa)\,.
\ee 
Plugging in \eqref{eq:Pn_OPA}, we find that the Fisher information's dependence on the amplifier's gain is given by 
\bal 
\mathcal{I}_{F,\rm PA}(G)=\frac{4M(G-1)G\kappa N_S(1+N_S)\mathrm{sin}^2\theta}{\overline N(1+\overline N)}\,.
\label{eq:optG_PA receiver}
\eal
The receiver's optimal gain is then found to be
\be 
G_{\rm opt}^{\rm PA}\equiv {\rm argmax}_G \mathcal{I}_{F,\rm PA}(G)=\max\{G^*,1\},
\ee 
where 
\be 
G^*=1+\frac{\sqrt{N_S \left(N_S+1\right) \left(N_B'-1\right) N_B'}+N_S \left(N_S+1\right)}{ \left(N_B'-N_S-1\right) \left(N_B'+N_S\right)}\,,
\ee
and $N_B'\equiv N_B+\kappa  N_S+1$.
Here it is necessary to take the maximum between $G^*$ and $1$, because when $N_S>N_B/(1-\kappa)$, the optimum $G^*$ falls below 1, which is not physical. As a result, when $N_S\ll 1$ we have $G^*=1+{\sqrt{N_S}}/{\sqrt{N_B(1+N_B)}}+O(N_S)$. In this regime, the optimum Fisher information is
$
\mathcal{I}_{F,\rm PA} ={4M\kappa N_S\sin^2\theta}/{(1+N_B)} +O(N_S^{3/2})\,.
$

In the communication scenario, let us consider BPSK modulation in which $\theta\in\{0,\pi\}$ with equal probability.  The conditional statistics of \eqref{eq:Pn_OPA} then lead to the unconditional statistics $P_N^{(M)}(n)\equiv \sum_{\theta\in\{0,\pi\}} P_{N|\theta,\kappa}^{(M)}(n|\theta,\kappa)/2$. Using these two distributions, we obtain the Shannon information
\be 
I(N;\theta)=H(N)-H(N|\theta)\,,
\label{eq:Shannon_supp}
\ee
where 
\bal 
H(N|\theta)&=\!-\!\!\sum_{\theta\in\{0,\pi\}}\frac{1}{2}\sum_{n=0}^\infty P_{N|\theta,\kappa}^{(M)}(n|\theta,\kappa)\log_2 P_{N|\theta,\kappa}^{(M)}(n|\theta,\kappa)\,,\\
H(N)&=-\sum_{n=0}^\infty P_N^{(M)}(n)\log_2 P_N^{(M)}(n)\,.
\eal 
In our simulations we chose $M=1000$ to match the PA receiver's $M$ value for quantum illumination, where the Gaussian approximation requires $M\gg 1$ for its validity. Indeed, we find that it achieves a performance almost identical to the optimum choice $M=1$. That optimality is due to the fact that data processing---here, summing over $M$ modal photon counts---can only decrease Shannon information.
% In our simulation, by assuming a large copy number $M=10/\sqrt{N_S}\gg 1$ \big(such that $\expval{(N-\expval{N})^3/M^3}\sim \left(\expval{(N-\expval{N})^2/M^2}\right)^{3/2}$\big), we approximate $P_{N|\theta(n|\theta)}$ \eqref{eq:Pn_OPA} to a Gaussian probability density function
% \be 
% P_{N|\theta}^{(M)}(n|\theta)=\frac{1}{\sqrt{2\pi\sigma^2(\theta)}}\exp{-\frac{(n-\mu(\theta))^2}{2\sigma^2(\theta)}}
% % \label{eq:Pn_OPA_Gauss}
% \ee
% with mean $\mu(\theta)$ of $N$ conditioned on the phase $\theta$ can be calculated as
% \be
% \mu(\theta)=\sum_{N=1}^\infty N P_{N|\theta}(n|\theta)=M \overline N
% \,,\ee
% and variance 
% \bal 
% \sigma^2(\theta)&=\sum_{N=1}^\infty [N-\mu(\theta)]^2 P_{N|\theta}(n|\theta)=M \overline N(\overline N+1)\,,
% \eal 

It is worth noting that after the PA receiver's phase-conjugation transform, the other output port's annihilation operators,  
$\hat d^{(m)}=\sqrt{G}\, \hat a_{R}^{(m)}+\sqrt{G-1}\,\hat a_{I}^{\dagger(m)}, 1\le m\le M$ have phase-sensitive cross correlations with their $\hat c^{(m)}$ mode counterparts, viz.,
\begin{align}
\expval{\hat c^{(m)}\hat d^{(m)}}=&GC_pe^{i\theta}+(G-1)C_pe^{-i\theta}
\nonumber
\\
&+\sqrt{G(G-1)}(1+N_S+N_B+\kappa N_S).
\end{align} 
The terms proportional to $C_p \equiv \sqrt{\kappa N_S(N_S+1)}$ imply that the PA receiver has failed to convert all of the incoming phase-sensitive cross correlation into phase-insensitive cross correlation.
\begin{figure}[t]
    \centering
    \includegraphics[width=0.3\textwidth]{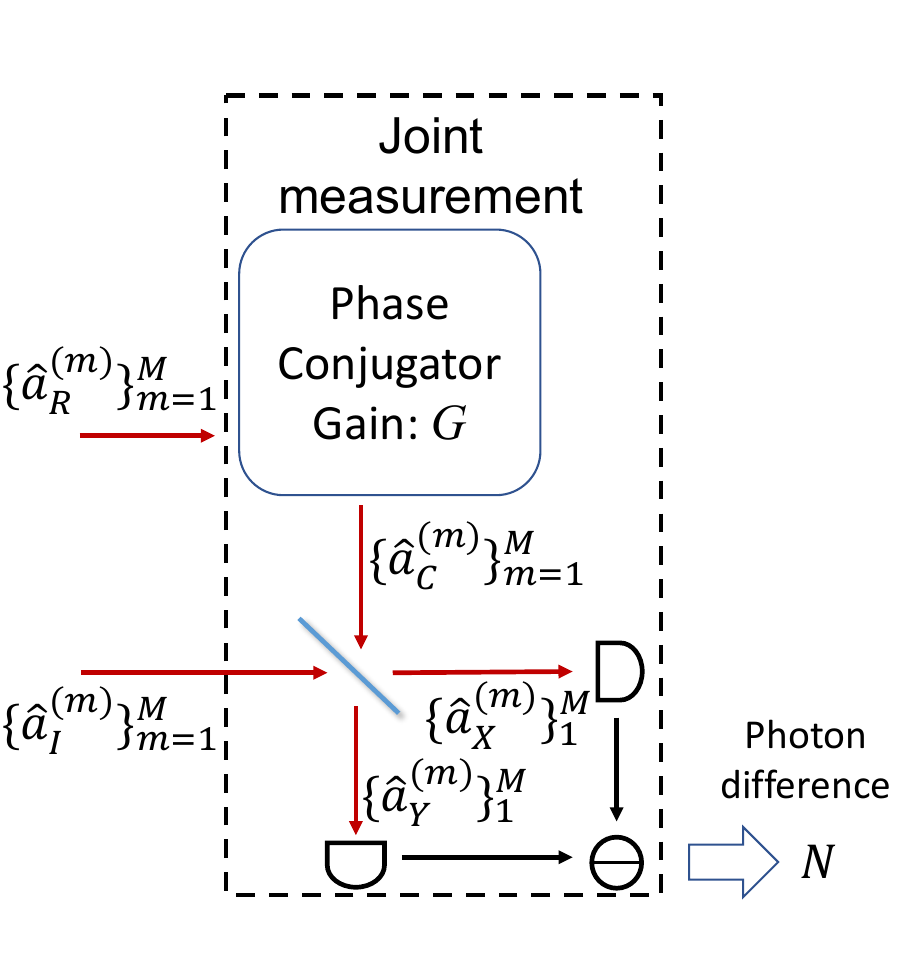}
    \caption{Schematic of the phase-conjugate (PC) receiver. $M$ iid return-idler mode pairs $\{\hat a_R^{(m)},\hat a_I^{(m)}\}_{m=1}^M$ are input to the receiver. The receiver uses a parametric interaction to phase conjugate the returned modes $\{\hat a_R^{(m)}\}$, yielding $\{\hat a_C^{(m)}\}$ with $\hat{a}_C^{(m)} = \sqrt{G}\,\hat{a}_v^{(m)} + \sqrt{G-1}\,\hat{a}_R^{(m)\dagger}$, where the $\{\hat{a}_v^{(m)}\}$ modes are in their vacuum states. It then interferes those conjugate modes with their idler companions on a 50--50 beam splitter and photodetects \jhs{the} resulting outputs The total photon number difference over the $M$ pairs is used to complete the EA sensing or communication task at hand. }
    \label{fig:schematic_PC receiver}
\end{figure}

\subsubsection{Phase conjugate (PC) receiver}

Supplementary Fig.~\ref{fig:schematic_PC receiver} shows a schematic of the phase-conjugate (PC) receiver. 
The inputs are $M$ iid return-idler mode pairs $\{\hat a_{R}^{(m)},\hat a_{I}^{(m)}\}$. A parametric interaction phase conjugates the $\{\hat a_{R}^{(m)}\}$ yielding $\{\hat a_C^{(m)}\}$, where $\hat{a}_C^{(m)} = \sqrt{G}\,\hat{a}_v^{(m)} + \sqrt{G-1}\,\hat{a}_{R}^{(m)\dagger}$ with the $\{\hat{a}_v^{(m)}\}$ modes being in their vacuum states. The receiver then interferes the $\{\hat a_C^{(m)}\}$ with their idler companions on a 50--50 beam splitter. The difference between the total photon counts from the $X,Y$ output ports, $\hat N= \hat N_X-\hat N_Y$ where $\hat N_X=\sum_{m=1}^M\hat a_X^{(m)\dagger} \hat a_X^{(m)} $, $\hat N_Y=\sum_{m=1}^M\hat a_Y^{(m)\dagger} \hat a_Y^{(m)}$, is the sufficient statistic for completing the EA sensing or communication task at hand. For $M\gg 1$, the central limit theorem allows us say that $N$ is approximately a Gaussian random variable with probabiity density function 
\be 
P_{N|\theta,\kappa}^{(M)}(n|\theta,\kappa)=\frac{1}{\sqrt{2\pi\sigma^2(\theta,\kappa)}}\exp{-\frac{\left(n-\mu\left(\theta,\kappa\right)\right)^2}{2\sigma^2\left(\theta,\kappa\right)}},
\label{eq:Pn_PC receiver}
\ee
whose mean and variance are
\bal 
\mu(\theta,\kappa)&=M\cdot 2C_{CI}\cos(\theta)\,,\\
\sigma^2(\theta,\kappa)&=M\cdot\left(N_I + 2 N_C N_I+  N_C+2 C_{CI}^2\cos(2\theta)\right)\,,
\label{eq:Pn_PC receiver_mean_var}
\eal 
where $N_C= ({G}-1)  ( \kappa N_S+{N_B}+1) , N_I= N_S$ and $C_{CI}=\sqrt{(G-1)\kappa N_S(1+N_S)} $. 

%Here we approximate $n$ as a continuous variable as $M\gg1$.

In the quantum illumination scenario the task is to discriminate between two channel hypotheses, $H_0:\Phi_{0,0}$ and $H_1:\Phi_{\kappa,0}$. The means and variances of $N$ differ under the two hypotheses: $\mu_0=\mu(0,0)$, $\sigma_0^2=\sigma^2(0,0)$ for $H_0$, and $\mu_1=\mu(0,\kappa)$, $\sigma_1^2=\sigma^2(0,\kappa)$ for $H_1$. Using the near-optimum threshold detector with threshold $N_{\rm th}\equiv\lceil M(\sigma_1\mu_0+\sigma_0\mu_1)/(\sigma_0+\sigma_1)\rceil$~\cite{Guha2009}, the target-detection error probability is
\be
P_{E,{\rm PC}} = \frac{1}{2}{\rm Erfc}\left(\sqrt{R_{\rm PC}^{\rm QI}M}\right),
\label{eq:pe_PC receiver}
\ee
where $R_{\rm PC}^{\rm QI} = \kappa N_S(N_S+1)/(2N_B+4N_SN_B+6N_S+4\kappa N_S^2+3\kappa N_S+2)$. In the $N_S\ll 1, \kappa\ll 1, N_B\gg 1$ asymptotic regime, we get $R_{\rm PC}^{\rm QI}\simeq \kappa N_S/(2N_B)$ matching the PA receiver's asymptotic-regime $R_{\rm PA}^{\rm QI}$. Outside that regime, however, the PC receiver typically has slightly better performance than the PA receiver.

In the phase estimation scenario, the PC \jhs{receiver's} Fisher information for estimating the $\Phi_{\kappa,\theta}$ channel's phase is 
\begin{align}
\mathcal{I}_{F,\rm PC}(G)&=\!\int_{-\infty}^\infty \! {\rm d}n \left[\partial_\theta \ln\left(P_{N|\theta,\kappa}^{(M)}(n|\theta,\kappa)\right)\right]^2 \! P_{N|\theta,\kappa}^{(M)}(n|\theta,\kappa)
\nonumber
\\
&=\left[\partial_\theta \mu(\theta,\kappa) \right]^2/[\sigma^2(\theta,\kappa)/M].
\end{align} 
Substituting from \eqref{eq:Pn_PC receiver_mean_var}, the Fisher information's dependence on the conjugator's gain $G$ is
\be
\mathcal{I}_{F,{\rm PC}}(G)= M\cdot \frac{4 ({G}-1) \kappa {}  {} {N_S} ({N_S}+1)\sin^2\theta}{(N_I+N_C)+\left(2N_CN_I+2C_{CI}^2\cos(2\theta)\right)}.
\ee
It is easy to check that $\mathcal{I}_{F,\rm PC}(G)$ monotonically increases with increasing $G$, while the gradient decays rapidly. As a result, one may regard the case of $G=2$ as almost saturating the high-gain limit, and obtain a performance sufficiently close to the optimum
\begin{align} 
&\mathcal{I}_{F,\rm PC}(G=2)=
\nonumber
\\
&\frac{4M \kappa  N_S \left(N_S+1\right)  \sin^2\theta }{ N_B(1+2N_S)+N_S \left(2\kappa  N_S+\kappa+3\right)+2 \kappa  \cos (2 \theta ) N_S \left(N_S+1\right)+1}.
\end{align}
In practice, the gain can be limited because the broadband parametric interaction is intrinsically weak. For $N_S\ll 1$, we can obtain a less stringent condition for $G$ to saturate the quantum advantage. Consider the weak-gain limit, $G-1\ll 1$, where we have
\bal 
&\mathcal{I}_{F,{\rm PC}}(G)=\frac{4M   \kappa  {N_S} ({N_S}+1)\sin^2\theta} {1+N_B+N_S/(G-1)+O(N_S)}.
\label{eq:FI_PC receiver_asym}
\eal 
The term $N_S/(G-1)$ in the denominator will be negligible when
\be
 (G-1)(1+N_B) \gg N_S
\label{eq:gain_cond}
\,.\ee
Indeed, as long as this condition holds, the PC receiver's Fisher information reduces to the zeroth-order asymptotic formula $\mathcal{I}_F(\hat{\rho}_{R\mid \theta})_{{\rm PC}}\simeq 4M\kappa N_S\sin^2\theta/(1+N_B)$, which saturates the optimum 3\,dB entanglement-assisted advantage over the classical coherent-state approach locally at $\theta=\pi/2$.

In the communication scenario, we consider the BPSK modulation such that $\theta\in\{0,\pi\}$ with equal probability. Then the conditional statistics of \eqref{eq:Pn_PC receiver} leads to the unconditional statistics $P_{N}^{(M)}(n)=\sum_{\theta\in\{0,\pi\}} P_{N|\theta,\kappa}^{(M)}(n|\theta,\kappa)/2$, and thereby the Shannon information is given by  \eqref{eq:Shannon_supp}, using $P_{N|\theta,\kappa}^{(M)}(n|\theta,\kappa)$ from  \eqref{eq:Pn_PC receiver}.
% \be 
% I(N;\theta)=S(\theta)-S(N|\theta)\,,
% \ee
% where 
% \bal 
% S(N|\theta)&=\\
%  -\!\sum_{\theta\in\{0,\pi\}}&\frac{1}{2}\int_{-\infty}^\infty dn P_{N|\theta,\kappa}^{(M)}(n|\theta,\kappa)\log_2 P_{N|\theta,\kappa}^{(M)}(n|\theta,\kappa)\,,\\
% S(\theta)&=-\sum_{\theta\in\{0,\pi\}} \frac{1}{2}P_{N}^{(M)}(n)\log_2 P_{N}^{(M)}(n)\,.
% \eal 
In our simulations we chose $M=1000$ ensure \eqref{eq:Pn_PC receiver}'s validity. As found for the PA receiver, the PC receiver's Shannon information numerical results did not decay significantly as $M$ increased up to 1000 in the parameter region of interest.

The unused output from the PC receiver's phase conjugator, i.e., 
\be 
\hat{a}_D^{(m)}=\sqrt{G}\,\hat{a}_R^{(m)}+\sqrt{G-1} \hat{a}_v^{(m)\dagger},
\ee 
has a phase-sensitive cross correlation with its idler companion given by
\be 
\expval{\hat{a}_D^{(m)} \hat{a}_I^{(m)}}=\sqrt{G-1}C_p e^{i\theta}.
\ee 
That this correlation is non-zero indicates that the PC receiver does not make full use of the phase-sensitive cross correlation between the returned radiation and the idler.

\section{Detailed analyses of ${\rm C}\to {\rm D}$ reception}
\label{sec:details_C2D}

In this section, we provide detailed analyses supporting  results presented in the main text.

\subsection{Error probability analyses}
\label{app:LB_exp}

Here we analyze ${\rm C}\to {\rm D}$ receiver's error probability for quantum illumination target detection. In particular, we obtain the upper bound on $\Pr(e)_{{\rm C}\to {\rm D}}$ in the main text's \eqref{PeCD} from which we obtain the lower bound on the error exponent $r_{{\rm C}\to {\rm D}}$ given in the main text's \eqref{r_LB}.

\begin{lemma}
\label{lemma:PCD_UB}
The ${\rm C}\to {\rm D}$ receiver's error probability for quantum-illumination target detection obeys
\jhs{\begin{eqnarray}
\lefteqn{\Pr(e)_{{\rm C}\to {\rm D}}} \nonumber \\[.05in]
&&\le \frac{1}{2}\min_{s\in[0,1]} \left(1+\frac{4\xi}{\Lambda_{s}(1+2N_S)+\Lambda_{1-s}(1+2E)}\right)^{-M},
\label{PCD_UB}
\end{eqnarray}}
where $\xi$ is defined in \eqref{xi_definition} and 
\be 
\Lambda_p(\nu) \equiv \frac{(\nu+1)^p+(\nu-1)^p}{(\nu+1)^p-(\nu-1)^p}.
\ee 
\end{lemma}

Before proving the lemma, we provide some discussion.
By choosing $s=1/2$, we obtain a slightly looser upper bound
\be 
\Pr(e)_{{\rm C}\to {\rm D}}\le \left(1+\frac{4\xi}{h(N_S)+h(E)}\right)^{-M},
\ee 
where $h(y)\equiv \Lambda_{1/2}(1+2y)=\left(\sqrt{y+1}+\sqrt{y}\right)^2$.  The desired lower bound on the ${\rm C}\to {\rm D}$ receiver's error exponent, $r_{{\rm C}\to {\rm C},{\rm LB}}$ now follows via the following argument  
\begin{align}
&r_{{\rm C}\to {\rm D}} 
\ge \max_{s\in[0,1]}\ln\left[1+\frac{4\xi}{\Lambda_{s}(1+2N_S)+\Lambda_{1-s}(1+2E)}\right]
\label{eq:C11}
\\
&\ge 
\ln \left[1+\frac{2}{h(N_S)+h(E)}\frac{ {\kappa  N_S \left(N_S+1\right)}}{N_B+\kappa  N_S+1}\right]
\label{eq:C12}
\\
&\ge 
\ln\left[1+(\sqrt{N_S+1}-\sqrt{N_S})^2\frac{ {\kappa  N_S \left(N_S+1\right)}}{N_B+\kappa  N_S+1}\right],
\label{eq:C13}
\end{align} 
where \jhs{we} used the fact \jhs{that $E\le N_S$}. A comparison of the above three lower bounds, normalized by the coherent-state Chernoff exponent
$ 
r_{\rm CS}=\kappa N_S(\sqrt{N_B+1}-\sqrt{N_B})^2
$~\cite{tan2008quantum}, is shown in Supplementary Fig.~\ref{fig:QCB_Nb}. The lower bounds are shown to be always close to each other, especially for $N_S\leq 10$, so the main text uses the one from \eqref{eq:C13}. 

\begin{figure}
    \centering
    \includegraphics[width=0.5\textwidth]{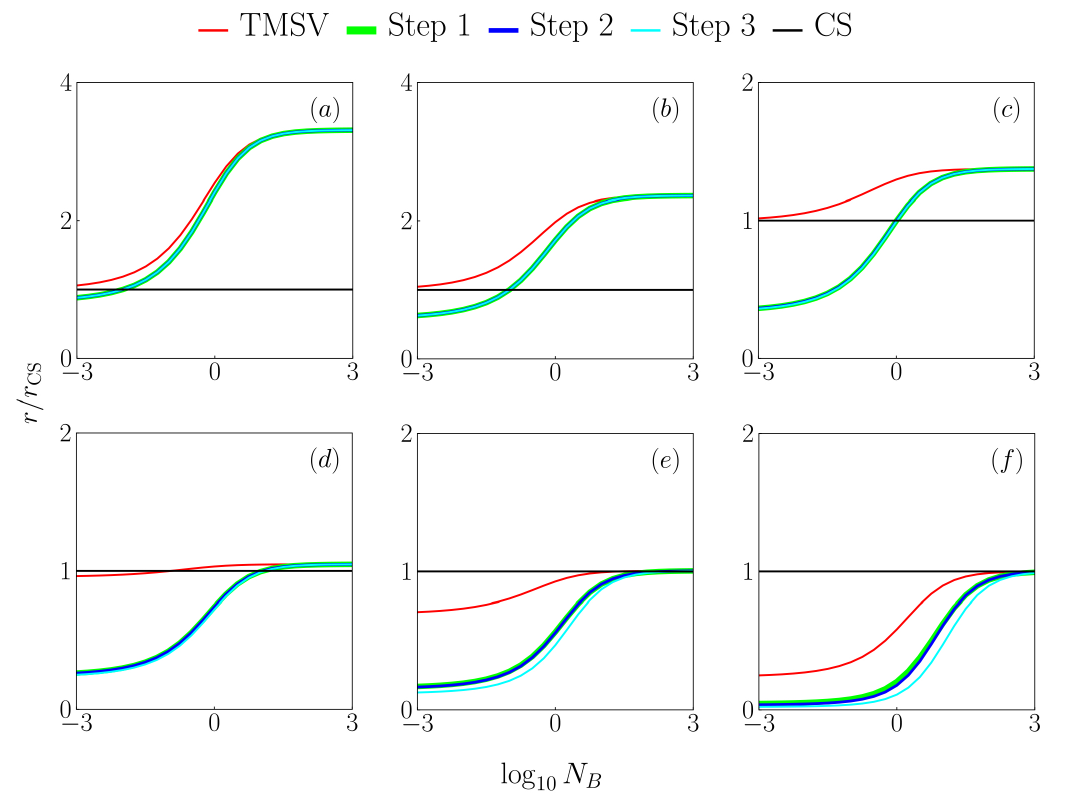}
    \caption{Comparison of the ${\rm C}\to {\rm D}$ receiver's error-exponent lower bounds for quantum-illumination target detection given in   \jhs{(\ref{eq:C11})--(\ref{eq:C13})} with that error exponent of that receiver's quantum Chernoff bound. All are  normalized by $R_{\rm CS}$, the error exponent of a coherent-state system.  Plots are versus $N_B$ with $\kappa = 0.01$, and, in (a-f), $N_S=0.01, 0.1,1,10,100,1000$. Normalized by the coherent-state error exponent $r_{\rm CS}$.  
    }
    \label{fig:QCB_Nb}
\end{figure}

Now we prove Lemma~\ref{lemma:PCD_UB}

\begin{proof}
To begin with, the Helstrom limit is upper bounded by the QCB for any number of mode pairs~\cite{Audenaert2007,Pirandola2008}, therefore, using $\hat{\rho}_{0,N_S}$ to denote a thermal state with average photon number $N_S$ and $\hat{\rho}_{\sqrt{x},E}$ to denote a displaced thermal state with displacement photon number $x$ and an average noise photon number $E$, we have
\begin{align} 
&P_{\rm H}(\hat{\rho}_{0,N_S},\hat{\rho}_{\sqrt{x},E})
\\
&\le \frac{1}{2}\inf_{s\in[0,1]}Q_s\left(\hat{\rho}_{0,N_S},\hat{\rho}_{\sqrt{x},E}\right)
\\
&=\frac{1}{2}\inf_{s\in[0,1]}\overline{Q}_{s}\exp{-\frac{1}{2}{\bm d}^T\left(\tilde{V}_1(s)+\tilde{V}_2(1-s)\right)^{-1}{\bm d}}
\\
&=\frac{1}{2}\inf_{s\in[0,1]}\overline{Q}_{s}\exp{-\frac{2x}{\Lambda_{s}(1+2N_S)+\Lambda_{1-s}(1+2E)}},
\end{align}
where we utilized the QCB definition from Sec.~\ref{app:QCB} with ${\bm d}=(2\sqrt{x},0)$ and 
$
\tilde{V}_1(s)= \Lambda_{s}(1+2N_S)\mathbb{I}, \tilde{V}_2(s)= \Lambda_{s}(1+2E)\mathbb{I}.
$ 

The remainder of the proof readily follows:
\begin{align}
&\Pr(e)_{{\rm C}\to {\rm D}}=\int {\rm d}x P_{\rm disp}^{(M)}(x) P_{\rm H}(\hat{\rho}_{0,N_S},\hat{\rho}_{\sqrt{x},E}) 
\\
&\le 
\int {\rm d}x P_{\rm disp}^{(M)}(x)
\frac{1}{2}\inf_{s\in[0,1]}\left[\overline{Q}_{s}\right. \nonumber \\
&\left.\exp{-\frac{2x}{\Lambda_{s}(1+2N_S)+\Lambda_{1-s}(1+2E)}}\right]
\\
&\le \frac{1}{2}\inf_{s\in[0,1]} \left[\overline{Q}_{s}
\int {\rm d}x P_{\rm disp}^{(M)}(x) \right. \nonumber\\
&\left.\exp{-\frac{2x}{\Lambda_{s}(1+2N_S)+\Lambda_{1-s}(1+2E)}}\right]
\\
&= \frac{1}{2}\inf_{s\in[0,1]} \overline{Q}_{s}\left(1+\frac{4\xi}{\Lambda_{s}(1+2N_S)+\Lambda_{1-s}(1+2E)}\right)^{-M}
\\
&\le 
\frac{1}{2}\min_{s\in[0,1]} \left(1+\frac{4\xi}{\Lambda_{s}(1+2N_S)+\Lambda_{1-s}(1+2E)}\right)^{-M}.
\end{align} 

\end{proof}

\begin{figure}
    \centering
    \includegraphics[width=0.5\textwidth]{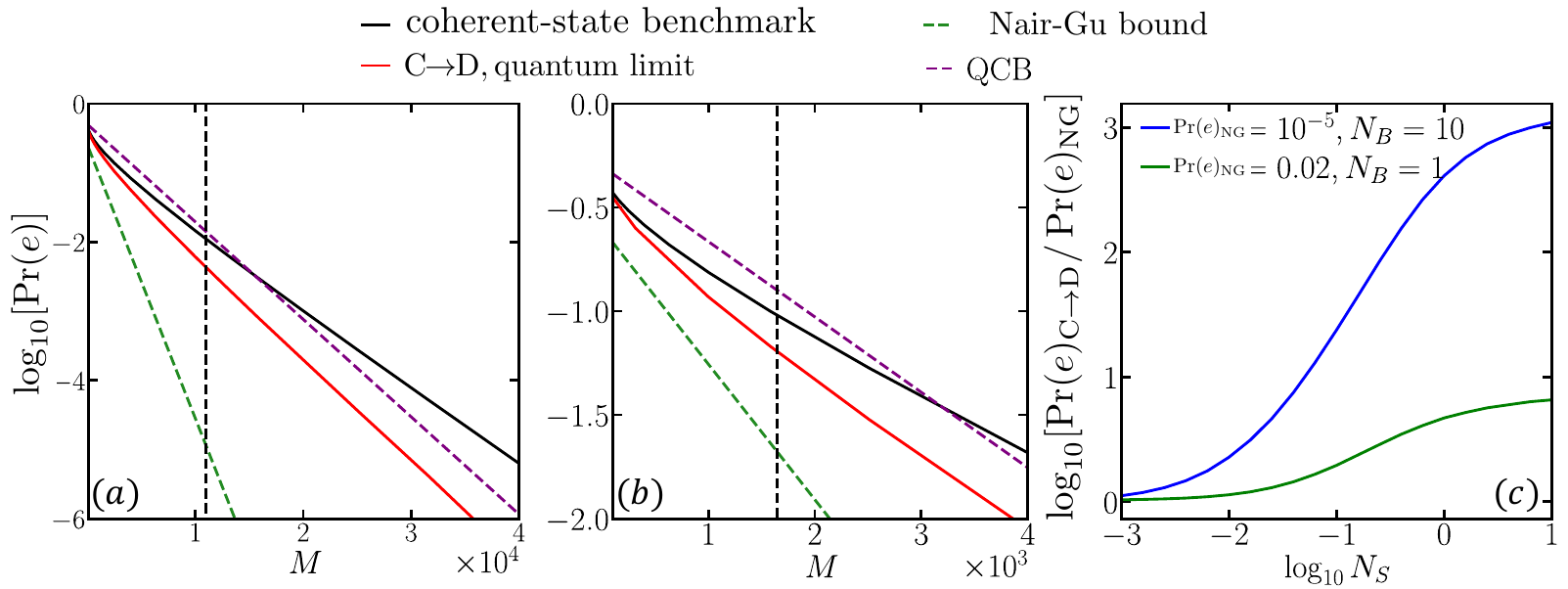}
    \caption{Quantum illumination error probability versus number of mode pairs $M$ with (a) $N_S=1$, $N_B=10$ and (b) $N_S=0.3, N_B=1$. Vertical dashed lines in (a) and (b) indicate the corresponding error probability with $\Pr(e)_{\rm NG}=10^{-5}$ and $\Pr(e)_{\rm NG}=0.02$, respectively. (c) Error-probability ratio $\Pr(e)_{{\rm C}\to  {\rm D}}/\Pr(e)_{\rm NG}$ versus $N_S$ for $\Pr(e)_{\rm NG}$ and $N_B$ chosen from (a) and (b). In all cases $\kappa=0.01$.}
    \label{fig:illumination_largeN}
    \centering
    \includegraphics[width=0.5\textwidth]{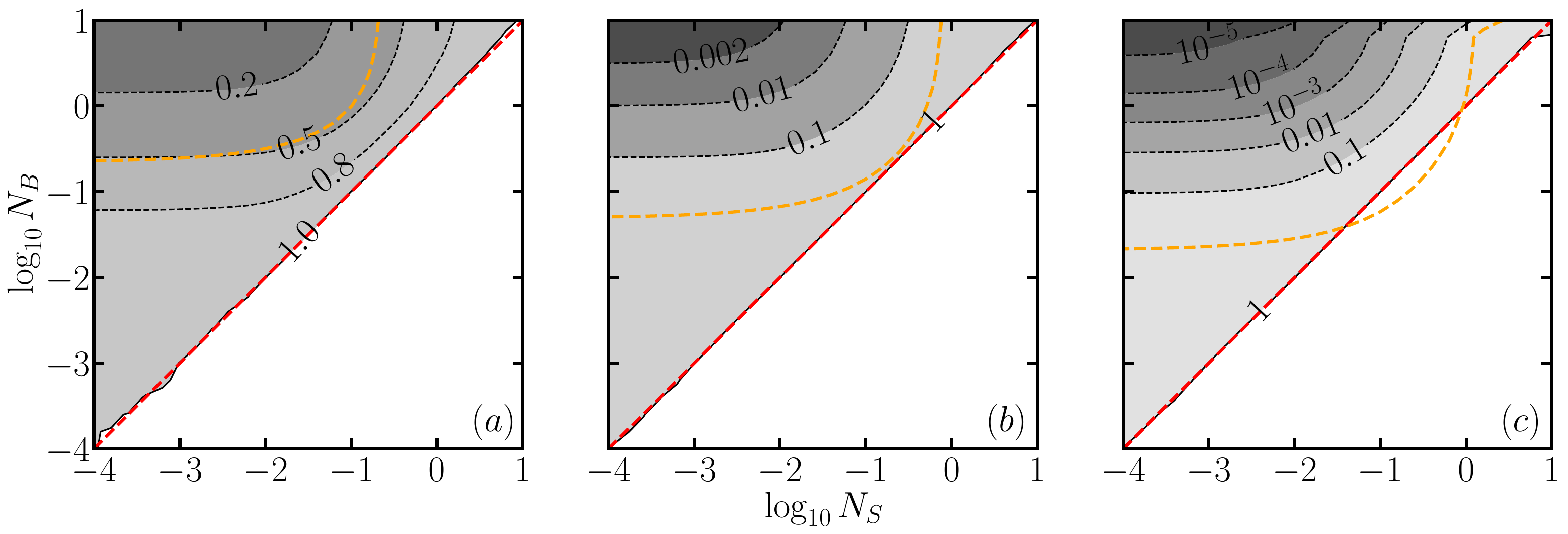}
    \caption{Quantum illumination error-probability ratio $\Pr(e)_{{\rm C}\to  {\rm D}}/P_{\rm H,CS}$ versus $N_S$, $N_B$ with, from left to right, $P_{\rm H,CS}=0.1, 0.01, 0.001$. Red dashed lines indicate the quantum-advantage boundary of${\rm C}\to {\rm D}$ reception, viz., $N_S\le N_B$ while the orange dashed curves -represent the quantum advantage boundary predicted by the QCB. The discontinuity of the QCB curve in the upper right of (c) is due to the requirement that the optimal $M$ value to hold $P_{\rm H,CS}=0.001$ be an integer. In all cases $\kappa=0.01$.}
    \label{fig:illumination_contours}
\end{figure}

To complete our error-probability analysis for ${\rm C}\to {\rm D}$ reception, we augment the main text's non-asymptotic cases by examining the error probability's behavior for neither $N_S \gg 1$, nor $N_B \gg 1$ but $N_S \le N_B$.  Supplementary Figs.~\ref{fig:illumination_largeN}(a) and (b) show the quantum advantage still exists. In the non-asymptotic region, we also see that quantum advantage advantage---found in our analysis of ${\rm C}\to {\rm D}$ reception---is not revealed by the QCB.   Note that at high brightness, there is a relatively large gap between the ${\rm C}\to {\rm D}$ receiver's error probability and the Nair-Gu bound from the main text's Eq.~(23), which is further confirmed in Supplementary~Fig.~\ref{fig:illumination_largeN}(c). 

In the main text, we show the error probability ratio with fixed $P_{\rm H,CS}=0.05$; here, we extend to $P_{\rm H, CS}=0.1, 0.01, 0.001$ in Fig.~\ref{fig:illumination_contours}(a)-(c) to cover both the non-asymptotic and asymptotic regions. With the conversion module, the error probability decreases with a smaller fixed $P_{\rm H,CS}$, indicating a larger quantum advantage. At the same time, the parameter region where quantum advantage can be predicted by QCB also increases. Note that there exist a region where $P_{\rm QCB}< P_{\rm H,CS}<\Pr(e)_{{\rm C}\to  {\rm D}}$ and quantum advantage can only be revealed by QCB, as shown in Fig.~\ref{fig:illumination_contours}(c). The advantage can also be seen, from Supplementary~Figs.~\ref{fig:QCB_Nb}(b) and (c), where, when $N_B\lessapprox N_S$ , the ${\rm C}\to {\rm D}$ receiver's error-exponent lower bound is lower than the coherent-state error exponent while the QCB error-exponent exceeds the coherent-state result.

\subsection{Quantum Fisher information for phase sensing}
\label{sec:Fisher}

In this section, we evaluate the quantum Fisher information (QFI) for phase sensing, utilizing the approach from Ref.~\cite{gao2014bounds}, where
the parameter of interest is the $\Phi_{\kappa,\theta}$ channel's phase shift $\theta$.

A displaced thermal state $\rho_{\sqrt{x}e^{i\theta},y}$ has the complex mean and covariance matrix (see Ref.~\cite{gao2014bounds}'s definitions)
\bal 
\bm d&=[\sqrt{x} e^{i\theta},\sqrt{x} e^{-i\theta}]^T,
\\
\Sigma&=\left(
\begin{array}{cc}
 0 &  y+1/2 \\
  y+1/2 & 0 \\
\end{array}
\right)\,.
\eal
The QFI for phase sensing is therefore 
\be 
\mathcal{I}_{F,\rm DTS}=\frac{4x}{1+2y}\,.
\label{QFI_DTS}
\ee
Now consider using $M$ independent identically distributed (iid) modes to estimate the phase shift of the $\Phi_{\kappa,\theta}$ channel, see the main text's Eq.~(7) for a definition of this channel,  with $N_B$ being the modal noise brightness and $N_S$ being the modal signal brightness. For a classical protocol using iid coherent-state mosdes, $\ket {\sqrt{N_S}}^{\otimes M}$, observe that the channel output $[\Phi_{\kappa,\theta}(\ketbra{\sqrt{N_S}}{\sqrt{N_S}})]^{\otimes M}$ is a product of displaced thermal states. Then the $M$ channel outputs can be combined into a single mode in a displaced thermal state by a balanced $M$-port beam splitter. This processing does not change the QFI, because the beam splitter transform is unitary and the output is again a product state, where the undisplaced (i.e., the noise) modes can be discarded. The output state has $x=M\kappa N_S$, $y=N_B$, thus
\be 
\mathcal{I}_{F,\rm CS}=\frac{4M\kappa N_S}{1+2N_B}\,.
\ee
Similarly, for an entanglement-assisted protocol using broadband TMSV radiation and ${\rm C}\to {\rm D}$ reception, the idler modes, after the heterodyned returned radiation is used to program a \jhs{mode selector}, are combined into a displaced thermal state, whose squared mean $x$ has the $\chi^2$ distribution $P_\kappa^{(M)}(x)$ defined in \eqref{p_overall}. Thus
\be 
\mathcal{I}_F(\hat{\rho}_{RI\mid \theta})_{{\rm C}\to {\rm D}}\equiv\int {\rm d}x P_{\rm disp}^{(M)}(x) \mathcal{I}_F(\hat{\rho}_{e^{i\theta}\sqrt{x},E})=\frac{8M\xi}{1+2E}\,,
\ee 
where $\mathcal{I}_F(\hat{\rho}_{e^{i\theta}\sqrt{x},E})=4x/(1+2E)$ is the QFI of the displaced thermal state conditioned on a specific $x$. Plugging the definitions of $\xi$, $E$ in the main text, we obtain
\be 
\mathcal{I}_F(\hat{\rho}_{RI\mid \theta})_{{\rm C}\to {\rm D}}=\frac{4 M\kappa  N_S \left(N_S+1\right)}{1+N_B+N_S \left(2 N_B+2-\kappa \right)}.
% \label{eq:qfi_c2d_supp}
\ee

In the neighborhood of true value, the QFI of a displaced thermal state is achieved by a homodyne measurement. This can be seen as follows. Suppose that one first applies an angle $-\theta_c$ phase rotation to the $\hat\rho_{\sqrt{x}e^{i\theta},y}$ state, so that it  becomes $\hat\rho_{\sqrt{x}e^{i(\theta-\theta_c)},y}$. Then we homodyne detect the imaginary quadrature, giving the random readout $Q$ with probability density
\be 
p_Q(q)=\frac{1}{\sqrt{2\pi\sigma^2}}\exp{-\frac{\left(Q-\sqrt{2}\Im\left(d\right)\right)^2}{2\sigma^2}},
\ee
where $\sigma^2=1/2+y$, $d=\sqrt{x}e^{i(\theta-\theta_c)}$.
Thus the Fisher information of homodyne measurement, depending on the phase compensation $\theta_c$, can be calculated from the distribution with the following result,
\be 
\mathcal{I}_{F,\rm hom}(x,E,\theta)=\frac{4x}{1+2y}\sin^2(\theta-\theta_c)\,.
% \label{eq:Fisher_hom_supp}
\ee
Now we see that homodyne measurement achieves the QFI in \eqref{QFI_DTS} locally, which is true only when the prior knowledge is sufficient such that $|\theta-\theta_c|\ll 1$, while its performance decays rapidly with increasing $|\theta-\theta_c|$. When prior knowledge is insufficient, an adaptive policy can be designed to approach the ideal compensation, as the number of available mode pairs becomes sufficiently high. %To improve the convergence speed of the adaptive policy, displaced photon counting~\cite{izumi2016optical} gives a wider working range by providing a smoother profile $\sim \cos^2(\theta/2)$. This may sharply improve the adaptive receiver in low signal-to-noise-ratio region, as the performance adaptive process depends nonlinearly on the information per step.
\begin{figure}[t]
    \centering
    \includegraphics[width=0.25\textwidth]{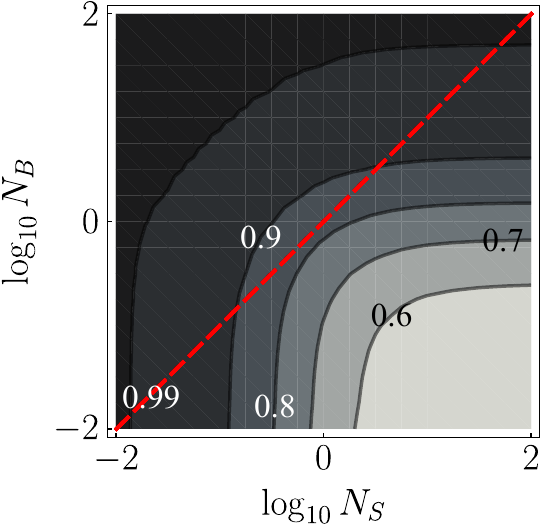}
     \caption{The ${\rm C}\to {\rm D}$ receiver's  quantum Fisher information, normalized by the TMSV limit, versus $N_S$ and $N_B$ with $\kappa = 0.01$. The red dashed diagonal line indicates the quantum-enhanced region's boundary.}
    \label{fig:qfi_limits_NbNs_TMSV}
\end{figure}

It is worthwhile to note that the ${\rm C}\to  {\rm D}$ receiver is optimal for TMSV-based phase estimation: it achieves the QFI of the channel output for TMSV sources \cite{shi2020practical}
\be 
\mathcal{I}_{F,\rm TMSV}=\frac{4 M\kappa  N_S \left(N_S+1\right)}{1+N_B(1+ 2N_S)+N_S(1-\kappa)} 
\label{eq:QFI_TMSV_supp}
\ee
in the limit of $N_B\gg 1$. This behavior is verified in the numerical evaluations shown in Fig.~\ref{fig:qfi_limits_NbNs_TMSV}.

\subsection{Entanglement-assisted communication rate analyses}
\label{sec:holevo}

In this section we evaluate the Holevo information~\cite{holevo1973bounds,wilde2013quantum} of the output ensemble from our $\rm C\to  D$ receiver, using a phase-encoded broadband TMSV source. The Holevo information is a tight upper bound on a channel's information rate given a specific encoding ensemble $\{p_\theta,\hat\rho_\theta\}$, which is achieved by the optimum receiver. The ultimate capacity can be obtained by optimizing the Holevo information over $\{p_\theta,\hat \rho_\theta\}$. In general $\theta$ can be an arbitrary parameter, while we specify it to be the phase shift in the main text and here. We consider repetition coding that yields $M$ iid copies of the output ensemble. Given $\theta$, the output state is a displaced thermal state $\hat\rho_{\sqrt{X}e^{i\theta},E}$, where $X$ is a displacement photon number whose probability density function is the $\chi^2$ distribution $P_{\rm disp}^{(M)}(x)$ defined in \eqref{p_overall}. Let the encoding phase be a random variable \jhs{$\Theta$ with} probability distribution $P_\Theta$. Denote the output quantum system as $O$. In the communication protocol, the displacements $X$ \jhs{and} the output quantum system $O$, along with the input symbol $\Theta$, are in a classical-quantum state
\bal 
&\hat{\sigma}_{XO\Theta}=\\
&\int \! {\rm d}\theta p_\Theta(\theta) \! \int  \! {\rm d}x P_{\rm disp}^{(M)}(x) \state{x}_X \otimes \left(\hat{\rho}_{\sqrt{x}e^{i\theta},E}\right)_O\otimes \ketbra{\theta}{\theta}_\Theta\,.
\eal 
The overall Holevo information about the input symbol is then 
\bal 
\chi_{\rm C\to  D}&\equiv \frac{1}{M} [S(XO)_{\hat \sigma}-S(XO|\Theta)_{\hat \sigma}]\\
&= \frac{1}{M}\!\int \! {\rm d}x p_X(x) \left[S(O|X=x)_{\hat \sigma} \! -\! S(O|\Theta,X=x)_{\hat \sigma}\right]\\
&= \frac{1}{M}\int {\rm d}x P_{\rm disp}^{(M)}(x) \chi\left(\{p_\Theta,\hat{\rho}_{\sqrt{x}e^{i\Theta},E}\}\right)\,,
\label{eq:overallHolevo_supp}
\eal
where the second equality is due to the joint entropy theorem \cite{nielsen2002quantum} given the orthogonality of $\{\ket{x}_X\}$. Here $\chi\left(\{p_\Theta,\hat{\rho}_{\sqrt{x}e^{i\Theta},E}\}\right)$ can be efficiently evaluated in the following example.

Consider continuous PSK (CPSK) modulation on TMSV sources with $P_\Theta(\theta)=1/2\pi, \theta\in [0,2\pi)$. The output ensemble of the $\rm C\to  D$ module yields the Holevo information
\bal
&\chi\left(\{P_\Theta,\hat{\rho}_{\sqrt{x}e^{i\Theta},E}\}\right)
\\
&=S(\int {\rm d}\theta P_\Theta(\theta)\hat{\rho}_{\sqrt{x}e^{i\theta},E})-\int {\rm d}\theta P_\Theta(\theta)S(\hat{\rho}_{\sqrt{x}e^{i\theta},E})
\\
&=H\left[\{P(n|X=x)\}\right]-g(E),
\label{eq:almost_closed_form_supp}
\eal 
where third line is justified as follows. The conditional states $\{\hat{\rho}_{\sqrt{x}e^{i\theta},E}\}$ are Gaussian with identical entropies,
$
S(\hat{\rho}_{\sqrt{x}e^{i\theta},E})=g(E),
$ 
where $g(n)=(n+1)\log_2(n+1)-n \log_2 n$ is the entropy of a thermal state with mean photon number $n$. Thus
$
\int {\rm d}\theta P_\Theta(\theta)S(\hat{\rho}_{\sqrt{x}e^{i\theta},E})=g(E).
$
The unconditional state $\int {\rm d}\theta P_\Theta(\theta)\hat{\rho}_{\sqrt{x}e^{i\theta},E}$ is completely dephased owing to the uniformly-distributed phase encoding, thus its eigenbasis is the Fock basis. Moreover, its distribution on that basis is
\be 
P(n|X=x)=E^n (E+1)^{-n-1} e^{-\frac{x}{E+1}} L_n\left(-\frac{x}{E (E+1)}\right),
\ee 
where $L_n(x)$ is the $n$th Laguerre polynomial. 
Thus the unconditional entropy reduces to the Shannon entropy of the photon-number distribution
\be 
H\left[\{P(n|X=x)\}\right]\equiv-\sum_{n=0}^\infty P(n|X=x)\log \left(P(n|X=x)\right).
\label{eq:S_app}
\ee
Combining Eqs.~(\ref{eq:overallHolevo_supp})and~(\ref{eq:almost_closed_form_supp}), we have the Holevo information for CPSK
\be 
\chi_{\rm C\to  D}^{\rm CPSK}=\frac{1}{M}\left[\int {\rm d}x P_{\rm disp}^{(M)}(x) H\left[\{P(n|X=x)\}\right]-g(E)\right].
\label{chi_CD_exact}
\ee 
We can use \eqref{chi_CD_exact} for efficient numerical evaluation. Below, we obtain some asymptotic results.

By the law of large number, $x$ converges to $2M\xi$ in as $M\rightarrow \infty$.  We then have a closed-form formula,
\be 
\chi_{{\rm C}\to  {\rm D}}^{\rm CPSK}=\frac{1}{M}\left[H\left[\{P(n|X=2M\xi)\}\right]-g(E)\right]
\ee
that expand, in the $N_S\rightarrow$ limit, to get
\ba
&\chi_{{\rm C}\to  {\rm D}}^{\rm CPSK}=
\frac{\kappa N_S \left[ \ln \left(\frac{1}{N_S }\right)+\calR_{\rm C\to  D}\right]}{ \left(N_B+1\right)\ln 2}+O(N_S^2)\\
   &=\frac{\kappa  N_S \ln(1/N_S)}{(N_B+1)\ln 2}+O(N_S)\,,
\label{eq:chi_CPSK_supp}
\ea 
where 
\jhs{\begin{eqnarray}
\lefteqn{\calR_{\rm C\to  D} =}\nonumber \\[.05in]
&&\frac{2 \left(-N_B+\kappa -1\right) }{\kappa M}\tanh ^{-1}\left(\frac{\kappa  M}{2 N_B-2 \kappa +\kappa  M+2}\right) \nonumber\\[.05in] 
&&+\,\,\left(\ln
   \left(N_B+1\right)+1\right)-\ln\left(N_B+\kappa  (M-1)+1\right),
\end{eqnarray}}   
and $|O(x)|/x<\infty $ as $x\to 0$.

Note that the above scaling at $N_S\to 0$ saturates the EA classical capacity in \eqref{CE_expansion}, and therefore is asymptotically optimal. At the same time, the information per mode is strictly higher than the case of $M\gg1$, therefore the optimal scaling applies to any finite $M$.\\

A similar route can be pursued to derive the information rate for the binary PSK (BPSK) case, where $P_\Theta(0)=P_\Theta(\pi)=1/2$. The conditional entropy is the same as that in the CPSK, 
$
S(\hat{\rho}_{\sqrt{x}e^{i\Theta},E})=g(E)
$. 
The evaluation of the unconditional entropy is more challenging: it is now a von Neumann entropy for which the  eigenvalues of the density operator $\hat{\overline{ \rho}}=\int {\rm d}\theta P_\Theta(\theta)\hat{\rho}_{\sqrt{x}e^{i\theta},E}$ must be found. Nevertheless, we find that a closed-form formula is still available \jhs{in} the limit \jhs{$M\rightarrow \infty$ (so} that $x\to 2M\xi$) and $N_S\to 0$. Indeed, the performance of BPSK is almost identical to the CPSK case in the parameter region of the main text's Fig.~\ref{fig:EACOMM_c2d_limits}. Below, we approximate the eigenvalues via matrix perturbation theory. We consider the Fock-basisrepresentation in  
\small 
\bal 
&\rho_{mn}=\expval{m|\hat{\overline{ \rho}}|n}
\\
&
=\frac{\sqrt{m!}  E^n \left| x\right|^{(m-n)/2} 
      ~_1\tilde{F}_1\left(m+1;m-n+1;\frac{x}{E^2+E}\right)
   e^{-\frac{x}{E}+i \theta (m-n)}}{(E+1)^{m+1}\sqrt{n!}},
\eal
\normalsize
where $_1\tilde{F}_1(a;b;z)$ is the regularized confluent hypergeometric function. In the numerical evaluation, we truncate $\rho$ to $d\times d$ dimensions.

In the final approximation to the eigenvalues, we expect to keep infinitesimal terms up to \jhs{$O(N_S)$, in which case} $d=3$ suffices, but we postpone the error analysis. Now, we apply a Taylor expansion to each matrix entry $\rho_{mn}$, viz.,  
\be 
\rho_{mn}=\tilde\rho_{mn}+\delta\rho_{mn}\,,
\ee
where the approximation $\tilde \rho_{mn}\sim O(N_S^{d})$ omits the higher-order term $\delta\rho_{mn} \sim O(N_S^{d+1})$. With $d=3$, the eigenvalues of $\tilde \rho$ can be solved for analytically, and thus we obtain the Holevo information
\bal 
&\chi_{{\rm C}\to  {\rm D}}^{\rm BPSK}=\frac{1}{M}\left[S( \rho)-g(E)\right]+O(\delta_d)\\
=&
\frac{\kappa N_S \left[ \ln \left(\frac{1}{N_S }\right)+\calR_{\rm C\to  D}\right]}{ \left(N_B+1\right)\ln 2}+O(\delta_{N_S})+O(\delta_d)\\
=&\frac{\kappa  N_S \ln(1/N_S)}{(N_B+1)\ln 2}+O(N_S)+O(\delta_{N_S})+O(\delta_d)\,,
\label{eq:chi_BPSK_supp}
\eal 
where $\delta_d$ is the maximum eigenvalues error from the matrix truncation and $\delta_{N_S}$ is from the matrix's Taylor expansion, and the $\calR_{\rm C\to  D}$ is the same as that defined earlier for the CPSK case below \eqref{eq:chi_CPSK_supp}.
The leading term of the matrix's Taylor expansion at the second equality coincides with \eqref{eq:chi_CPSK_supp}.

At last, we analyze the errors $\delta_d, \delta_{N_S}$ in \eqref{eq:chi_BPSK_supp}. First, let us consider $\delta_d$. Define the true eigenvalues of the operator $\hat {\overline{ \rho}}$ to be $\mu_1\ge \mu_2\ge \ldots $ and let the eigenvalues of the truncated representation $\rho$ be $\lambda_1\ge \lambda_2\ge \ldots\ge \lambda_d$. Then $\delta_d \equiv \max_{i\in [d]}|\mu_i-\lambda_i|$. Using Theorem~4.14 from Ref.~\cite{stewart1990matrix}, we have
\be 
|\delta_d| =||{\rm diag} \left(\mu_i-\lambda_i\right)||_2\le  || X||_2\,,
\ee 
%\QZ{**to be checked**}
where $ X= \rho^{(\infty)}  I_d- I_d  \rho$, $ I_d$ is a $d\times d$ matrix that implements the cutoff, and $ \rho^{(\infty)}$ is the exact infinite-dimensional matrix representation of $\hat {\overline{\rho}}$. Here the matrix 2-norm is defined using vector 2-norm: for a $d\times d$ matrix $A\in \mathbb R^{d\times d}$, $||A||_2\equiv \sup_{x\neq 0} ||Ax||_2/||x||_2, \forall x\in \mathbb R^{ d}$. 
Observe that the 2-norm $|| X||_2=O(N_S^{d/2})$. Thus $d=3$ is sufficient to suppress the error to $O(N_S^{3/2})$.
Next, we consider $\delta_{N_S}$. Define the eigenvalues of ${\rho}$ as $\{\lambda_i\}_{i=1}^{d}$, and the eigenvalues of ${\tilde{\rho}}$ as $\{\tilde \lambda_i\}_{i=1}^{d}$. The eigenvalues' error  is equal to the Hausdorff distance $|\delta_{N_S}|\equiv\max_i |\tilde\lambda_i-\lambda_i|= hd(\rho,{\tilde{\rho}})$, when the perturbation is small such that the eigenvalues are still pairwise matched: $j=\argmin_{j'}|\tilde \lambda_{j'}-\lambda_j|$.
Note that $||\delta{\rho}||_2=O(N_S^{d+1})$. According to Elsner's theorem~\cite{stewart1990matrix}, the error is bounded above by
\be 
|\delta_{N_S}|\le (||{\rho}||_2+||\tilde{{\rho}}||_2)^{1-1/d}||\delta {\rho}||_2^{1/d}=O(N_S^{1+1/d})\,,
\ee
which is much smaller than $O(N_S)$.
% $o(x)/x\to 0$ as $x\to 0$.
Finally, we see that the overall error $|\delta_d|+|\delta_{N_S}|\ll O(N_S)$ when $d=3$, $N_S\to 0$. 

%\QZ{use small o notation?}\hw{I found small o only needed here so I chose to avoid another definition}

%
%\subsection{Kennedy receiver for thermal coherent state inputs}
%\label{app:Kennedy}
%Consider the discrimination of two thermal coherent state inputs, $H_0: \hat \rho_{0,\overline N}$, $H_1: \hat \rho_{\alpha,\overline N}$. Here $\hat \rho_{\alpha,\overline N}$ is a displaced thermal state with complex displacement $\alpha$ and thermal photon number $\overline N$. The displacement photon number is $|\alpha|^2$. For direct photon detection, the distribution of the photon count $n$ of a displaced thermal state is
%\be 
%P(n|\alpha,\overline N)=\overline N^n (\overline N+1)^{-n-1} e^{-\frac{|\alpha|^2}{\overline N+1}} L_n\left(-\frac{|\alpha|^2}{\overline N (\overline N+1)}\right)\,.
%\ee
%
%The Kennedy receiver \cite{Kennedy_1972} makes decision based on the photon count readout, according to the decision rule is
%\be 
%\tilde H=
%\begin{cases}
%H_0, \quad & n=0\\
%H_1, &{\rm otherwise}
%\end{cases}
%\ee
%Thus the average error rate, assuming equal likelihood of $H_0,H_1$, is
%\be 
%Pr(e)=\frac{1}{2}[1-P(0|0,\overline N)]+\frac{1}{2}P(0|\alpha,\overline N)\,.
%\ee

\subsection{Simulation of the noisy Dolinar receiver}

Here, we generalize the Dolinar receiver described in Sec.~\ref{app:receivers_summary} to the noisy coherent-state case and describe the numerical evaluation of its  performance. 
When there is noise $N_{B,h}$ under both hypotheses, the original candidate states become displaced thermal states, $\hat{\rho}_0=\hat{\rho}_{0,N_{B,0}}$ and $\hat{\rho}_1=\hat{\rho}_{\alpha,N_{B,1}}$. For simplicity, we only consider the case with equal thermal noises $N_{B,0}\simeq N_{B,1} = N_B$. 

When one divides the input state into many slices, as in the Dolinar receiver presented in Sec.~\ref{app:receivers_summary}, the thermal noises in different slices are no longer \jhs{independent, creating} a challenge in numerical performance evaluation. To overcome this problem, we make use of the fact that $\hat{\rho}_h$ has a positive $P$-function, and therefore can be realized by generating random coherent states, $\ket{\alpha_h}$, where $\alpha_0 = r_0$ and $\alpha_1 = \alpha + r_1$ with $\{r_h\}_{h=0}^1$ are complex-valued random numbers whose magnitudes, $|r_h|$, are Rayleigh distributed with mean $\sqrt{\pi N_B/4}$ and whose phases are uniformly distribute on $[0,2\pi]$~\cite{Lachs_1965}. As the measured states are still coherent states, the probability distribution of measured photons also follows a Poisson distribution $N^{(k)}\sim p_N(n,g|h)$,
\begin{equation}
     p_N(n,g|h)= \begin{cases}
    {\rm Pois}(n;|\frac{\alpha_h}{\sqrt{S}}-\gamma+u^{(k)}|^2), & \mbox{if $g=0$}\\[.05in]    {\rm Pois}(n;|\frac{\alpha_h}{\sqrt{S}}-\gamma-u^{(k)}|^2), & \mbox{otherwise}.
    \end{cases}
\end{equation}
Recall that photon number probability distribution for a displaced thermal state $\hat{\rho}_{\alpha,N_B}$ is~\cite{Lachs_1965}
\begin{align}
    P_{\alpha,N_B}(n)
    =& e^{-\frac{|\alpha|^2}{N_B+1}}\frac{N_B^n}{(1+N_B)^{n+1}}
    \nonumber
    \\
    &\quad \times {}_1F_1\left(-n,1,-\frac{|\alpha|^2}{N_B(N_B+1)}\right),
\end{align}
where ${}_1F_1(a,b;z)$ is the confluent hypergeometric function of the first kind. The conditional Bayesian probability of getting $N^{(k)}$ photons is thus
\begin{equation}
    p(N^{(k)},g|h) = \begin{cases}
    P_{\gamma-u^{(k)},N_B/S}(N^{(k)}), & \mbox{if $g=h$}\\[.05in]
    P_{\gamma+u^{(k)},N_B/S}(N^{(k)}), & \mbox{otherwise.}
    \end{cases}
\end{equation}

\subsection{Analyses of channel pattern classification}
\label{general_pattern}

Below we prove the main text's Theorem~1.
Consider an ensemble of composite channels $\Phi_{\bm \kappa^{(h)}, \bm \theta^{(h)}}=\otimes_{\ell=1}^K \Phi_{\kappa_\ell^{(h)}, \theta_\ell^{(h)}}$ with index $1\le h \le r$, each with prior probability $p_\ell$. Before going to detailed analyses, we quote an upper bound on Helstrom limit's error exponent that we use in our proof.

The error exponent of multiple-hypothesis testing is given by the worst-case of binary hypothesis testing between any two of hypotheses involved~\cite{li2016discriminating,nussbaum2011asymptotic,Audenaert2007}.
More precisely, this comes from the following upper bound on the Helstrom limit for hypothesis testing between states $\rho_1^{\otimes n},\cdots, \hat\rho_r^{\otimes n}$ with priors $p_1,\cdots,p_r$) that is true for any number $n\ge 1$ of iid states~\cite{li2016discriminating}:
\begin{align}
&P_H(\{p_1\hat \rho_1^{\otimes n},\cdots, p_r \hat\rho_r^{\otimes n}\})
\le 10(r-1)^2 C_r^2 (n+1)^{2d}
\nonumber
\\
& \quad \times \max\{p_1,\cdots,p_r\}
\max_{i,j}\left({\rm inf}_{s\in[0,1]}Q_s\left(\hat{\rho}_i,\hat{\rho}_j\right)\right)^n,
\label{PH_precise_multi}
\end{align}
which is asymptotically tight in the error exponent when $n\gg1$. Here $C_r^2$ is the binomial coefficient of $r$ choose 2. Although originally derived for finite $d$-dimensional systems, we can choose a cut-off to approximate each mode as a qudit and any finite cutoff will not change the error exponent. Since we are considering Gaussian states, the convergence of such cut-off is guaranteed.

\subsubsection{Classical performance}
Owing to the convexity
of both the Helstrom limit and the quantum Chernoff bound, see Ref.~\cite{zhuang2021quantum}'s supplemental materials, the optimal classical strategy is to utilize a product of coherent states $\hat{\rho}_{\rm C}=\otimes_{\ell=1}^K\ket{\alpha_\ell}$ as the composite channel's inputs, which leads to their outputs being a product of displaced thermal states
$
\otimes_{\ell=1}^K \hat{\rho}_{\exp\left(i\theta^{(h)}_\ell\right)\sqrt{\kappa^{(h)}_\ell}\alpha_\ell, N_B}
$ 
after going through the channel $\Phi_{\bm \kappa^{(h)}, \bm \theta^{(h)}}$.
When one has a large number of input  coherent states, the iid nature of the output states allows us to focus on the quantum channel discrimination (QCD) between the worst pair of two channels $\Phi_{\bm \kappa^{(h)}, \bm \theta^{(h)}}$, with $h=1,2$. For the two displaced thermal state, we have the mean
\begin{align}
    \overline{\bm x}_h &= \left(q_1^{(h)},p_1^{(h)},\dots,q_K^{(h)},p_K^{(h)}\right)^T,
\end{align}
where $q_\ell^{(h)}=2\Re{e^{i\theta_\ell^{(h)}}\sqrt{\kappa_\ell^{(h)}}\alpha_\ell}$ and $p_\ell^{(h)}=2\Im{e^{i\theta_\ell^{(h)}}\sqrt{\kappa_\ell^{(h)}}\alpha_\ell}$. The covariance matrix is diagonal $B\mathbb{I}$ where $B\equiv 2N_B+1 $. Now we evaluate the quantum Chernoff bound according to Section~\ref{app:QCB}. First, the quantity
\begin{equation}
\begin{split}
    Q_s &= \overline{Q}_se^{-\frac{1}{2}{\bm d}^T[\tilde{V}_1(s)+\tilde{V}_2(1-s)]^{-1}{\bm d}}\\
    &= \overline{Q}_s e^{-\frac{1}{2(\Lambda_s(B)+\Lambda_{1-s}(B))}{\bm d}^T{\bm d}}\\
    &= \overline{Q}_s e^{-\frac{1}{2(\Lambda_s(B)+\Lambda_{1-s}(B))}\sum_{\ell=1}^K (q_\ell^{(1)}-q_\ell^{(2)})^2+(p_\ell^{(1)}-p_\ell^{(2)})^2}\\
    &= \overline{Q}_s e^{-\frac{2}{\Lambda_s(B)+\Lambda_{1-s}(B)}\sum_{\ell=1}^K |e^{i\theta_\ell^{(1)}}\sqrt{\kappa_\ell^{(1)}}\alpha_\ell-e^{i\theta_\ell^{(2)}}\sqrt{\kappa_\ell^{(2)}}\alpha_\ell|^2}\\
    &= \overline{Q}_s e^{-\frac{2}{\Lambda_s(B)+\Lambda_{1-s}(B)}\sum_{\ell=1}^K \delta_\ell|\alpha_\ell|^2},
    \label{eq:Qs}
\end{split}
\end{equation}
where we define $\delta_\ell \equiv|e^{i\theta^{(1)}_\ell}\sqrt{\kappa^{(1)}_\ell}-e^{i\theta^{(2)}_\ell}\sqrt{\kappa^{(2)}_\ell}|^2$ and
\begin{equation}
    \overline{Q}_s = 2^K\left(\frac{ G_s(B)G_{1-s}(B)}{\Lambda_s(B)+\Lambda_{1-s}(B)}\right)^K = 1.
    \label{eq:Qs_2}
\end{equation}
Therefore, the quantum Chernoff bound is
\begin{equation}
\begin{split}
    P_{\rm QCB}^{\rm QCD} &= \frac{1}{2}{\rm inf}_{s\in[0,1]}Q_s\\
    & = \frac{1}{2}Q_{1/2}\\
    &= \frac{\overline{Q}_{1/2}}{2}\exp\left[-\frac{1}{\Lambda_{1/2}(B)}\sum_{\ell=1}^K \delta_\ell|\alpha_\ell|^2\right]\\
    &= \frac{1}{2}\exp\left[-\frac{1}{\Lambda_{1/2}(B)}\sum_{\ell=1}^K \delta_\ell|\alpha_\ell|^2\right]\\
    &= \frac{1}{2}\exp\left[-\sum_{\ell=1}^K \delta_\ell|\alpha_\ell|^2(\sqrt{N_B+1}-\sqrt{N_B})^2\right], \label{eq:channel_error}
\end{split}
\end{equation}
where in the second line we utilized the fact that the minimum of $Q_s$ takes place at $s=1/2$. This is so because $\overline{Q_s}$ in \eqref{eq:Qs_2} is independent of $s$ and the exponent $1/(\Lambda_s(B)+\Lambda_{1-s}(B))$ in \eqref{eq:Qs} is symmetric in $s$ and strictly concave as its \jhs{second derivative} is negative
\begin{align}
    &\partial_s^2 \left(\frac{1}{\Lambda_s(B)+\Lambda_{1-s}(B)}\right)\nonumber \\ 
    &= -\frac{1}{4}\frac{(B+1)^{2s-1}+ (B-1)^{2s-1}}{(B^2-1)^{s-1}}\log ^2\left(\frac{B-1}{B+1}\right) < 0,
\end{align}
due to $B > 1$. Therefore $\max_{s\in[0,1]}\{1/(\Lambda_s(B)+\Lambda_{1-s}(B))\} = 1/2\Lambda_{1/2}(B)$ where ${\rm inf}_{s\in[0,1]}Q_s = Q_{1/2}$.

To conclude, we have that the error exponent for discrimination between any two channels via coherent-state inputs follows from
\jhs{\begin{eqnarray}
    \lefteqn{-\ln\left[P_{\rm CS}^{\rm QCD}\right]/M \sim }\nonumber \\[.05in]
    &-&\left[\sum_\ell \delta_\ell |\alpha_\ell|^2\left(\sqrt{N_B+1}-\sqrt{N_B}\right)^2\right]/M \nonumber \\[.05in] 
    &&\,\,\sim -\left[\sum_\ell \delta_\ell |\alpha_\ell|^2/4N_B\right]/M,
    \label{CS:exponent_pattern}
\end{eqnarray}}
where we have used 
\jhs{\begin{eqnarray}
\left(\sqrt{N_B+1}-\sqrt{N_B}\right)^2 &=& N_B(\sqrt{1+1/N_B}-1)^2 \nonumber\\[.05in] 
&=& N_B(1/2N_B+O(N_B^2))^2 \nonumber \\[.05in]
&\simeq& 1/4N_B, \mbox{ for $N_B \gg 1$.}
\end{eqnarray}} 
Note that this error exponent is asymptotically tight for both $M\rightarrow 
\infty$ of the input coherent states, or equivalently for  $|\alpha_\ell|^2 \gg 1$.

\subsubsection{Entanglement-assisted performance}
For the entangled strategy, one inputs a product of TMSV states, with each signal and idler having mean photon number $ N_{S,\ell}=|\alpha_\ell|^2/M$, such that the total energy in $M$ iid copies matches that of the classical input. For each channel, we have $M$ identical TMSV mode pairs. Combining the $K$ channels, the overall output state has the identical copy form of $\{\rho_h^{\otimes M}\}_h$, where each $\rho_h$ has $K$ modes. Once we apply the ${\rm C}\to {\rm D}$ conversion to the output state, the output states are no longer because the measurement outcomes on differnt mode pairs are statistically independent. The way out of such a dilemma is the following: 
when provided with $M$  output states, $\{\rho_h^{\otimes M}\}_h$, from the quantum channel, perform ${\rm C}\to {\rm D}$ conversion on each sub-channel output, thus producing a product of displaced thermal states $\{\hat\sigma_h\}_h$, where each state has $KM$ modes and
\be 
\hat \sigma_h=\otimes_{\ell=1}^K \left[\otimes_{m=1}^M \hat{\rho}_{\zeta_\ell^{(h)} e^{ i \theta_\ell^{(h)} }{\calM_{\ell,m}}^{*},E_\ell^{(h)}}\right]
\label{sigma_h}
\ee   
is conditioned on the measurement results $\{\calM_{\ell,m}, 1\le \ell\le K, 1\le m \le M\}$. The constant's definition is
\begin{align}
\zeta_\ell^{(h)}\equiv   \frac{\sqrt{\kappa_\ell^{(h)}  N_{S,\ell} \left(N_{S,\ell}+1\right)}}{N_B+\kappa_\ell^{(h)}  N_{S,\ell}+1},
\end{align} 
and the mean thermal photon number is 
\be 
E_\ell^{(h)}=\frac{N_{S,\ell} \left(N_B-\kappa_\ell^{(h)} +1\right)}{ N_B+\kappa_\ell^{(h)}  N_{S,\ell}+1}.
\ee
The measurement result ${\calM_{\ell,m}}=(q_{\calM_{\ell,m}}+ip_{\calM_{\ell,m}})/2$, with each quadrature output having a zero-mean Gaussian distribution with variance 
$
(N_B + \kappa_\ell^{(h)} N_S+1)/2\simeq N_B/2
$.
We consider the $N_B\gg1$ limit, then $E_\ell^{(h)}\simeq N_{S,\ell}\simeq N_S$ is a constant noise background.

Because ${\rm C}\to {\rm D}$ conversion is a quantum process, we have the following Helstrom limit for the entangled-input case: 
\begin{align}
&P_H(\{p_h\rho_h^{\otimes M}\})
\le 
\mathbb{E}[P_H(\{p_h\sigma_h\})]
\le 10(r-1)^2 C_r^2 2^{2d}
\nonumber
\\
& \quad \times \max\{p_1,\cdots,p_r\}
\mathbb{E}\max_{i,j}
\left[{\rm inf}_{s\in[0,1]}Q_s\left(\hat{\sigma}_i,\hat{\sigma}_j\right)\right],
\label{PH_UB}
\end{align}
where the expectation is over the measurement statistics. In the last step, we applied (\ref{PH_precise_multi}) with the $M=1$ case and provide an upper bound on the error probability directly. Then we can still reduce the calculation to the Chernoff exponents for the displaced thermal states. However, now each state $\hat\sigma_h$ has $KM$ modes and are dependent on the measurement result.

Now we consider the binary exponent ${\rm inf}_{s\in[0,1]}Q_s\left(\hat{\sigma}_i,\hat{\sigma}_j\right)$ in (\ref{PH_UB}).
Conditioned on the measurement result, following \eqref{eq:channel_error}, we have
\small
\begin{equation}
\begin{split}
    &{\rm inf}_{s\in[0,1]}Q_s\left(\hat{\sigma}_i,\hat{\sigma}_j\right)
   \simeq 
   \\
   & \exp\left[- \frac{ \sum_\ell N_{S,\ell}(N_{S,\ell}+1)\delta_\ell^{(i,j)}\sum_{m=1}^M|{\calM_{\ell,m}}^*|^2}{(N_B+1)^2(\sqrt{N_{S,\ell}+1}+\sqrt{N_{S,\ell}})^2}\right]
   \\
   &\simeq \exp\left[- n\frac{ \sum_\ell N_{S,\ell}\delta_\ell^{(i,j)}}{N_B}\right],
\end{split}
\end{equation}
\normalsize
where $\delta_\ell^{(i,j)} \equiv|e^{i\theta^{(i)}_\ell}\sqrt{\kappa^{(i)}_\ell}-e^{i\theta^{(j)}_\ell}\sqrt{\kappa^{(j)}_\ell}|^2$. In the asymptotic limit of $M\gg1$, $\sum_{m=1}^M|{\calM_{\ell,m}}^*|^2$ converges to its mean $ M N_B$, and therefore we have the last step of our approximation. We have also applied the asymptotic limit $N_B\gg1$ and $N_S\ll1$.

Then,  from \eqref{PH_UB} we have
\begin{align}
&P_H(\{p_h\rho_h^{\otimes M}\})
\le 
10(K-1)^2 C_K^2 2^{2d}
\nonumber
\\
& \quad \times \max\{p_1,\cdots,p_K\}
\max_{i,j}
\exp\left[- M\frac{ \sum_\ell N_{S,\ell}\delta_\ell^{(i,j)}}{N_B}\right].
\end{align}

Comparing \eqref{CS:exponent_pattern} to the above result with $MN_{S,\ell}=|\alpha_\ell|^2$ to match the signal energies used by the quantum and classical sources, we have proved the main text's Theorem~1: broadband TMSV-enabled EA channel classification enjoys a 6\,dB error-exponent advantage over its classical counterpart of the same transmitted energy.

\subsection{Near-Optimal Mode Selection}
\label{sec:mode_selection}

Here we will derive the filtering efficiencies and average output-noise photon numbers for the filter and chirp-disperse-gate near-optimal mode selectors considered in the main text's Sec.~\ref{SelectorDesign}.

For both architectures, $N_I$, the average displacement photon number at their inputs is
\begin{align}
N_I &= \zeta^2\langle N_R\rangle = \zeta^2\int_{\mathcal{T}_0}\!{\rm d}t\,\langle |E_R(t)|^2\rangle =  \zeta^2\int_{\mathcal{T}_0}\!{\rm d}t\,\langle A^2_R(t)\rangle \\[.05in]
&= \frac{M\kappa N_S(N_S+1)}{\kappa N_S+N_B + 1}.
\end{align}
The filter architecture's average displacement photon number at its output is easily seen to be
\begin{align}
N_d &=  \int_{-\pi W'}^{\pi W'}\!\frac{{\rm d}\omega}{2\pi}\int_{\mathcal{T}_0}\!{\rm d}u\int_{\mathcal{T}_0}\!{\rm d}v\,\zeta^2\langle A_R(u)A_R(v)\rangle  e^{i\omega(u-v)}\\[.05in]
&=   \int_{\mathcal{T}_0}\!{\rm d}u\int_{\mathcal{T}_0}\!{\rm d}v\,\zeta^2\langle A_R(u)A_R(v)\rangle W'\,{\rm sinc}[\pi W'(u-v)].
\label{filterstart}
\end{align}
From Eq.~(9.22) of Ref.~\cite{middleton1960introduction} we have
\begin{align}
\langle A_R(u)A_R(v)\rangle &= \frac{\pi (\kappa N_S + N_B + 1)W}{4} \nonumber \\[.05in]
& \times {}_2F_1[-1/2,-1/2;1;{\rm sinc}^2[\pi W(u-v)]],
\end{align}
where ${}_2F_1[-1/2,-1/2;1; {\rm sinc}^2[\pi W(u-v)]]$ is a hypergeometric function.  Substituting this result into Eq.~(\ref{filterstart}), changing the integration variables to $z \equiv (u+v)/2$ and $\tau \equiv u-v$, we can do the $z$ integral and obtain $N_d = \eta N_I$ with
\begin{align}
\eta &= \frac{\pi}{4}\int_{-T}^T\!{\rm d}\tau\,(1-|\tau|/T)\,{}_2F_1[-1/2,-1/2;1;{\rm sinc}^2(\pi W\tau)] \nonumber \\[.05in]
& \times W'\,{\rm sinc}^2(\pi W'\tau).
\end{align}

For the filter architecture's $N_n$, the average noise photon number at its output, we can parallel what we just did.  We have
\jhs{\begin{align}
N_n &=  \int_{-\pi W'}^{\pi W'}\!\frac{{\rm d}\omega}{2\pi}\int_{\mathcal{T}_0}\!{\rm d}u\int_{\mathcal{T}_0}\!{\rm d}v\,\langle \hat{E}^{\prime \dagger}_n(u)\hat{E}'_n(v)\rangle  \langle \rangle  \nonumber \\[.05in]
& \times e^{-i[\phi_R(u)-\phi_R(v)]}e^{i\omega(u-v)}\\[.05in]
&=   \int_{\mathcal{T}_0}\!{\rm d}u\int_{\mathcal{T}_0}\!{\rm d}v\ N_{I\mid R}W\,{\rm sinc}[\pi W(u-v)]\langle \rangle \nonumber \\[.05in]
&\times e^{-i[\phi_R(u)-\phi_R(v)]}W'\,{\rm sinc}[\pi W'(u-v)].
\label{filternoise}
\end{align}}
Now, using Eq.~(9.32) from Ref.~\cite{middleton1960introduction}, we have
\begin{align}  
&\langle e^{-i\phi_R(u)+i\phi_R(v)}\rangle \nonumber\\[.05in]
&=\int_0^{2\pi}\!{\rm d}\phi \int_0^{2\pi}\!{\rm d}\phi' p_{\phi_R(u),\phi_R(v)}(\phi,\phi')\exp{-i(\phi-\phi')}, 
\end{align}
where
\begin{align}
p_{\phi_R(u),\phi_R(v)}(\phi,\phi')
&=
\frac{1-{\rm sinc}^2[\pi W(u-v)]}{4\pi^2} (1-\beta^2)^{-3/2}\nonumber \\[.05in]
&\times\left(\!\beta \sin^{-1}(\beta)+\!\frac{\pi\beta}{2}\!+\!\sqrt{1-\beta^2}\right),
\end{align}
with $\beta \equiv {\rm sinc}[W(u-v)]\cos(\phi-\phi')$, is the joint probability density function (pdf) of $\phi_R(u)$ and $\phi_R(v)$.  Because this pdf is an even function of $\phi-\phi'$ and depends only on $\tau = u-v$, we can simplify Eq.~(\ref{filternoise}) to
\begin{align}
N_n &= MN_{I\mid R}\int_{-T}^T\!{\rm d}\tau\, (1-|\tau|/T)\,{\rm sinc}(\pi W\tau)\nonumber \\[.05in]
&\times R_{\phi_R\phi_R}(\tau)W'\,{\rm sinc}(\pi W'\tau),
\end{align} 
where 
\begin{equation}
R_{\phi_R\phi_R}(\tau) = \int_{-2\pi}^{2\pi}\!{\rm d}\phi\,\cos(\phi)p_{\Delta\phi_R(\tau)}(\phi),
\end{equation}
with
\begin{align}
p_{\Delta\phi_R(\tau)}(\phi)
&=
\left(1-\frac{|\phi|}{2\pi}\right)\!\frac{1-{\rm sinc}^2(\pi W\tau)}{4\pi^2} (1-\beta'^2)^{-3/2}\nonumber \\[.05in]
&\times\left(\!\beta' \sin^{-1}(\beta')+\!\frac{\pi\beta'}{2}\!+\!\sqrt{1-\beta'^2}\right),
\end{align}
for $\beta'= {\rm sinc}(\pi W\tau)\cos(\phi)$, being the pdf of $\Delta\phi_R(\tau) = \phi_R(t+\tau)-\phi_R(t)$.

Our final tasks are to find the chirp-disperse-gate architecture's $\eta$ and $N_n$.  The starting points for those evaluations are
\begin{align}
N_d &=  \frac{W_c}{T}\int_{-T'/2}^{T'/2}\!{\rm d}t\int_{\mathcal{T}_0}\!{\rm d}u\int_{\mathcal{T}_0}\!{\rm d}v\,\zeta^2\langle A_R(u)A_R(v)\rangle\nonumber \\[.05in]  
& \times e^{-i2\pi W_c(u-v)t/T}\\[.05in]
&=  \frac{W_cT'}{T} \int_{\mathcal{T}_0}\!{\rm d}u\int_{\mathcal{T}_0}\!{\rm d}v\,\zeta^2\langle A_R(u)A_R(v)\rangle \nonumber \\[.05in]
& \times {\rm sinc}[\pi W_cT'(u-v)/T],
\label{chirpstart}
\end{align}
and
\begin{align}
N_n &= \frac{W_c}{T} \int_{-T'/2}^{T'/2}\!{\rm d}t\,\int_{\mathcal{T}_0}\!{\rm d}u\int_{\mathcal{T}_0}\!{\rm d}v\,\langle \hat{E}^{\prime \dagger}_n(u)\hat{E}'_n(v)\rangle   \nonumber \\[.05in]
& \times \langle e^{-i[\phi_R(u)-\phi_R(v)]}\rangle e^{-i2\pi W_c(u-v)t/T} \\[.05in]
&= \frac{W_cT'}{T} \int_{\mathcal{T}_0}\!{\rm d}u\int_{\mathcal{T}_0}\!{\rm d}v\,N_{I\mid R}W\,{\rm sinc}[\pi W(u-v)] \nonumber \\[.05in]
& \times \langle e^{-i[\phi_R(u)-\phi_R(v)]}\rangle\, {\rm sinc}[\pi W_cT'(u-v)/T],
\end{align}
where we have assumed $\mathcal{T}_0 = [-T/2,T/2]$ in setting the limits of the ${\rm d}t$ integral.  These results are identical to the filter architecture's Eqs.~(\ref{filterstart}) and (\ref{filternoise}) when $W_cT'/T = W'$, i.e., when both architectures filter down to the same time-bandwidth product, i.e., $TW'$ equals $ T'W_c$.

\end{document}